\documentclass[12pt]{iopart}

\expandafter\let\csname equation*\endcsname\relax
\expandafter\let\csname endequation*\endcsname\relax
\usepackage{amsmath}

\usepackage{iopams}
\usepackage{graphicx,subcaption,amsmath,amssymb}
\usepackage{physics, mathtools, braket}
\usepackage{mathdots}
\usepackage{kbordermatrix}
\renewcommand{\kbldelim}{(}
\renewcommand{\kbrdelim}{)}

\usepackage[final]{showkeys}

\newcommand{\ii}{\mathrm{i}}
\newcommand{\ee}{\mathrm{e}}
\newcommand{\tot}{\textrm{tot}}
\newcommand{\ntot}{n^\tot}

\newcommand{\PT}{\mathcal{PT}}

\usepackage{xcolor}

\begin{document}

\title{Quantum transport on Bethe lattices with non-Hermitian sources and a drain}

\author{Naomichi Hatano$^1$\footnote{Corresponding Author: \texttt{hatano@iis.u-tokyo.ac.jp}}, Hosho Katsura$^{2,3,4}$, and Kohei Kawabata$^{5}$}

\address{$^1$ Institute of Industrial Science, The University of Tokyo, Kashiwanoha, Kashiwa, Chiba 277-8574, Japan}
\ead{hatano@iis.u-tokyo.ac.jp}

\address{$^2$ Department of Physics, The University of Tokyo, Hongo, Bunkyo-ku, Tokyo 113-0033, Japan}
\address{$^3$ Institute for Physics of Intelligence, The University of Tokyo, 
Hongo, Bunkyo-ku, Tokyo 113-0033, Japan}
\address{$^4$ Trans-scale Quantum Science Institute, The University of Tokyo, 
Hongo, Bunkyo-ku, Tokyo 113-0033, Japan}
\ead{katsura@phys.s.u-tokyo.ac.jp}

\address{$^5$ Institute for Solid State Physics, University of Tokyo, Kashiwa, Chiba 277-8581, Japan}
\ead{kawabata@issp.u-tokyo.ac.jp}

\begin{abstract}
We consider quantum transport in a tight-binding model on the Bethe lattice of finite generation, or the Cayley tree, which we expect to be the first step toward analyzing electronic transport in a light-harvesting molecule.
We seek conditions under which the electronic current from the peripheral light-harvesting sites to the central site reaches its maximum.
As a new feature for analyzing quantum transport, we add complex potentials for sources at peripheral sites and a drain at the central site,
and solve a non-Hermitian eigenvalue problem, instead of simulating an initial-value problem. 
Solving the eigenvalue problem clearly reveals which electronic channels contribute most to the quantum transport.
Indeed, we find that the number of eigenstates that can penetrate from the peripheral sites to the central site is quite limited among the total number of eigenstates.
All the other eigenstates are localized around the peripheral sites and cannot reach the central site.
The former eigenstates can carry current, reducing the problem to quantum transport on a parity-time ($\PT$)-symmetric tight-binding chain. 
We find that the current has a maximum with respect to the strengths of the sources and the drain.
Counterintuitively, the current decreases as we increase the strengths beyond the maximum and vanishes in the limit of infinite strength.
Moreover, we find that the current maximum is given by a zero mode (a zero-energy eigenstate).
When the number of links is common to all generations, the current takes the maximum value at the exceptional point where two eigenstates coalesce to a zero mode, which emerges because of the non-Hermiticity due to the $\PT$-symmetric complex potentials.
By introducing randomness either into the hopping amplitude or the number of links in each generation of the tree, we obtain a random-hopping tight-binding model, and find that the current reaches its maximum not exactly, but approximately, for a zero mode, although it is no longer located at an exceptional point in general.
\end{abstract}
\maketitle

\section{Introduction}
\label{sec1}

Quantum transport has been one of the central issues in non-equilibrium physics; see \textit{e.g.} Ref.\ \cite{NoneqReview24}. 
The Landauer~\cite{Landauer57, Datta95} formula well describes the conductance due to one-electron transport through various structures attached to leads as a potential-scattering problem.
Studies based on the Landauer formula have often been conducted in one-dimensional or quasi-one-dimensional models, because the research motivation has mostly focused on quantum transport through semiconductor structures, such as quantum leads with quantum dots.

Partly motivated by observations of exciton transfer in natural light-harvesting molecules~\cite{Engel07, Panitchayangkoon10, Collini10, Mohseni14}, an article~\cite{Novo16} considered quantum transport through the Bethe lattice, or the Cayley tree.
In fact, quantum transport on graphs and networks is gathering more attention based on various motivations; see \textit{e.g.} Refs.~\cite{Mulken06, Rebentrost09, Jackson12, MaquinBatalha22, Silva24}.

Stimulated by these studies, we here consider a tight-binding model on a tree-like network; see Fig.~\ref{fig1}.
We seek conditions under which the electronic current from the peripheral light-harvesting sites to the central site reaches its maximum.
The system may appear simplistic, but we claim that it is a good starting point for studying more complex systems, as it is exactly solvable.

A new feature of the present study is that the model has sources at the peripheral sites and a drain at the central site, represented by complex potentials, making the system non-Hermitian.
Another new feature of the present paper is that we solve the eigenvalue problem rather than simulate the initial-value problem.
Thanks to the model's simplicity, we find all eigenstates almost analytically, allowing us to clearly see which channels produce the most current.

In fact, analysis of the eigenvalue equation of the system in Fig.~\ref{fig1} shows that the problem itself is an interesting mathematical-physical one in its own right. 
We find that the eigenstates that can penetrate from the peripheral sites to the central site are quite limited; most eigenstates cannot reach the central site.
The current is carried by the former eigenstates.
This may be common to electronic conduction on fractal systems.

We discover the following two points.
First, as we introduce and increase the complex potentials representing the sources and the drain, the current from the peripheral light-harvesting sites to the central site initially increases, then reaches a maximum, beyond which it converges to zero counterintuitively.
This is also observed in the calculation of electronic conduction using the Landauer formula in a molecular junction~\cite{Toroker09}. 
This suggests that this is a general feature of electronic conduction in various types of molecular systems, and should be observed experimentally.

The second point that we discover is that the current maximum is attained by a zero mode (the zero-eigenvalue state).
When the number of links is the same for all generations, the current takes the maximum value at an exceptional point where two eigenstates coalesce to a zero mode. 
The zero mode occupies a central position in various areas of physics due to symmetry.
The present model is yet another example of the zero mode playing an important role.
These two findings suggest that these can be general features of electronic transfer in even more complex structures.

The paper is structured as follows.
Section~\ref{sec2} introduces the model and the notations.
Section~\ref{sec3} proves that most of the eigenstates are localized on the side of the peripheral sites and do not reach the central site.
Section~\ref{sec4} then presents the eigenvalues and eigenstates of the remaining extended states.
Based on the results, we finally compute the current carried by the extended conducting states in Sec.~\ref{sec5}.
We further investigate electronic conduction in Sec.~\ref{sec6}, introducing randomness into the system.
Section~\ref{sec7} provides a summary.
In \ref{appA}, we consider the definition of the expectation value, particularly regarding the usage of the left eigenvector. 
We claim that the definition depends on whether we regard the non-Hermitian system as an open quantum system or a closed non-Hermitian system.
We consider the former throughput the paper.
In \ref{appB}, we present a solution to the scattering problem through a non-Hermitian dot, which may justify introducing the non-Hermitian potentials.

\section{Tight-binding model on a tree-like network}
\label{sec2}
\begin{figure}
\centering
\includegraphics[width=0.8\textwidth]{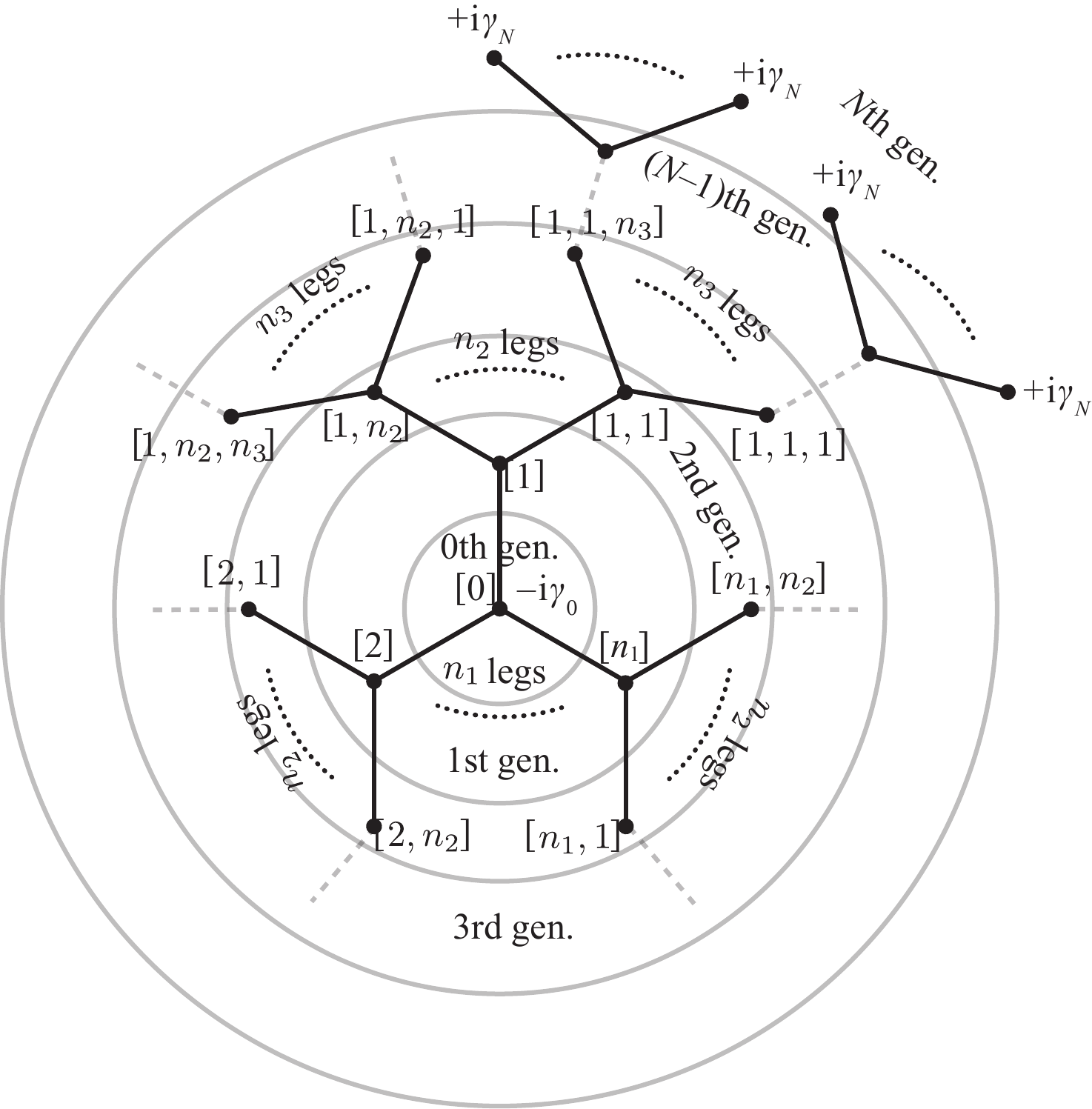}
\caption{Schematic view of the tree-like network. The source potentials $+\ii\gamma_N$ with $\gamma_N>0$ are added to the peripheral sites, while the drain potential $-\ii\gamma_0$ with $\gamma_0>0$ is added to the central site $0$.}
\label{fig1}
\end{figure}

The model that we consider in the present paper is the tight-binding model on a Bethe lattice of a finite generation, 
or the Cayley tree, which is schematically shown in Fig.~\ref{fig1} in a general case.
Let us fix the hopping elements to $-1$ to make it the unit of energy.
We also set $\hbar=1$ hereafter.

We use the following naming convention, unless otherwise noted.
There is only one origin site, $[0]$, in the zeroth generation.
The drain potential $-\ii\gamma_0$ with $\gamma_0>0$ is applied to the origin site.
For consistency, we let the total number of sites in the zeroth generation be denoted by $\ntot_0=n_0$, which is equal to unity.
Connected to the origin site in the zeroth generation are the $n_1$ sites in the first generation, which are referred to as
$[1]$, $[2]$, $[3]$, $\ldots$, $[n_1]$.
We express the total number of sites in the first generation as $\ntot_1\coloneqq n_0n_1=n_1$.
Connected to the sites in the first generation, $[\mu]$ (for $\mu=1,2,3,\ldots, n_1$), 
are $n_2$ pieces of sites in the second generation, which are referred to as $[\mu,1]$,\ \ \ $[\mu,2]$,\ \ \  $[\mu,3]$, \ \ \ $\ldots$,\ \ \  $[\mu,n_2]$.
Note that, in the most general case, the number of links for each site can vary across sites, but we set them equal in each generation in the present paper.
The total number of sites in the second generation is hence given by
$\ntot_2=\ntot_1n_2=n_0n_1n_2$.
%
Connected to the second generation sites, $[\mu,\nu]$ (for $\mu=1,2,3,\ldots, n_1$ and $\nu=1,2,3,\ldots,n_2$), are $n_3$ pieces of the third generation sites, named $[\mu,\nu,1]$, $[\mu,\nu,2]$, $\ldots$, $[\mu,\nu,n_3]$.
The total number of sites in the second generation is given by
$\ntot_3=\ntot_2n_3=n_0n_1n_2n_3$.

In short, each site is denoted by a sequence of numbers that uniquely determines the path from the origin to the site in question, such as $[\mu,\nu,\kappa,\lambda,\rho,\sigma,\ldots]$.
The number of links that connect each site in the $(\ell-1)$th generation and the sites in the $\ell$th generation is denoted by $n_\ell$. The total number of sites in the $\ell$th generation is given by
\begin{align}
\ntot_\ell=\prod_{m=0}^\ell n_m.
\end{align}
We continue this process up to the $N$th generation, and make them the peripheral sites.
On each of the peripheral sites in the $N$th generation, we apply the source potential $\ii\gamma_N$ with $\gamma_N>0$.
We let
\begin{align}\label{eq40}
\ntot\coloneqq\sum_{\ell=0}^N \ntot_\ell
\end{align}
denote the total number of sites in the lattice.

The tight-binding Hamiltonian on the network up to the $N$th generation is therefore given 
by
\begin{align}\label{eq50}
H_\tot\coloneqq&-\ii\gamma_0\dyad{[0]}
\nonumber\\
&-\sum_{\mu=1}^{n_1}\qty\Big(\dyad{[0]}{[\mu]}+\dyad{[\mu]}{[0]})
-\sum_{\mu=1}^{n_1}\sum_{\nu=1}^{n_2}\qty\Big(\dyad{[\mu]}{[\mu,\nu]}+\dyad{[\mu,\nu]}{[\mu]})
\cdots
\nonumber\\
&-\sum_{\mu_1=1}^{n_1}\sum_{\mu_2=1}^{n_2}\sum_{\mu_3=1}^{n_3}\cdots\sum_{\mu_{N-1}=1}^{n_{N-1}}\sum_{\mu_N=1}^{n_N}
\qty\Big(\dyad{[\mu_1,\ldots,\mu_{N-1}]}{[\mu_1,\ldots,\mu_{N-1},\mu_N]} + \textrm{H.c.}) \nonumber\\
&+\ii\gamma_N\sum_{\mu_1=1}^{n_1}\sum_{\mu_2=1}^{n_2}\sum_{\mu_3=1}^{n_3}\cdots\sum_{\mu_{N-1}=1}^{n_{N-1}}\sum_{\mu_N=1}^{n_N}\dyad{[\mu_1,\ldots,\mu_{N-1},\mu_N]},
\end{align}
where we always assume $\gamma_0> 0$ and $\gamma_N>0$. 
The complex potentials on the peripheral sites $\ii\gamma_N$ and the central site $-\ii\gamma_0$ can emerge when we regard this system as an open quantum system embedded in a much larger system, often called the environment.
Tracing out the degrees of freedom of the environments attached to the peripheral sites and the central site yields effective complex potentials as a result of the Feshbach projection formalism~\cite{Hatano14}; see \ref{appA}. 
The complex potentials generally depend on the environment's energy, which reflects the non-Markovianity of the dynamics~\cite{Hatano21}.
The present constant complex potentials can arise
after eliminating the energy dependence of the effective potential, which in turn means the Markov approximation of the dynamics.
Another way to justify the constant effective potential is as follows.
If we approximate the dispersion relation of the Fermi particles in the environment as a linear one around the Fermi level with the bottom of the energy band far away from the energy range of interest, the resulting effective potential does not depend on energy~\cite{Nakabayashi25} and the dynamics becomes Markovian~\cite{Nishino24, Taira24}.

The model~\eqref{eq50} is a one-body model, and hence the total number of eigenstates of the Hamiltonian~\eqref{eq50} is equal to the total number of sites, $\ntot$ in Eq.~\eqref{eq40}.
In Sec.~\ref{sec3}, we first show that $(\ntot_N-\ntot_{N-1})$ pieces of eigenstates are strictly localized on the peripheral sites (the $N$th generation).
We then show that $2\qty(\ntot_{N-1}-\ntot_{N-2})$ pieces of eigenstates have amplitudes only on the $N$th and $(N-1)$th generations.
This goes on; 
we generally show that $(N+1-\ell)\qty(\ntot_{\ell}-\ntot_{\ell-1})$ pieces of eigenstates have amplitudes from the $N$th generation up to the $\ell$th generation with $\ell=N,N-1,N-2,\ldots,2,1$.
In other words, the total number of eigenstates that have null amplitude at the origin site (the zeroth generation) is given by
\begin{align}\label{eq300-1}
\sum_{\ell=1}^N (N+1-\ell)\qty(\ntot_{\ell}-\ntot_{\ell-1})&=
\sum_{\ell=1}^N (N+1-\ell)\ntot_\ell - \sum_{\ell=0}^{N-1} (N-\ell) \ntot_\ell
\notag\\
&=\sum_{\ell=1}^{N-1} \ntot_\ell +\ntot_N - N\ntot_0
=\ntot-\qty(N+1),
\end{align}
where we used the notation $\ntot_0=1$ and Eq.~\eqref{eq40}.
Since the total number of eigenstates is equal to the total number of sites, $\ntot$, as we stated above, Eq.~\eqref{eq300-1} means that we miss
$N+1$ pieces of eigenstates. 
In fact, we find in Sec.~\ref{sec4} that the remaining $N+1$ pieces of eigenstates penetrate up to the origin site and contribute to the quantum transport.
This completes the Hilbert space of the tight-binding model on the Bethe lattice.
These $N+1$ delocalized eigenstates are obtained from the eigenstates of the tight-binding model on a linear chain with $N+1$ sites.

\section{Localized eigenstates}
\label{sec3}

In this section, we list all localized states with the null element at the origin.
We start with the ones localized on the $N$th generation, then move to the ones localized on the $N$th and $(N-1)$th generations, and finally generalize the argument.

\subsection{Eigenstates localized on the peripheral sites}
\label{subsec3.1}

\begin{figure}
\centering
\includegraphics[width=0.4\textwidth]{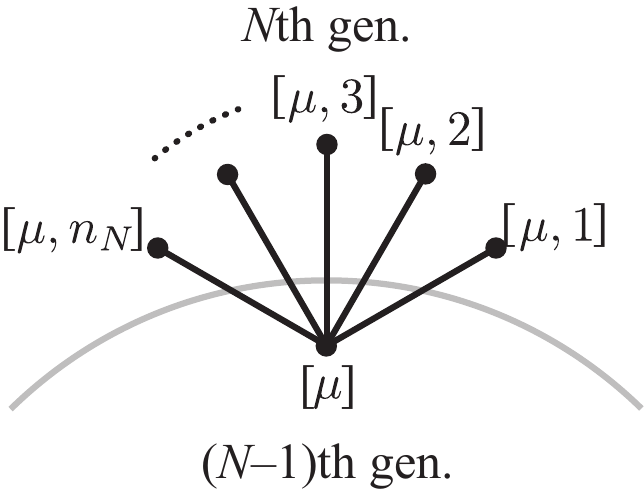}
\caption{One branch of the tree lattice with the root site $\mu$ in the $(N-1)$th generation and $n_N$ pieces of peripheral sites in the $N$th generation.}
\label{fig2}
\end{figure}

Let us first consider one site in the $(N-1)$th generation $[\mu_1,\mu_2,\ldots,\mu_{N-1}]$ and the attached peripheral sites $[\mu_1,\mu_2,\ldots,\mu_{N-1},\mu_N]$ with $\mu_N=1,2,3,\ldots,n_N$ in the $N$th generation; see Fig.~\ref{fig2}.
For brevity, let us refer to them as $[\mu]$ and $[\mu,\nu]$ with $\nu=1,2,3,\ldots,n_N$ in the present subsection.
For a fixed value of $\mu$ in the range of $1\le\mu\le n_{N-1}$, the nonzero matrix elements of the Hamiltonian of the portion indicated in Fig.~\ref{fig1} 
are given by
\begin{align}\label{eq20}
\braket{[\mu]|H_{N-1}|[\mu,\nu]}&=\braket{[\mu,\nu]|H_{N-1}|[\mu]}=-1
\quad\mbox{and}\quad
\braket{[\mu,\nu]|H_{N-1}|[\mu,\nu]}=+\ii\gamma_N
\end{align}
for $1\le\nu\le n_N$. For readers' reference, let us also give
the matrix representation of the Hamiltonian of this portion:
\begin{align}\label{eq20mat}
&\begin{matrix}
&&
[\mu] &
[\mu,1] &
[\mu,2] &
\cdots &
[\mu,n_N]
\end{matrix}
\notag\\
H_{N-1} =
\begin{matrix}
[\mu] \\
[\mu,1] \\
[\mu,2] \\
\vdots \\
[\mu,n_N]
\end{matrix}
&\left(\begin{array}{c|cccc}
0 & -1 & -1 & \cdots & -1 \\\hline
-1 &  +i\gamma_N &  &  &  \\
-1 &   & +i\gamma_N &  &  \\
\vdots &   &  & \ddots &  \\
-1 &   &  &  & +i\gamma_N
\end{array}\right).
\end{align}

The following $(n_N-1)$ pieces of vectors, $\ket{\psi_n}$ for $n=1,2,\ldots, n_N-1$, are all eigenvectors of the partial Hamiltonian $H_{N-1}$ with the degenerate eigenvalue $+\ii\gamma_N$:
\begin{align}\label{eq100}
\braket{[\mu]|\psi_n}&=0
\quad\mbox{and}\quad
\braket{[\mu,\nu]|\psi_n}=\frac{1}{\sqrt{n_N}}\ee^{\ii \nu n \theta_N}\quad\mbox{for $\nu=1,2,\ldots,n_N$},
\end{align}
where $\theta_N\coloneqq 2\pi/n_N$.
Let us also give the vector representation of the eigenvectors:
\begin{align}\label{eq100vec}
\begin{pmatrix}
0 \\ \hline  \ee^{\ii\theta_N} \\  \ee^{2\ii\theta_N}  \\  \ee^{3\ii\theta_N}  \\ \vdots \\  \ee^{(n_N-1)\ii\theta_N}  \\  \ee^{n_N\ii\theta_N} 
\end{pmatrix},
\quad
\begin{pmatrix}
0 \\ \hline  \ee^{2\ii\theta_N} \\  \ee^{4\ii\theta_N}  \\  \ee^{6\ii\theta_N}  \\ \vdots \\  \ee^{(n_N-1)2\ii\theta_N}  \\  \ee^{n_N2\ii\theta_N} 
\end{pmatrix},\quad
\begin{pmatrix}
0 \\ \hline \ee^{3\ii\theta_N} \\  \ee^{6\ii\theta_N}  \\  \ee^{9\ii\theta_N}  \\ \vdots \\  \ee^{(n_N-1)3\ii\theta_N}  \\  \ee^{n_N3\ii\theta_N} 
\end{pmatrix},\quad
\ldots,
\quad
\begin{pmatrix}
0 \\ \hline  \ee^{(n_N-1)\ii\theta_N} \\  \ee^{2(n_N-1)\ii\theta_N}  \\  \ee^{3(n_N-1)\ii\theta_N}  \\ \vdots \\  \ee^{(n_N-1)(n_N-1)\ii\theta_N}  \\  \ee^{n_N(n_N-1)\ii\theta_N} 
\end{pmatrix}
\end{align}
with normalization constant $1/\sqrt{n_N}$. 
To confirm the eigenvalue equation $H_{N-1}\ket{\psi_n}=+\ii\gamma_N\ket{\psi_n}$, note the following:
\begin{align}\label{eq70-1}
\sum_{\nu=1}^{n_N}\braket{[\mu]|H_{N-1}|[\mu,\nu]}\braket{[\mu,\nu]|\psi_n}
=-\sum_{\nu=1}^{n_N} \ee^{\ii \nu n\theta_N}=0
\end{align}
for $n=1,2,\ldots,n_N-1$.
In other words, for each state $\ket{\psi_n}$ of $1\leq n\leq n_N-1$, the amplitudes 
at all the peripheral sites $1\leq \nu\leq n_N$, upon hopping to the root site $[\mu]$, destructively interfere with each other there, amounting to zero. This is similar to the mechanism by which localized eigenstates arise from destructive interference in a class of tight-binding models with flat bands~\cite{sutherland1986localization, bergman2008band, tasaki2020physics}.
The same destructive interference has also been revealed to produce many zero-eigenvalue eigenstates in complex networks~\cite{Bueno20}.

By putting all the other sites on the older generations $0\le\ell\le N-2$ to zero in $\ket{\psi_n}$, as in
\begin{align}
\braket{[\mu_1,\mu_2,\ldots,\mu_{N-2}]|\psi_n}\equiv0
\end{align}
for all $\mu_1$, $\mu_2$, \ldots, $\mu_{N-2}$,
we can make the state $\ket{\psi_n}$ an eigenstate of the total Hamiltonian~\eqref{eq50} with the same eigenvalue $+\ii\gamma_N$.

Since the amplitudes on the root site $[\mu]$ in the $(N-1)$th generation and all the sites in the older generations vanish, these eigenvectors are all localized on the peripheral sites only.
There are $(n_N-1)$ pieces of such localized eigenvectors in Eq.~\eqref{eq100} for every root site $[\mu]$ in the $(N-1)$th generation.
Since the number of the root sites in the $(N-1)$th generation is $\ntot_{N-1}$,
the total number of these localized eigenvectors 
is
$\ntot_{N-1}(n_N-1)=\ntot_N-\ntot_{N-1}$.
We can omit all these localized eigenvectors when computing the current through the system, since they have no amplitudes at the inner sites.

The only state that can have a finite amplitude on the root site $[\mu]$, not localizing on the $N$th generation, must take the form
\begin{align}
\label{eq140}
\braket{[\mu]|\psi_n}&\neq0
\quad\mbox{and}\quad
\braket{[\mu,\nu]|\psi_n}=\alpha\quad\mbox{for $1\le\nu\le n_N$},
\end{align}
where $\alpha$ is a constant including a proper normalization constant.
In the vector representation, it is proportional to
\begin{align}
\Bigl(
\begin{array}{c|ccccc}
1 & \alpha &\alpha &\cdots & \alpha & \alpha
\end{array}^T
\Bigr)
\end{align}
We will use this in the next subsection.

Since the eigenvalue of all the states in Eq.~\eqref{eq100} is $E=\ii\gamma_N$, each state grows in time as in $\exp(-\ii Et)=\exp(\gamma_N t)$.
We can interpret this growth as follows.
The energy flux is added to every peripheral site at the rate of $\ii\gamma_N$.
Since this energy flux does not penetrate into the inner generations, it stays
at the peripheral sites, and hence the amplitude grows exponentially.

\subsection{Eigenstates localized on the outer two generations}
\label{subsec3.2}

\begin{figure}
\centering
\includegraphics[width=0.5\textwidth]{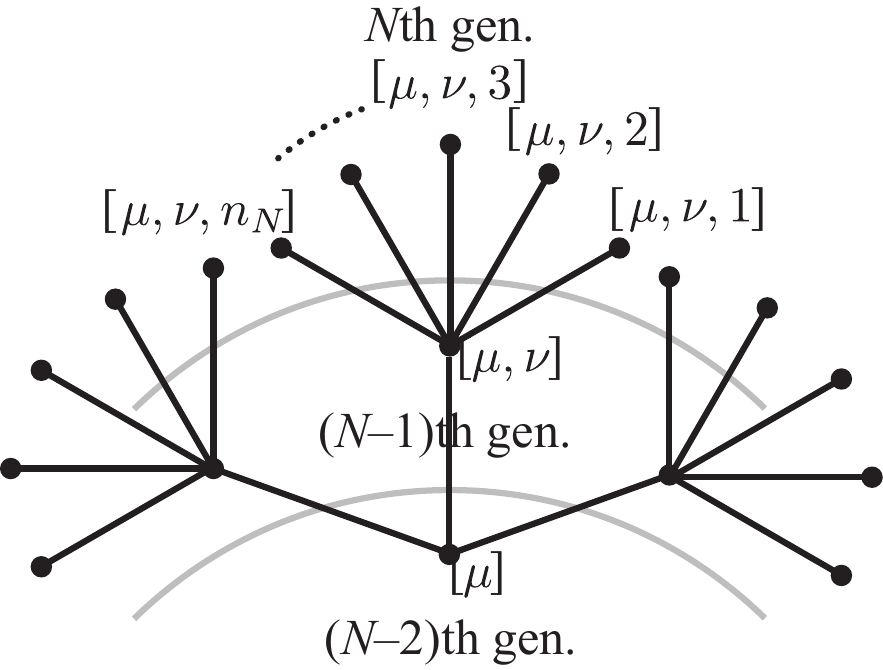}
\caption{One branch of the tree lattice with the root site $\mu$ in the $(N-2)$th generation and the subsequent sites in the $(N-1)$th generation and the peripheral sites in the $N$th generation.}
\label{fig3}
\end{figure}

Let us next consider one site in the $(N-2)$th generation and the branches growing from the root site; see Fig.~\ref{fig3}.
The standard notation would be $[\mu_1,\mu_2,\ldots,\mu_{N-2}]$ for the root site, 
$[\mu_1,\mu_2,\ldots,\mu_{N-2},\mu_{N-1}]$ with $\mu_{N-1}=1,2,3,\ldots,n_{N-1}$ for the
connecting sites in the $(N-1)$th generation, and 
$[\mu_1,\mu_2,\ldots,\mu_{N-2},\mu_{N-1},\mu_N]$ with $\mu_N=1,2,3,\ldots,n_N$ for the further connecting sites on the periphery, but for brevity let us refer to them as $[\mu]$ for the root site, $[\mu,\nu]$ with $\nu=1,2,3,\ldots,n_{N-1}$ for the connecting sites in the $(N-1)$th generation, and $[\mu,\nu,\kappa]$  with $\kappa=1,2,3,\ldots,n_N$ for the peripheral sites.

We now construct eigenvectors with non-zero amplitudes only on the $(N-1)$th and $N$th generations.
To achieve this, we employ the same principle as the one used in the previous subsection~\ref{subsec3.1}.
We set the elements of the eigenvectors, $\ket{\psi_n^s}$ for $1\le n \le n_{N-1}-1$, as follows:
\begin{align}\label{eq100-2}
\braket{[\mu]|\psi_n^s}=0
\quad\mbox{and}\quad
\braket{ [\mu,\nu] |{\psi_n^s}}=\frac{1}{\sqrt{\mathcal N}}\ee^{\ii \nu n\theta_{N-1}}
\quad\mbox{for $1\le \nu \le n_{N-1}$}.
\end{align}
where 
$\theta_{N-1}\coloneqq2\pi/n_{N-1}$ and the normalization constant $\mathcal{N}$ is to be determined below.
The amplitudes on the sites on the $N$th generation $\braket{[\mu,\nu,\kappa]|\psi_n^s}$ will be determined below, too.
It will also be revealed that the superscript $s$ can take the values $s=\pm$.
In addition, we put the amplitudes of all the sites in the older generations $0\le\ell\le N-3$  to zero:
\begin{align}
\braket{[\mu_1,\mu_2,\ldots,\mu_{N-3}]|\psi_n^s}\equiv0
\end{align}
for all $\mu_1$, $\mu_2$, \ldots, $\mu_{N-3}$.
In the same principle as in the previous subsection, the amplitudes on all the sites $[\mu,\nu]$ for all $\nu$ on the $(N-1)$th generations, upon hopping to the root site $[\mu]$,  destructively interfere with each other, which results in the zero amplitude at the site $[\mu]$.

The only state that can have a finite amplitude on the root site $[\mu]$, not localizing on the $(N-1)$th and $N$th generations, must take the form
\begin{align}
\label{eq140-10}
\braket{[\mu]|\psi_n}&\neq0
\quad\mbox{and}\quad
\braket{[\mu,\nu]|\psi_n}=\beta\quad\mbox{for $\nu=1,2,\ldots,n_N$},
\end{align}
where $\beta$ is a constant including a proper normalization constant.
We will use this in the next subsection.

Let us now determine the amplitudes on the peripheral sites, $\braket{[\mu,\nu,\kappa]|\psi_n^s}$, so that the vectors $\ket{\psi_n^s}$ in Eq.~\eqref{eq100-2} can be eigenvectors of the total Hamiltonian. 
Since the amplitude at the root site $[\mu]$ in the $(N-2)$th generation vanishes, the branch dangling out of each link of $[\mu,\nu]$ ($1\le\nu\le n_{N-1}$) is isolated.
Each branch 
has the partial Hamiltonian specified by the nonzero elements
\setcounter{MaxMatrixCols}{20}
\begin{subequations}\label{eq240-1}
\begin{align}
\braket{[\mu,\nu]|H_{N-1}|[\mu,\nu,\kappa]}&=\braket{[\mu,\nu,\kappa]|H_{N-1}|[\mu,\nu]}=-1,
\\
\braket{[\mu,\nu,\kappa]|H_{N-1}|[\mu,\nu,\kappa]}&=+\ii\gamma_N,
\end{align}
\end{subequations}
both for $1\le\kappa\le n_N$.

According to the observation in the previous subsection, the amplitudes of all the peripheral sites $[\mu,\nu,\kappa]$ ($1\le\kappa\le n_N$) must be equal to each other as in Eq.~\eqref{eq140} in order for the state to penetrate to the $(N-1)$th generation.
We thereby take the Ansatz for the eigenvector in the form
\begin{align}\label{eq140-1}
\braket{[\mu,\nu,\kappa]|\psi_n^s}=\frac{\ee^{\ii\nu n \theta_{N-1}}}{\sqrt{\mathcal N}}\alpha
\end{align}
for all $\kappa$, where the prefactor comes from the element $\braket{[\mu,\nu]|\psi_n^s}$ in Eq.~\eqref{eq100-2} and $\alpha$ is a complex number to be determined.

To compute the eigenvalues and fix the constant $\alpha$, let us use a trick introduced originally in Ref.~\cite{Mahan01}.
We introduce a state
\begin{align}
\ket{(\mu,\nu)}\coloneqq\frac{1}{\sqrt{n_N}}\sum_{\kappa=1}^{n_N}
\ket{[\mu,\nu,\kappa]}.
\end{align}
Since all amplitudes on the $N$th generation summed up here are equal to each other as in Eq.~\eqref{eq140-1}, we find
\begin{align}\label{eq16}
\braket{[\mu,\nu,\kappa]|\psi_n^s}=\frac{1}{\sqrt{n_N}}\braket{(\mu,\nu)|\psi_n^s}
\end{align}
for all $\kappa$.

A straightforward algebra produces
\begin{subequations}
\begin{align}
H_{N-1}\ket{[\mu,\nu]}&=-\sqrt{n_N}\ket{(\mu,\nu)},\\
H_{N-1}\ket{(\mu,\nu)}&=-\sqrt{n_N}\ket{[\mu,\nu]}+\ii\gamma_N\ket{(\mu,\nu)},
\end{align}
\end{subequations}
which means that the two-dimensional subspace spanned by $\ket{[\mu,\nu]}$ and $\ket{(\mu,\nu)}$ is closed under the action of $H_{N-1}$.
In other words, the total Hamiltonian has a $2\times2$ block for the subspace in the form
\begin{align}
H^\textrm{sub}_2\coloneqq
\begin{pmatrix}
\braket{[\mu,\nu]|H_{N-1}|[\mu,\nu]} & \braket{[\mu,\nu]|H_{N-1}|(\mu,\nu)} \\ 
\braket{(\mu,\nu)|H_{N-1}|[\mu,\nu]} & \braket{(\mu,\nu)|H_{N-1}|(\mu,\nu)}
\end{pmatrix}=
\begin{pmatrix}
0 & -\sqrt{n_N} \\
-\sqrt{n_N} & \ii\gamma_N
\end{pmatrix},
\end{align}
which gives the two eigenvalues of the total Hamiltonian in the form
\begin{align}\label{eq260}
E^\pm\coloneqq\frac{\ii\gamma_N\pm\sqrt{4n_N-{\gamma_N}^2}}{2}.
\end{align}
The corresponding eigenvectors of the effective Hamiltonian are given by
\begin{align}
\begin{pmatrix}
\braket{[\mu,\nu]|\psi_n^\pm} \\
\braket{(\mu,\nu)|\psi_n^\pm}
\end{pmatrix}\propto
\begin{pmatrix}
1 \\ -E^\pm/\sqrt{n_N}
\end{pmatrix},
\end{align}
where we have assigned $\pm$ to the superscript $s$.
Using Eq.~\eqref{eq16}, we obtain $\braket{[\mu,\nu,\kappa]|\psi_n^\pm}=(-E^\pm/n_N)\braket{[\mu,\nu]|\psi_n^\pm}$,
which means $\alpha=-E^\pm/n_N$ for Eq.~\eqref{eq140-1}.
Note that $\abs{E^\pm}^2=n_N$.
After the normalization, we arrive at
\begin{subequations}\label{eq100-2-1}
\begin{align}
\braket{ [\mu,\nu] |{\psi_n^\pm}}&=\frac{\ee^{\ii \nu n\theta_{N-1}}}{\sqrt{2n_{N-1}}},
\\
\braket{ [\mu,\nu,\kappa] |{\psi_n^\pm}}&=-\frac{\ee^{\ii \nu n\theta_{N-1}}}{\sqrt{2n_{N-1}}}\frac{E^\pm}{n_N}
\end{align}
\end{subequations}
for $1\le \nu \le n_{N-1}$, $1\le\kappa\le n_N$ and $1\le n\le n_{N-1}-1$,
with all other elements set to zero.

Let us now count the number of these eigenstates.
There are $2(n_{N-1}-1)$ pieces of localized eigenvectors in Eq.~\eqref{eq100-2-1} for every root site in the $(N-2)$th generation;
the factor $2$ comes from the superscript $s=\pm$ and the the factor $(n_{N-1}-1)$ comes from the range of the subscript $n$.
The number of the root sites in the $(N-2)$th generation is $\ntot_{N-2}$.
Therefore, the total number of these localized eigenvectors is $2\ntot_{N-2}(n_{N-1}-1)=2\qty(\ntot_{N-1}-\ntot_{N-2})$.
We can omit all these localized eigenvectors when computing the current through the system, since they do not have amplitudes at the inner sites.

Since the imaginary parts of the eigenvalues of all these states are positive as in Eq.~\eqref{eq260}, each state grows in time.
We can interpret this growth in the same manner as in the previous subsection.
These localized states do not carry the input at the peripheral sites to the drain on the root site, and hence the amplitudes accumulate at the sites of the $N$ and $(N-1)$th generations.
The growth rate, particularly for $4n_N>{\gamma_N}^2$, is halved to $\ii\gamma_N/2$ from the one in the previous subsection~\ref{subsec3.1} because the amplitudes can spread up to the $(N-1)$th generation.

\subsection{Eigenstates localized on the outer three generations}
\label{subsec3.3}

To ensure that we find the general rule, let us analyze one more case.
We now consider one site in the $(N-3)$th generation (which we now call the root site in this subsection) and the branch growing from it.
For brevity, here in the present subsection, we refer to them as $[\mu]$ for the root site, $[\mu,\nu]$ with $\nu=1,2,3,\ldots,n_{N-2}$ for the connecting sites in the $(N-2)$th generation, $[\mu,\nu,\kappa]$ with $\kappa=1,2,3,\ldots,n_{N-1}$ for the connecting sites in the $(N-1)$th generation, and $[\mu,\nu,\kappa,\rho]$  with $\rho=1,2,3,\ldots,n_N$ for the peripheral sites.

We now construct eigenvectors with non-zero amplitudes only on the $(N-2)$th, $(N-1)$th, and $N$th generations.
To achieve this, we set the amplitudes of the eigenvectors on the root site $[\mu]$ and $n_{N-2}$ pieces of sites $[\mu,\nu]$ on the $(N-2)$th generation to 
\begin{align}\label{eq100-3}
\braket{[\mu]|\psi_n^s}=0
\quad\mbox{for}\quad
\braket{[\mu,\nu]|\psi_n^s}=\frac{1}{\sqrt{\mathcal N}}\ee^{\ii \nu n\theta_{N-2}}
\quad\mbox{for $1\le \nu\le n_{N-2}$}
\end{align}
and for the subscript $1\le n\le n_{N-2}-1$, where $\theta_{N-2}\coloneqq2\pi/n_{N-2}$.
Again the normalization constant $\mathcal{N}$ as well as the amplitudes of $\braket{[\mu,\nu,\kappa]|\psi_n^s}$ and $\braket{[\mu,\nu,\kappa,\rho]|\psi_n^s}$ can be determined below, but we will not list them because the expressions are too complicated.
This time the superscript will have three values, which we will express as $s=1,2,3$.
The amplitude of the root site can vanish in the same principle above; the hopping amplitudes from all the sites in the $(N-2)$th generation to the root site $[\mu]$ in the $(N-3)$th generation destructively interfere.

Since the amplitude on the root site $\mu$ in the $(N-3)$th generation vanishes, each branch dangling out of each link $[\mu,\nu]$ is isolated, and hence we can find the eigenvalues and the eigenvectors by considering only the partial Hamiltonian of the dangling branch.
According to the argument in the previous subsections~\ref{subsec3.1} and~\ref{subsec3.2}, the amplitudes on the $(N-1)$th generation must be equal to each other as in Eq.~\eqref{eq140-10} and those on the $N$th generation likewise in Eq.~\eqref{eq140}.

\begin{figure}
\begin{subfigure}{0.45\textwidth}
\includegraphics[width=\textwidth]{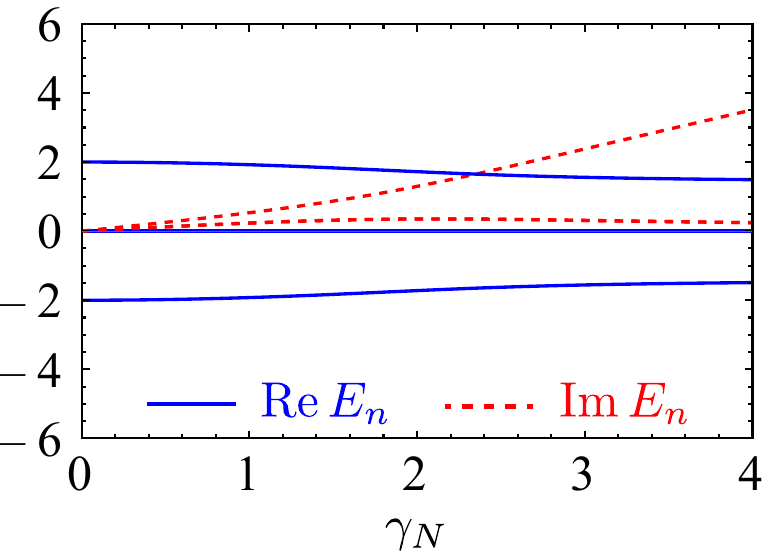}
\caption{$n_{N-1}=n_N=2$}
\end{subfigure}
\hfill
\begin{subfigure}{0.45\textwidth}
\includegraphics[width=\textwidth]{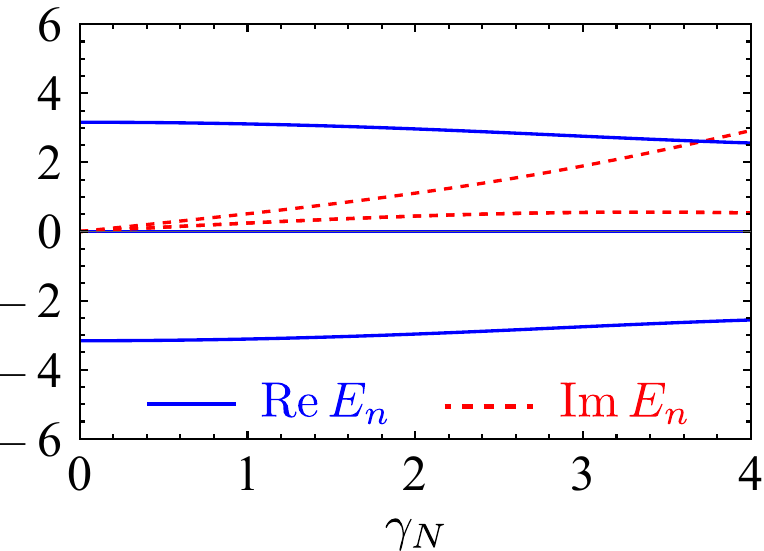}
\caption{$n_{N-1}=n_N=5$}
\end{subfigure}
\\[\baselineskip]
\begin{subfigure}{0.45\textwidth}
\includegraphics[width=\textwidth]{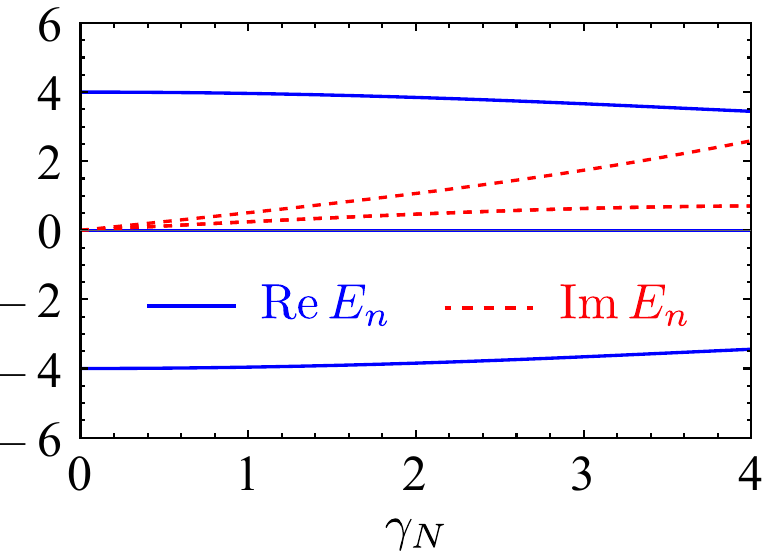}
\caption{$n_{N-1}=n_N=8$}
\end{subfigure}
\hfill
\begin{subfigure}{0.45\textwidth}
\includegraphics[width=\textwidth]{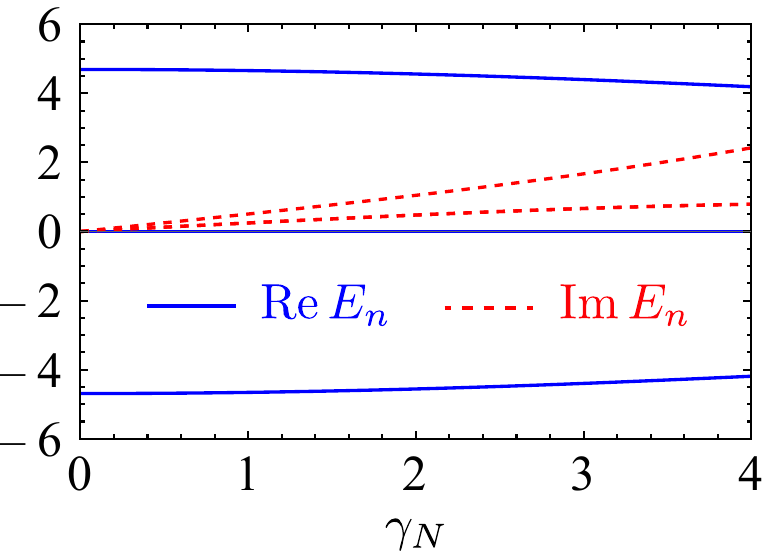}
\caption{$n_{N-1}=n_N=11$}
\end{subfigure}
\caption{Real and imaginary parts of the three eigenvalues of Eq.~\eqref{eq320}.
One eigenvalue is pure imaginary with the largest imaginary part, while the other two eigenvalues have a common imaginary part and the real parts of the same magnitude with opposite signs.}
\label{fig4}
\end{figure}

To compute the eigenvalues, we again use the trick introduced in Ref.~\cite{Mahan01}.
We introduce 
\begin{subequations}
\begin{align}
\ket{(\mu,\nu)}
&\coloneqq\frac{1}{\sqrt{n_{N-1}}}\sum_{\kappa=1}^{n_{N-1}}
\ket{[\mu,\nu,\kappa]},
\\
\ket{\{\mu,\nu\}}
&\coloneqq\frac{1}{\sqrt{n_{N-1}n_N}}\sum_{\kappa=1}^{n_{N-1}}\sum_{\rho=1}^{n_N}
\ket{[\mu,\nu,\kappa,\rho]};
\end{align}
\end{subequations}
note again that the amplitudes summed up here are equal to each other as shown in Eqs.~\eqref{eq140} and~\eqref{eq140-10},
and hence
\begin{subequations}
\begin{align}
\braket{[\mu,\nu,\kappa]|\psi_n^s}&=\frac{1}{\sqrt{n_{N-1}}}\braket{(\mu,\nu)|\psi_n^s},\\
\braket{[\mu,\nu,\kappa,\rho]|\psi_n^s}&=\frac{1}{\sqrt{n_{N-1}n_N}}\braket{\{\mu,\nu\}|\psi_n^s}.
\end{align}
\end{subequations}
Let $H_{N-2}$ denote the partial Hamiltonian of the branch dangling out of $[\mu,\nu]$.
A straightforward algebra produces
\begin{align}
H_{N-2}\ket{[\mu,\nu]}&=-\sqrt{n_{N-1}}\ket{(\mu,\nu)},
\\
H_{N-2}\ket{(\mu,\nu)}&=-\sqrt{n_{N-1}}\ket{[\mu,\nu]}-\sqrt{n_N}\ket{\{\mu,\nu\}},
\\
H_{N-2}\ket{\{\mu,\nu\}}&=-\sqrt{n_N}\ket{(\mu,\nu)}+\ii\gamma_N\ket{\{\mu,\nu\}}.
\end{align}
Similarly to the previous subsection~\ref{subsec3.2}, 
The subspace spanned by $\ket{[\mu,\nu]}$, $\ket{(\mu,\nu)}$, and $\ket{\{\mu,\nu \}}$ is closed under the action of $H_{N-2}$. 
The effective Hamiltonian in this subspace reads
\begin{align}\label{eq320}
H^\textrm{sub}_3=
\begin{pmatrix}
0 & -\sqrt{n_{N-1}} & 0 \\
-\sqrt{n_{N-1}} & 0 & -\sqrt{n_N} \\
0 & -\sqrt{n_N} & \ii\gamma_N
\end{pmatrix}.
\end{align}
Therefore, the eigenvalues are given by the third-order secular equation
\begin{align}\label{eq325}
E^3-\ii\gamma_N E^2-(n_{N-1}+n_N)E+\ii\gamma_N n_{N-1}=0.
\end{align}
The numerical plots in Fig.~\ref{fig4} show 
that one eigenvalue is pure imaginary with a relatively large imaginary part of the order of $\ii\gamma_N$, while the other two eigenvalues have a common, relatively small imaginary part;
for each of the three eigenvalues, the imaginary part is non-negative. 

We prove the non-negativity of the imaginary parts of the eigenvalues as follows.
Let us decompose the $3 \times 3$ non-Hermitian matrix $H^\textrm{sub}_3$ in Eq.~\eqref{eq320} as 
\begin{align}
    H^\textrm{sub}_3 = h + \ii \gamma_{N}  g^\dag g
\end{align}
with a Hermitian matrix $h$ and $g = \mathrm{diag} \left( 0,0,1\right)$.
From the eigenequation $H^\textrm{sub}_3 \ket{\psi_n} =  E_n \ket{\psi_n}$, we have
\begin{align}
    \braket{\psi_n|h|\psi_n}
    +\ii \gamma_{N} \braket{\psi_n|g^\dag g|\psi_n}
    =E\braket{\psi_n|\psi_n}.
\end{align}
Since all three matrix elements must be real, we find the imaginary part of $E$ in the form
\begin{align}\label{eq350}
    \mathrm{Im}\,E_n = \frac{\gamma_{N} \braket{\psi_n | g^{\dag} g | \psi_n}}{\braket{\psi_n | \psi_n}} \geq 0.
\end{align}
One can arrive at the same conclusion by applying Bendixon's theorem~\cite{loghin2006bounds}, which states that the imaginary parts of the eigenvalues $E_n$ of a non-Hermitian Hamiltonian $H$ are bounded as $E^{\rm A}_{\rm min} \le \Im E_n \le E^{\rm A}_{\rm max}$, where $E^{\rm A}_{\rm min}$ and $E^{\rm A}_{\rm max}$ are the smallest and largest eigenvalues of $H^{\rm A}:=(H-H^\dagger)/(2\ii)$, respectively. We note in passing that the matrix $H^\textrm{sub}_3$ respects
\begin{align}
    C \left( H^\textrm{sub}_3 \right)^{*} C^{-1} = - H^\textrm{sub}_3 
    \qquad \mbox{where} \quad C \coloneqq 
    \begin{pmatrix}
        1 & 0 & 0 \\
        0 & -1 & 0 \\
        0 & 0 & 1
    \end{pmatrix},
\end{align}
which makes the complex spectrum symmetric about the imaginary axis.
This symmetry can be considered as a non-Hermitian extension of particle-hole symmetry (particle-hole symmetry$^{\dag}$ in the terminology of Ref.~\cite{Kawabata19})].

There are $3(n_{N-2}-1)$ pieces of localized eigenvectors in Eq.~\eqref{eq100-3} for every root site in the $(N-3)$th generation.
The number of the root site in the $(N-3)$th generation is $\ntot_{N-3}$.
Therefore, the total number of these localized eigenvectors is $3\ntot_{N-3}(n_{N-2}-1)=3\qty(\ntot_{N-3}-\ntot_{N-2})$.
We can omit all these localized eigenvectors when computing the current through the system, since they have no amplitudes at the inner sites.

Only an eigenvector with the elements on the $(N-2)$th generation equal to each other can penetrate to the $(N-3)$ generation.
The argument thus goes on to the last one presented in the next subsection.

\subsection{Further `localized' eigenstates}
\label{subsec3.4}
We similarly construct eigenvectors with amplitudes only on generations younger than a specific one.
The last eigenstates that reach the sites of the first generation but fall short of the origin site are given by
\begin{align}\label{eq100-4}
\braket{[0]|\psi_n^s}=0
\quad\mbox{and}\quad
\braket{[\mu_1]|\psi_n^s}=\frac{1}{\sqrt{\mathcal N}}\ee^{\ii \mu_1 n \theta_1}
\quad\mbox{for $1\le\mu_1\le n_1$}
\end{align}
and for the subscript $1\le n  \le n_1-1$, where $\theta_{1}\coloneqq2\pi/n_{1}$.
Note that for each eigenvector, the amplitudes of the sites on each generation lower than the first are equal to each other, and hence we can use the same trick by Ref.~\cite{Mahan01} to find all the amplitudes of the eigenvectors in the younger generations after diagonalizing an $N\times N$ effective Hamiltonian in a subspace, and hence the superscript $s$ takes the values in the range $1\le s\le N$.
We can also prove the non-negativity of the imaginary parts of all complex eigenvalues in the same manner as in Eq.~\eqref{eq350} or by Bendixon's theorem~\cite{loghin2006bounds}.

As we described at the end of Sec.~\ref{sec2}, all the localized states given in the present section amount to the number $\ntot-(N+1)$, while the total number of eigenstates is $\ntot$.
We show in the next section~\ref{sec4} that the remaining $(N+1)$ pieces of eigenstates have amplitudes at the origin site and hence do contribute to the quantum transport from the peripheral sites to the origin site.
In these extended eigenvectors, the amplitudes of the sites of each generation, even of the first generation, must be equal to each other.

\section{Extended eigenstates}
\label{sec4}
We now use the trick in Ref.~\cite{Mahan01} one last time to find the final $N+1$ pieces of eigenstates that are extended from the peripheral sites up to the origin site $[0]$.
From all the discussions in the previous subsections, we can safely assume that for the extended eigenstates, the amplitudes at all sites of each generation are identical.
We hence define in the following forms the $(N+1)$ pieces of states: 
\begin{align}
\label{eq330}
\ket{(0)}&\coloneqq\ket{[0]},
\quad
\ket{(1)}\coloneqq\frac{1}{\sqrt{\ntot_1}}\sum_{\mu=1}^{n_1}\ket{[\mu]},
\quad
\ket{(2)}\coloneqq\frac{1}{\sqrt{\ntot_2}}\sum_{\mu=1}^{n_1}\sum_{\nu=1}^{n_2}\ket{[\mu,\nu]},\cdots
\nonumber\\
\ket{(N)}&\coloneqq\frac{1}{\sqrt{\ntot_{N}}}
\sum_{\mu_1=1}^{n_1}\sum_{\mu_2=1}^{n_2}\sum_{\mu_3=1}^{n_3}\cdots\sum_{\mu_{N-1}=1}^{n_{N-1}}\sum_{\mu_N=1}^{n_N}\ket{[\mu_1, \mu_2, \mu_3, \cdots,\mu_{N-1},\mu_N]}.
\end{align}
Applying the total Hamiltonian $H_\tot$ to these states, we indeed show that they form a closed subspace of the total Hilbert space and hence that the total Hamiltonian has an $(N+1)\times(N+1)$ block.
The effective Hamiltonian in this subspace is an $(N+1)\times(N+1)$ tridiagonal matrix with non-zero values only in the following elements:
\begin{align}\label{eq460-1}
&\braket{(0)|H^\textrm{eff}_N|(0)}=-\ii\gamma_0,
\qquad
\braket{(N)|H^\textrm{eff}_N|(N)}=+\ii\gamma_N,
\nonumber\\
&\braket{(\ell)|H^\textrm{eff}_N|(\ell+1)}=\braket{(\ell+1)|H^\textrm{eff}_N|(\ell)}=-\sqrt{n_\ell}
\quad\mbox{for $1\le\ell\le N$}.
\end{align}

At this point, let us focus on the following simplest case:
\begin{subequations}\label{eq430}
\begin{align}\label{eq430a}
&n:=n_1=n_2=n_3=\cdots=n_N, \\
\label{eq430b}
&\gamma:=\gamma_0=\gamma_N,
\end{align}
\end{subequations}
where we assume $\gamma>0$.
Then, the effective Hamiltonian is reduced to $\tilde{H}^\textrm{eff}_{N}\coloneqq H^\textrm{eff}_{N}/\sqrt{n}$, which is given in the simple form
\begin{align}\label{eq420}
\tilde{H}^\textrm{eff}_{N}&\coloneqq\frac{H^\textrm{eff}_{N}}{\sqrt{n}}
=\begin{pmatrix}
-\ii\tilde{\gamma} & -1 &&& \\
-1 & 0 & -1 && \\
& -1 & \ddots & \ddots & \\
&& \ddots & 0 & -1 \\
&&& -1 & \ii\tilde{\gamma}
\end{pmatrix},
\end{align} 
with
$\tilde{\gamma} \coloneqq \gamma/\sqrt{n}$.
Note that the dependence on $n$ is scaled out and only implicitly contained in $\tilde{\gamma}$ hereafter.

The Hamiltonian~\eqref{eq420} has parity-time symmetry, or 
$\PT$ symmetry~\cite{Bender98,Bender05,BenderBook}, in the following sense.
For the parity transformation, consider the parity operator $\mathcal{P}$ with
\begin{align}
\braket{(\ell)|\mathcal{P}|(N-\ell)}=1
\quad\mbox{for $0\le \ell \le N$},
\end{align}
while for the time-reversal transformation $\mathcal{T}$, consider 
complex conjugation.
The Hamiltonian~\eqref{eq420} goes back to itself after the simultaneous operation of $\mathcal{P}$ and $\mathcal{T}$.
The parity operator $\mathcal{P}$ makes the mirror image of the system, flipping the left and right of the system, and hence exchanges $-\ii\tilde{\gamma}$ at $\ell=0$ and $+\ii\tilde{\gamma}$ at $\ell=N$;
then the time-reversal operator $\mathcal{T}$ takes complex conjugation, flipping the sign of $\pm\ii\tilde{\gamma}$ back to the original one.
We therefore have $[\tilde{H}^\textrm{eff}_{N},\PT]=0$.

A consequence of $\PT$ symmetry, and more generally, the product of a linear operator, such as $\mathcal{P}$, and an anti-linear operator, such as $\mathcal{T}$, is the following.
Suppose that the Hamiltonian~\eqref{eq420} has an eigenvalue $E$, as in $\tilde{H}^\textrm{eff}_{N}\ket{\psi}=E\ket{\psi}$, without any degeneracy.
Applying $\PT$ from the left on both sides and using the commutation relation 
$[\tilde{H}^\textrm{eff}_{N},\PT]=0$, we have
\begin{align}
\tilde{H}^\textrm{eff}_{N}\qty(\PT\ket{\psi})=E\qty(\PT\ket{\psi})
\end{align}
\textit{if and only if} the eigenvalue $E$ is real.
Then, the $\PT$-reversed state $\PT\ket{\psi}$ is also an eigenvector of $\tilde{H}^\textrm{eff}_{N}$ with eigenvalue $E$, and hence $\PT\ket{\psi}$ is parallel to $\ket{\psi}$.
This means that this eigenvector of the Hamiltonian is also $\PT$ symmetric, and hence it is called the $\PT$-unbroken phase of this particular state.

On the other hand, if and only if $E$ is complex, we have
\begin{align}
\tilde{H}^\textrm{eff}_{N}\qty(\PT\ket{\psi})=E^\ast\qty(\PT\ket{\psi})
\end{align}
with $E\neq E^\ast$. 
The eigenvector $\PT\ket{\psi}$ is the one for the eigenvalue $E^\ast$ and hence is no longer parallel to $\ket{\psi}$.
This means that the eigenvector $\PT\ket{\psi}$ 
spontaneously
breaks $\PT$ symmetry of the Hamiltonian, and hence it is called the $\PT$-broken phase of the state.
Note that the commutation relation $[\tilde{H}^\textrm{eff}_{N},\PT]=0$ does not necessarily yield that the eigenvector of $\tilde{H}^\textrm{eff}_{N}$ is also one of $\PT$ because $\mathcal{T}$ is not a linear operator, but an anti-linear one.

An important comment regarding the eigenvectors is in order.
Because the $\PT$-symmetric Hamiltonian is generally non-Hermitian, it has a right eigenvector $\ket{\psi_n}$ and a left eigenvector $\bra{\phi_n}$ for each eigenvalue $E_n$ whenever it is diagonalizable;
see the demonstration in the case of $N=1$ below.
The set of right eigenvectors $\{\ket{\psi_n}\}$ is generally not orthogonal to each other; they are bi-orthogonal to the set of left eigenvectors $\{\bra{\phi_n}\}$ as in $\braket{\phi_m|\psi_n}=\delta_{mn}$.
There are hence two possibilities in defining the expectation value of a physical quantity $\hat{Q}$: one possibility is to use a right eigenvector and its Hermitian conjugate $\braket{\psi_n|\hat{Q}|\psi_n}$, as is done in the standard quantum mechanics; the other is to use the set of left and right eigenvectors, as in $\braket{\phi_n|\hat{Q}|\psi_n}$~\cite{Brody02, Brody14, Brody16}.
As is emphasized in \ref{appA}, we make a clear distinction between them;
the former definition evaluates the quantity in an open quantum system, which is a part of the entire Hermitian system, whereas the latter evaluates it in a closed non-Hermitian system.
Since we regard the present system as a light-harvesting molecule embedded in a large environmental system, we use the former definition of the expectation value here.
In fact, the expectation value of the current in the latter definition is identically zero, which is plausible because the system is regarded as a closed one in the latter.

Note that the above is about the eigenvectors of a non-Hermitian Hamiltonian.
The appearance of complex eigenvalues, including, but not limited to, those in the $\PT$-broken region, in open quantum systems still has physical significance; see \ref{appA} and Refs.~\cite{Hatano08, Hatano14, Hatano21}.

In passing, $\tilde{H}^\textrm{eff}_{N}$ further respects other symmetries.
As a prime example, $\tilde{H}^\textrm{eff}_{N}$ is invariant under transposition:
\begin{align}
    (\tilde{H}^\textrm{eff}_{N})^T = \tilde{H}^\textrm{eff}_{N}.
\end{align}
This is considered to be time-reversal symmetry$^{\dag}$ in the terminology of Ref.~\cite{Kawabata19} and, for example, forbids the non-Hermitian skin effect~\cite{Lee16, YaoWang18, Kunst18}.

\subsection{Case of $N=1$}
\label{subsec4.1}

As a tutorial example, let us first present an exact solution for $N=1$, for which the effective Hamiltonian~\eqref{eq420} reads
\begin{align}\label{eq500}
\tilde{H}^\textrm{eff}_1=
\begin{pmatrix}
-\ii\tilde{\gamma} & -1 \\
-1 & \ii\tilde{\gamma}
\end{pmatrix}.
\end{align}
The eigenvalues are given by
\begin{align}\label{eq510}
E_1^\pm=\pm\sqrt{1-\tilde{\gamma}^2},
\end{align}
which are both real for $\tilde{\gamma}< 1$ and both pure imaginary for $\tilde{\gamma}>1$.
The former 
corresponds to
the $\PT$-unbroken region and the latter the $\PT$-broken region.
The two eigenvalues coalesce at $\tilde{\gamma}_\textrm{ex}=1$, which is an exceptional point;
not only do the eigenvalues become equal to each other, but the eigenvectors also become parallel to each other, and the matrix rank drops to unity, as we will see below.
Here and hereafter, we use the common notation $\tilde{\gamma}_\textrm{ex}$ for the exceptional point, although it can take different values.

In the $\PT$-unbroken region $\tilde{\gamma}<\tilde{\gamma}_\textrm{ex}=1$, the eigenvectors for $E_1^\pm$ are given by
\begin{align}\label{eq520-2}
\ket{\psi_1^+}=\frac{1}{\sqrt{2}}
\begin{pmatrix}
1 \\
-\ii\tilde{\gamma}-\sqrt{1-\tilde{\gamma}^2}
\end{pmatrix},
\qquad
\ket{\psi_1^-}=\frac{1}{\sqrt{2}}
\begin{pmatrix}
\ii\tilde{\gamma}+\sqrt{1-\tilde{\gamma}^2}\\
1
\end{pmatrix},
\end{align}
respectively.
In the $\PT$-broken region $\tilde{\gamma}>\tilde{\gamma}_\textrm{ex}=1$, on the other hand, the eigenvectors for the redefined eigenvalues
\begin{align}
E_1^\pm=\pm\ii\sqrt{\tilde{\gamma}^2-1}
\end{align}
are given by
\begin{align}\label{eq520-0}
\ket{\psi_1^+}=\frac{1}{\sqrt{\mathcal{N}}}
\begin{pmatrix}
1 \\
-\ii\tilde{\gamma}-\ii\sqrt{\tilde{\gamma}^2-1}
\end{pmatrix},
\qquad
\ket{\psi_1^-}=\frac{1}{\sqrt{\mathcal{N}}}
\begin{pmatrix}
\ii\tilde{\gamma}+\ii\sqrt{\tilde{\gamma}^2-1}\\
1
\end{pmatrix}
\end{align}
with a different normalization constant $\mathcal{N}\coloneqq2\tilde{\gamma}\qty(\tilde{\gamma}+\sqrt{\tilde{\gamma}^2-1})$.
The crucial difference between the eigenvectors in Eqs.~\eqref{eq520-2} and~\eqref{eq520-0} lies in the amplitude distribution.
For the eigenvectors~\eqref{eq520-2} in the $\PT$-unbroken region, the amplitudes of the first and second elements are equal to each other as a direct consequence of $\PT$ symmetry.
For those~\eqref{eq520-0} in the $\PT$-broken region, the amplitude of $\ket{\psi_1^+}$ is larger in the second element while that of $\ket{\psi_1^-}$ is larger in the first element.
We will see in Subsec.~\ref{subsec4.2} that this is a general feature for any $N$.
The eigenstates are extended quite uniformly over the system in the $\PT$-unbroken region, while each of them exponentially increases towards one of the edges in the $\PT$-broken region.

Note that these eigenvectors are not orthogonal to each other as in $\braket{\psi_1^-|\psi_1^+}\neq0$ because the Hamiltonian~\eqref{eq500} is not Hermitian.
For a non-Hermitian matrix, we can define left eigenvectors~\cite{Brody14} as opposed to the right eigenvectors given in Eq.~\eqref{eq520-0}.
In the case of the symmetric matrix~\eqref{eq500}, each left eigenvector is the transpose of the corresponding right eigenvector:
\begin{align}\label{eq520-1}
\bra{\phi_1^+}=\frac{1}{\sqrt{2}}
\begin{pmatrix}
1 &
-\ii\tilde{\gamma}-\sqrt{1-\tilde{\gamma}^2}
\end{pmatrix},
\qquad
\bra{\phi_1^-}=\frac{1}{\sqrt{2}}
\begin{pmatrix}
\ii\tilde{\gamma}+\sqrt{1-\tilde{\gamma}^2}&
1
\end{pmatrix}
\end{align}
in the $\PT$-unbroken region and
\begin{align}\label{eq520-3}
\bra{\phi_1^+}=\frac{1}{\sqrt{\mathcal{N}}}
\begin{pmatrix}
1 &
-\ii\tilde{\gamma}-\ii\sqrt{\tilde{\gamma}^2-1}
\end{pmatrix},
\qquad
\bra{\phi_1^-}=\frac{1}{\sqrt{\mathcal{N}}}
\begin{pmatrix}
\ii\tilde{\gamma}+\ii\sqrt{\tilde{\gamma}^2-1}&
1
\end{pmatrix}
\end{align}
in the $\PT$-broken region.
We can indeed confirm the biorthogonality $\braket{\phi_1^+|\psi_1^-}=\braket{\phi_1^-|\psi_1^+}=0$.
Nonetheless, as we commented above, we use in the present paper $\bra{\psi_1^\pm}\coloneqq\ket{\psi_1^\pm}^\dag$ to define the expectation value as $\braket{\psi_1^\pm|\hat{Q}|\psi_1^\pm}$ for a quantity $\hat
{Q}$.
We do not utilize the left eigenvectors $\bra{\phi_1^\pm}$ because we attribute the model's non-Hermiticity to the coupling to the external environment, as described in~\ref{appA}.

At the exceptional point $\tilde{\gamma}=\tilde{\gamma}_\textrm{ex}=1$, both the eigenvalues~\eqref{eq510} reduce to $E_1^\textrm{EP}=0$, and the eigenvectors~\eqref{eq520-0} become parallel to each other:
\begin{align}
\ket{\psi_1^\textrm{EP}}=\frac{1}{\sqrt{2}}
\begin{pmatrix} 1 \\ -\ii \end{pmatrix}.
\end{align}
The rank of the matrix~\eqref{eq500} drops to one.
The matrix is not diagonalizable; it is only transformed into a Jordan block.

\subsection{General case}
\label{subsec4.2}

For general $N$, we assume an Ansatz
\begin{align}\label{eq470}
\psi_N(\ell)\coloneqq\braket{(\ell)|\psi}=A\ee^{\ii\ell k}+B\ee^{-\ii\ell k}
\end{align}
for $0\le \ell\le N$ and $0\le k\le \pi$.
The eigenvalue equation
\begin{align}\label{eq53}
\braket{(\ell)|\tilde{H}^\textrm{eff}_N|\psi}=\tilde{E}\braket{(\ell)|\psi}
\end{align}
for $1\leq \ell\le N-1$ reads
\begin{align}\label{eq460}
-\psi_N(\ell-1)-\psi_N(\ell+1)=\tilde{E}_N\psi_N(\ell).
\end{align}
Using the Ansatz~\eqref{eq470}, we find the scaled eigenvalue
\begin{align}
\tilde{E}_N\coloneqq\frac{E_N}{\sqrt{n}}=-2\cos k,
\end{align}
where $E_N$ denotes the eigenvalue of the unscaled Hamiltonian $H^\textrm{eff}_{N}$ in Eq.~\eqref{eq420}.
The eigenvalue equation~\eqref{eq53} for $\ell=0$ and $\ell=N$ reads
\begin{subequations}
\begin{align}
-\ii\tilde{\gamma}\psi_N(0)-\psi_N(1)&=\tilde{E}_N\psi_N(0),\\
-\psi_N(N-1)+\ii\tilde{\gamma}\psi_N(N)&=\tilde{E}_N\psi_N(N).
\end{align}
\end{subequations}
We compare them with the equations generalized from Eq.~\eqref{eq460} to the cases $\ell=0$ and $\ell=N$, finding
$-\ii\tilde{\gamma}\psi_N(0)=-\psi_N(-1)$ and
$+\ii\tilde{\gamma}\psi_N(N)=-\psi_N(N+1)$,
which are cast into the matrix equation
\begin{align}\label{eq570}
\begin{pmatrix}
\ee^{-\ii k}-\ii\tilde{\gamma} & \ee^{\ii k}-\ii\tilde{\gamma} \\
\ee^{\ii Nk}\qty(\ee^{\ii k}+\ii\tilde{\gamma}) & \ee^{-\ii Nk}\qty(\ee^{-\ii k}+\ii\tilde{\gamma})
\end{pmatrix}
\begin{pmatrix}A\\B\end{pmatrix}=\begin{pmatrix}0\\0\end{pmatrix}.
\end{align}
The condition that we do not have the trivial solution $A=B=0$ is given by the zero of the determinant of the matrix in Eq.~\eqref{eq570}, which is
\begin{align}\label{eq575}
-2\ii \sin[(N+2)k]-2\ii \tilde{\gamma}^2\sin (Nk)=0.
\end{align}

For $\tilde{\gamma}=0$, we have the obvious solutions of Eq.~\eqref{eq575} given by $\sin[(N+2)k]=0$, which are $k=m\pi/(N+2)$ for $m=1,2,\ldots,N+1$.
For $\tilde{\gamma}=1$, it is also exactly solvable, since it reduces to
\begin{align}
\sin[(N+1)k]\cos k=0,
\end{align}
which gives $k=\pi/2$ and $k=m\pi/(N+1)$ for $m=1,2,\ldots.N$.
In both cases, the energy eigenvalues are given by $E_N=-2\sqrt{n}\cos k$.

\begin{figure}
\begin{subfigure}{0.42\textwidth}
\includegraphics[width=\textwidth]{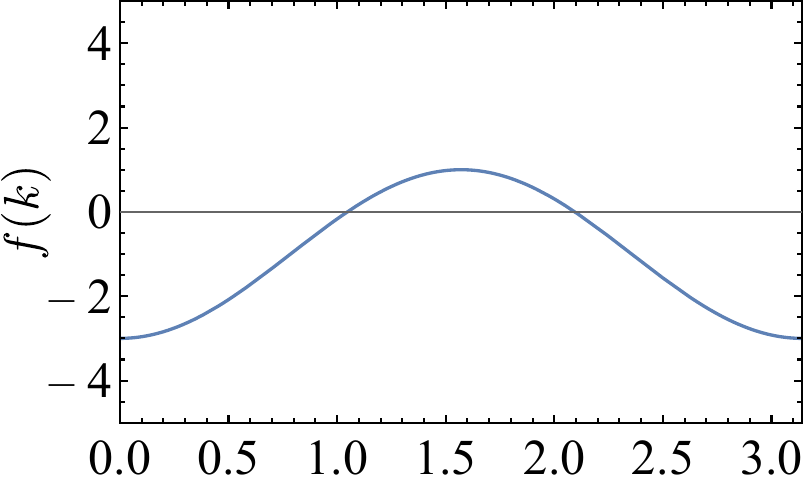}
\caption{$N=1$}
\end{subfigure}
\hfill
\begin{subfigure}{0.42\textwidth}
\includegraphics[width=\textwidth]{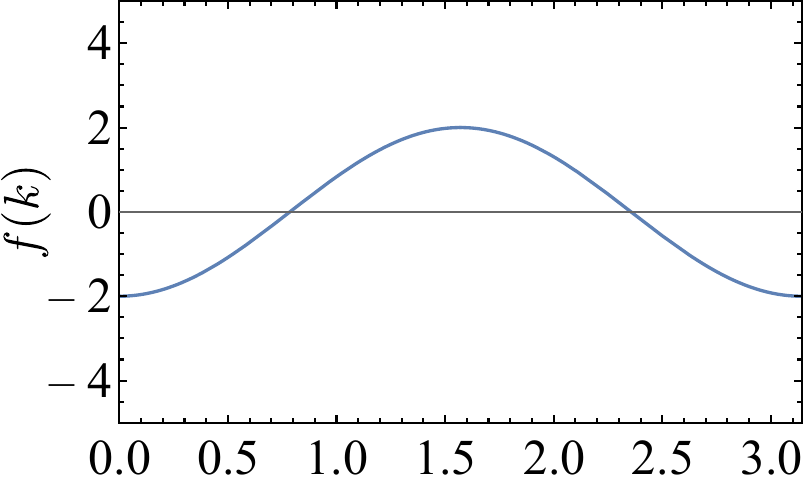}
\caption{$N=2$}
\end{subfigure}
\\[\baselineskip]
\begin{subfigure}{0.42\textwidth}
\includegraphics[width=\textwidth]{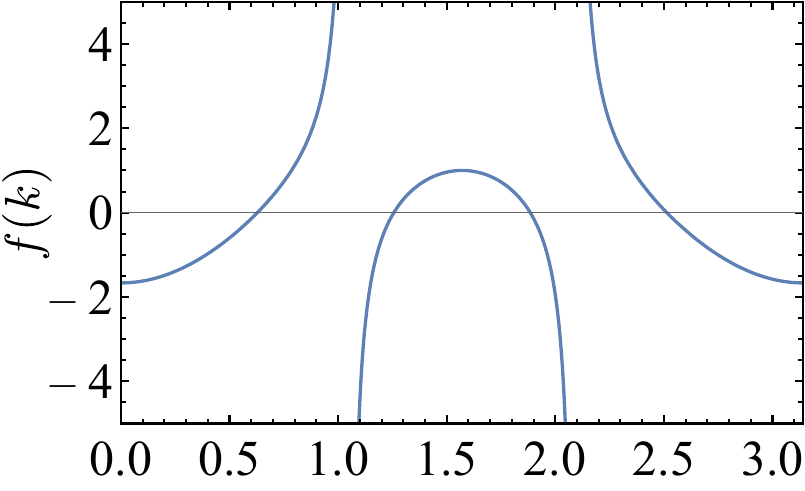}
\caption{$N=3$}
\end{subfigure}
\hfill
\begin{subfigure}{0.42\textwidth}
\includegraphics[width=\textwidth]{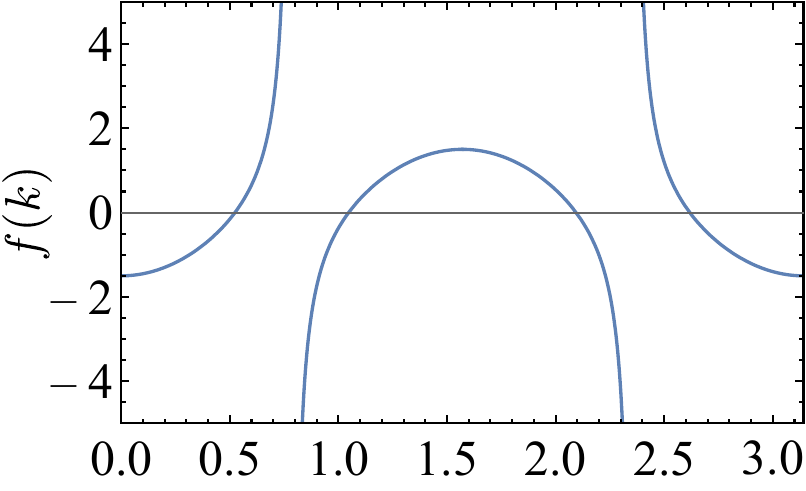}
\caption{$N=4$}
\end{subfigure}
\\[\baselineskip]
\begin{subfigure}{0.42\textwidth}
\includegraphics[width=\textwidth]{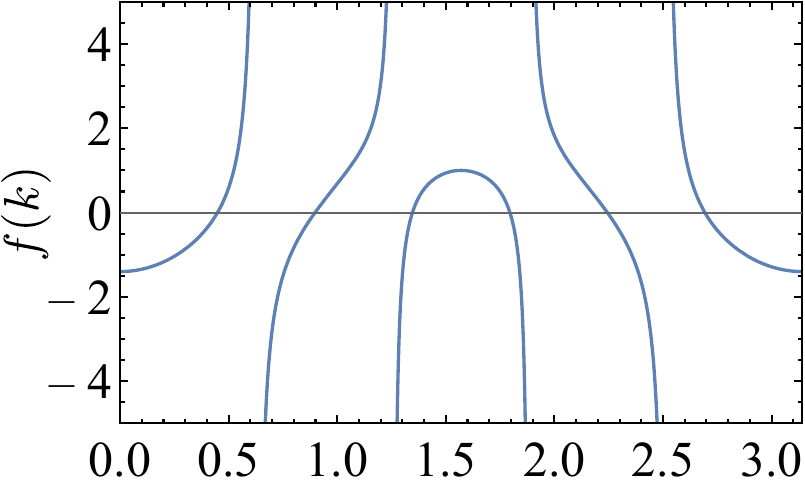}
\caption{$N=5$}
\end{subfigure}
\hfill
\begin{subfigure}{0.42\textwidth}
\includegraphics[width=\textwidth]{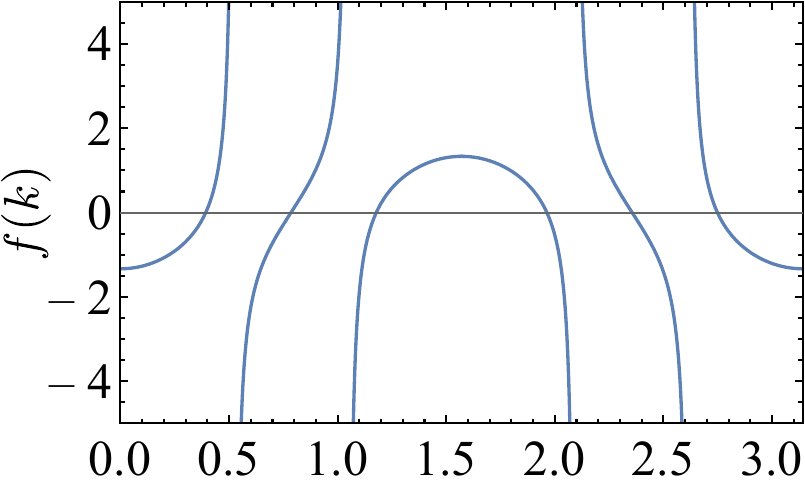}
\caption{$N=6$}
\end{subfigure}
\\
\caption{Numerical calculation of the function $f(k)$ in the left-hand side of Eq.~\eqref{eq580} for $1\leq N\leq6$.}
\label{fig5}
\end{figure}

For other values of $\tilde{\gamma}$, the solutions for $k$ are numerically obtained as the crossing points of $\tilde{\gamma}^2$ and the function
\begin{align}\label{eq580}
f(k)\coloneqq-\frac{\sin[(N+2)k]}{\sin (Nk)},
\end{align}
which is exemplified in Fig.~\ref{fig5}, except for the case
\begin{align}\label{eq680}
\sin (Nk) = \sin[ (N+2) k]=0,
\end{align}
which is satisfied by $k=0$ and $k=\pi$ for any $N$ and $k=\pi/2$ for even $N$.
However, each of the solutions $k=0$ and $k=\pi$ yields the trivial solution $\psi_N(\ell)\equiv0$, and should therefore be removed.
We will consider the other general solutions below.

\begin{figure}
\begin{subfigure}{0.45\textwidth}
\includegraphics[width=\textwidth]{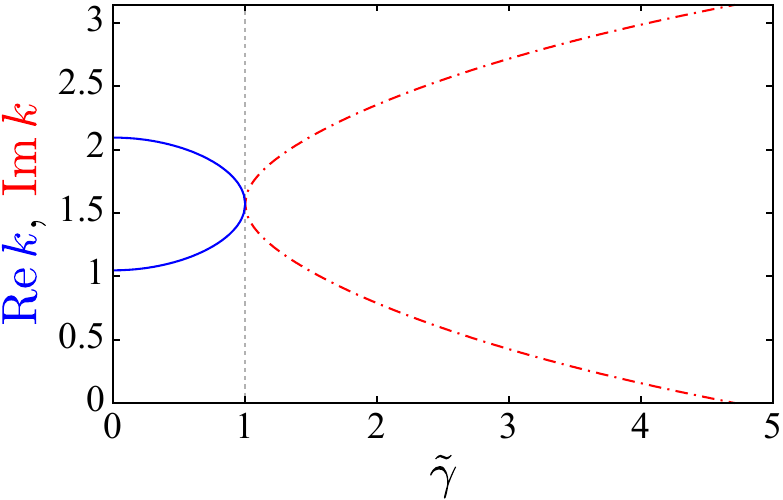}
\caption{$N=1$}
\end{subfigure}
\hfill
\begin{subfigure}{0.45\textwidth}
\includegraphics[width=\textwidth]{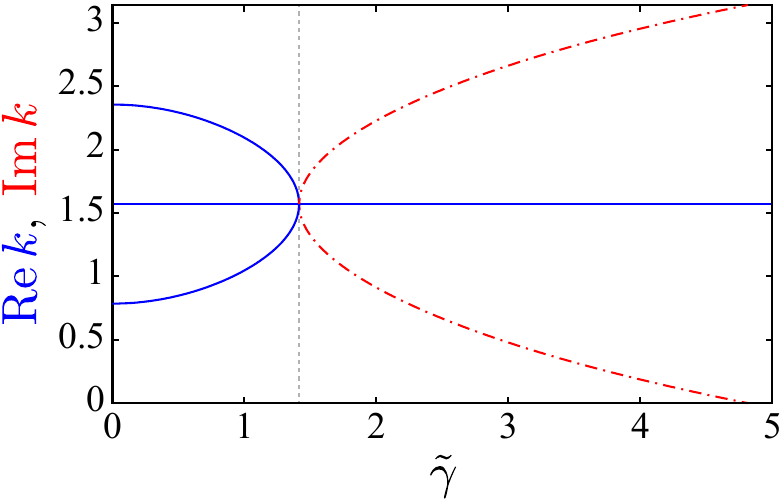}
\caption{$N=2$}
\end{subfigure}
\\[\baselineskip]
\begin{subfigure}{0.45\textwidth}
\includegraphics[width=\textwidth]{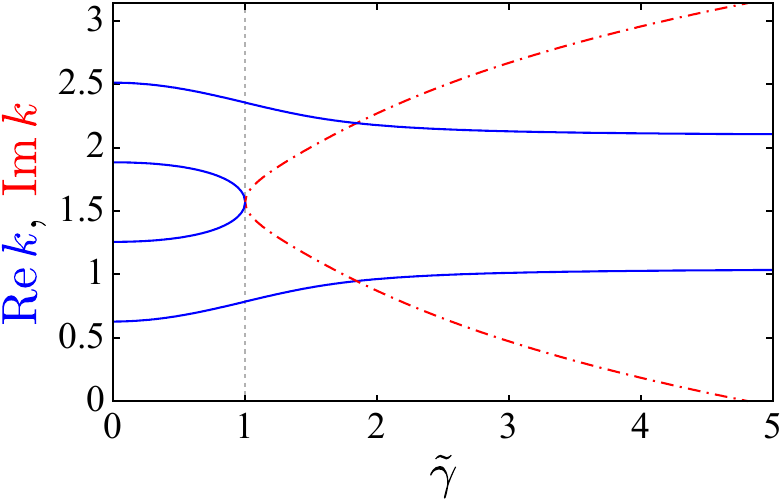}
\caption{$N=3$}
\end{subfigure}
\hfill
\begin{subfigure}{0.45\textwidth}
\includegraphics[width=\textwidth]{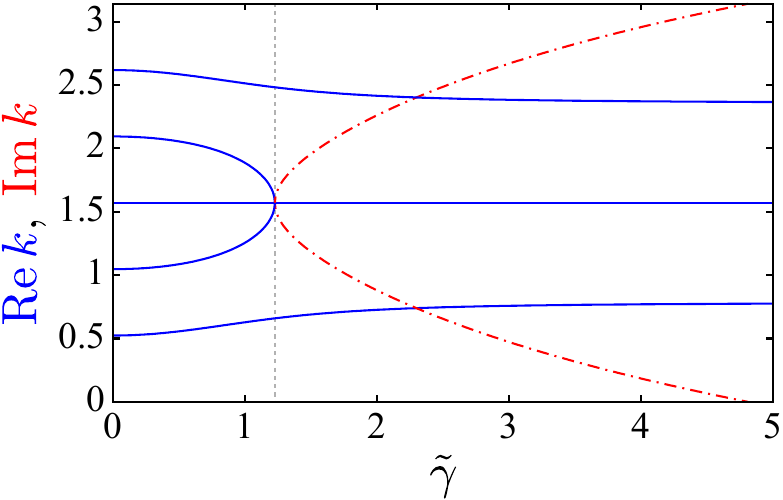}
\caption{$N=4$}
\end{subfigure}
\\[\baselineskip]
\begin{subfigure}{0.45\textwidth}
\includegraphics[width=\textwidth]{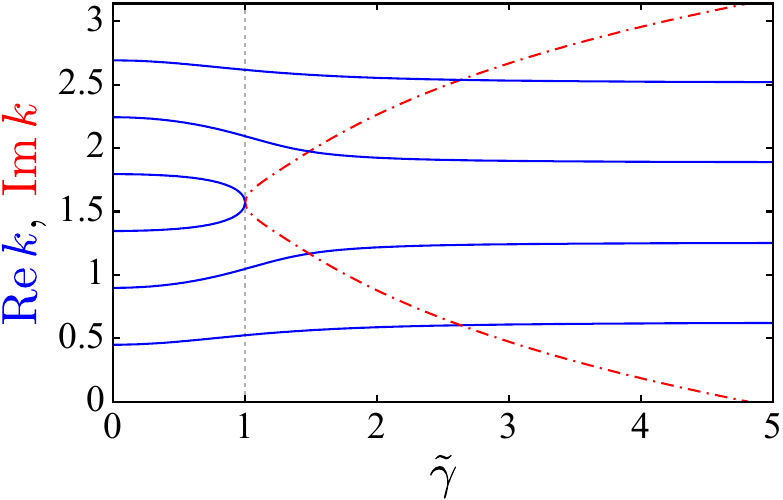}
\caption{$N=5$}
\end{subfigure}
\hfill
\begin{subfigure}{0.45\textwidth}
\includegraphics[width=\textwidth]{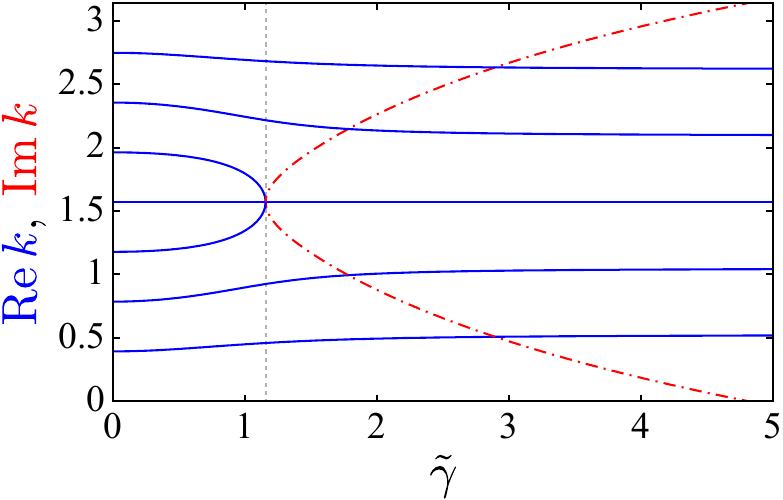}
\caption{$N=6$}
\end{subfigure}
\\
\caption{The $\tilde{\gamma}$-dependence of the real and imaginary parts of the solutions of $k$ for $1\leq N\leq6$.
The blue solid curves indicate the real solutions of $k$, while the red chain curves indicate the imaginary parts (shifted upwards by $+\pi/2$) of the complex solutions of $k$ with their real part $k=\pi/2$.}
\label{fig6}
\end{figure}

Upon switching on $\tilde{\gamma}^2$ in Fig.~\ref{fig5}, the real solutions for $\tilde{\gamma}=0$ start to move, but they remain real for small $\tilde{\gamma}^2$.
As we increase $\tilde{\gamma}^2$ further, the pair of solutions in the middle collide at $k=\pi/2$, that is, at the zero mode $E_N=0$, and become complex in the form $k=\pi/2\pm\ii\kappa$.
We can also see from Fig.~\ref{fig5} that these two are the only solutions that become complex; all other solutions remain real, as confirmed in Fig.~\ref{fig6}.

The solutions behave differently for odd and even $N$, as shown in Fig.~\ref{fig6}.
For odd $N$, there are $(N+1)$ pieces of real solutions with $0<k<\pi$ for $0\leq\tilde{\gamma}<1$,
two of which collide at $k=\pi/2$ for $\tilde{\gamma}=1$ and become complex for $\tilde{\gamma}>1$.
The point of the collision in the parameter space, $\tilde{\gamma}=\tilde{\gamma}_\textrm{ex}=1$, forms a second-order exceptional point.
The region beyond it is the $\PT$-broken phase for the two eigenvalues.
We analyze the $\PT$-broken region by inserting the Ansatz $k=\pi/2\pm\ii\kappa$ into Eq.~\eqref{eq580}.
For odd $N$, the equation reduces to
\begin{align}\label{eq710}
\frac{\cosh[(N+2)\kappa]}{\cosh(N\kappa)}=\tilde{\gamma}^2,
\end{align}
whose left-hand side converges to unity in the limit $\kappa\to0$ and hence $\kappa=0$ satisfies the equation with $\tilde{\gamma}=\tilde{\gamma}_\textrm{ex}=1$.
This is the same as the tutorial example for $N=1$ given in Sec.~\ref{subsec4.1}.
For large $\kappa$, both the denominator and numerator of the left-hand side of Eq.~\eqref{eq710} reduce to exponential functions, and hence we conclude $\tilde{\gamma}\simeq \exp\kappa$, or $\kappa \simeq \ln\tilde{\gamma}$ for large $\tilde{\gamma}$.
The red chain curves in Fig.~\ref{fig6}(e), for example, are consistent with this estimate.

For even $N$, in addition to the $N$ pieces of real solutions with $0<k<\pi$ for $0\leq\tilde{\gamma}<\sqrt{(N+2)/N}$, there is a solution $k=\pi/2$ of Eq.~\eqref{eq680} with $E_N=0$.
Two of the former solutions collide at $k=\pi/2$ and $\tilde{\gamma}=\sqrt{(N+2)/N}$ together with the additional solution of $k=\pi/2$ and become complex for $\tilde{\gamma}>\sqrt{(N+2)/N}$.
The point $\tilde{\gamma}=\tilde{\gamma}_\textrm{ex}=\sqrt{(N+2)/N}$ in the parameter space forms a third-order exceptional point.
Precisely speaking, we need to confirm that the eigenvectors for the three solutions are identical, which we indeed find in the following.
By inserting the Ansatz $k=\pi/2\pm\ii\kappa$ for the $\PT$-broken region into Eq.~\eqref{eq580} for even $N$, Eq.~\eqref{eq580} reduces to
\begin{align}
\frac{\sinh[(N+2)\kappa]}{\sinh(N\kappa)}=\tilde{\gamma}^2,
\end{align}
whose left-hand side indeed converges to $(N+2)/N$ in the limit $\kappa\to0$, and hence $\kappa=0$ satisfies the equation with $\tilde{\gamma}=\tilde{\gamma}_\textrm{ex}=\sqrt{(N+2)/N}$.
For large $\kappa$, we again find $\kappa \simeq \ln\tilde{\gamma}$.
The red chain curves in Fig.~\ref{fig6}(f), for example, are again consistent with this estimate.

\begin{figure}
\begin{subfigure}{0.45\textwidth}
\includegraphics[width=\textwidth]{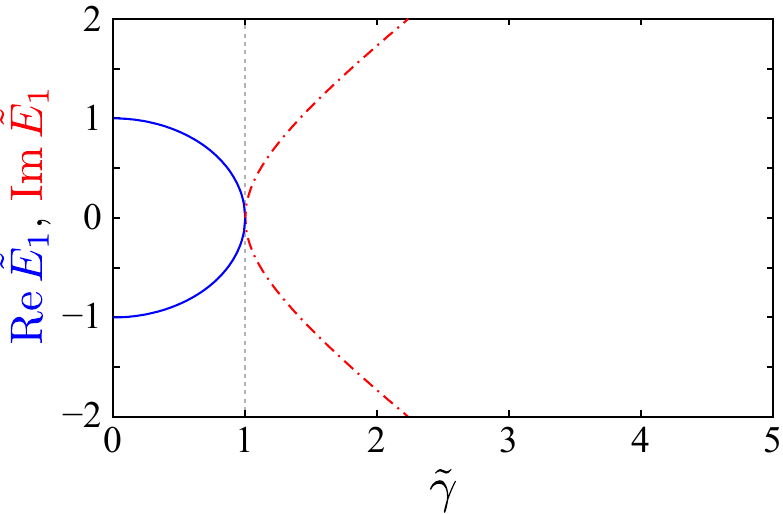}
\caption{$N=1$}
\end{subfigure}
\hfill
\begin{subfigure}{0.45\textwidth}
\includegraphics[width=\textwidth]{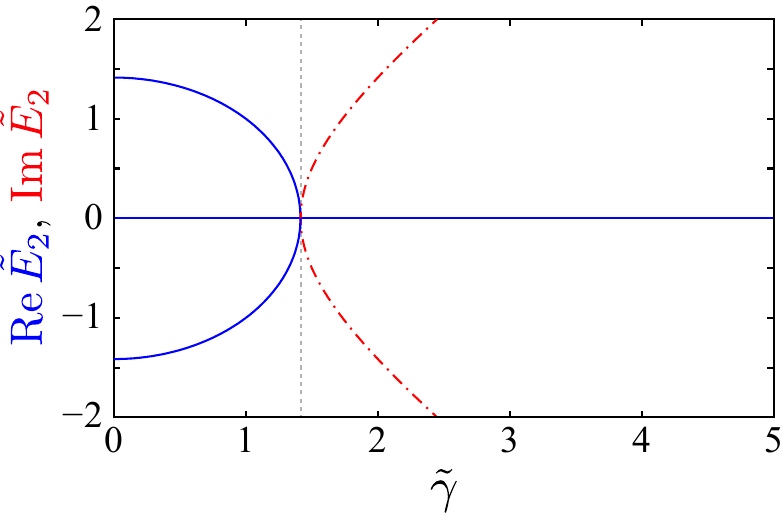}
\caption{$N=2$}
\end{subfigure}
\\[\baselineskip]
\begin{subfigure}{0.45\textwidth}
\includegraphics[width=\textwidth]{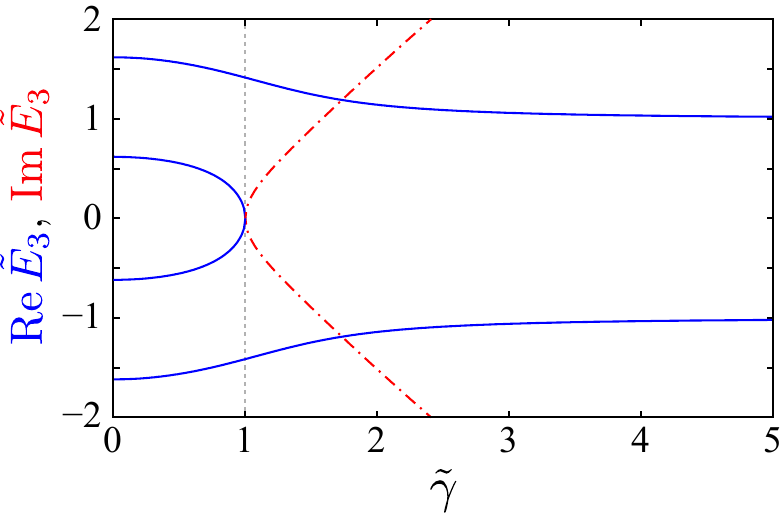}
\caption{$N=3$}
\end{subfigure}
\hfill
\begin{subfigure}{0.45\textwidth}
\includegraphics[width=\textwidth]{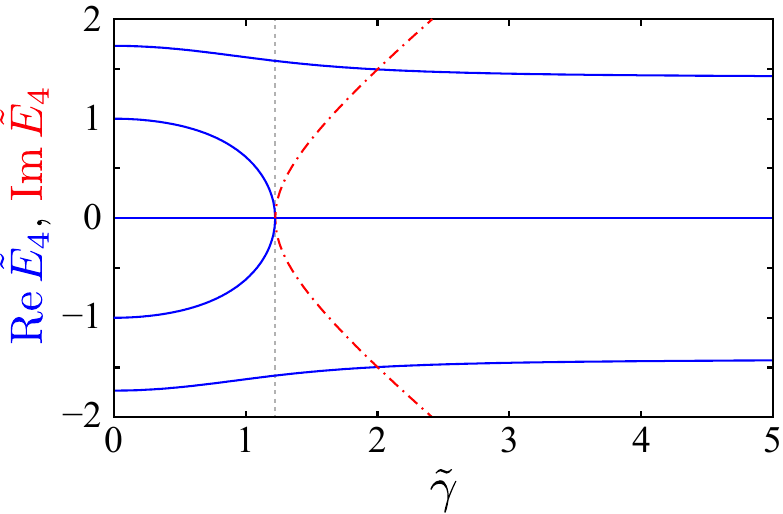}
\caption{$N=4$}
\end{subfigure}
\\[\baselineskip]
\begin{subfigure}{0.45\textwidth}
\includegraphics[width=\textwidth]{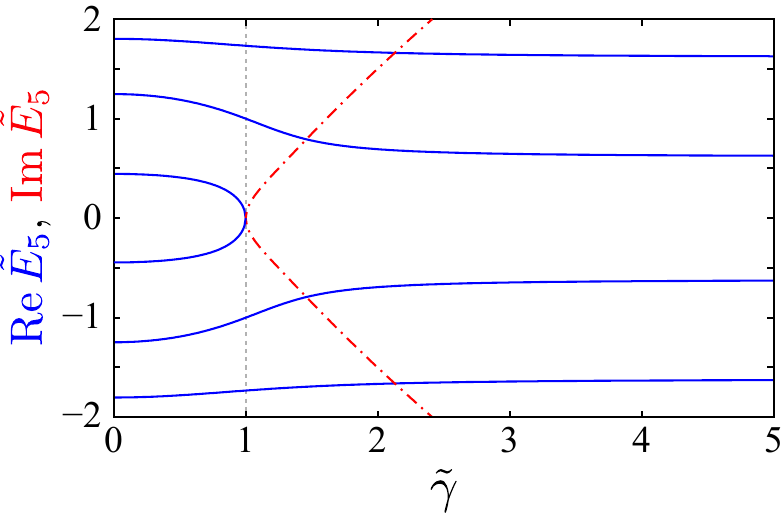}
\caption{$N=5$}
\end{subfigure}
\hfill
\begin{subfigure}{0.45\textwidth}
\includegraphics[width=\textwidth]{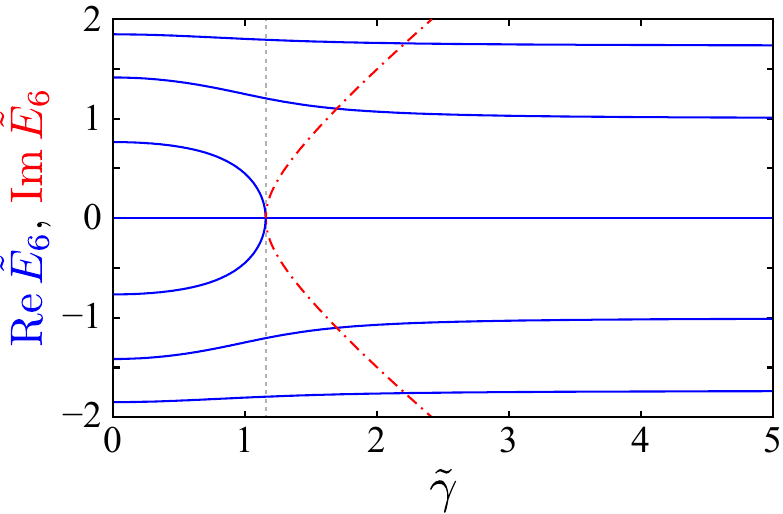}
\caption{$N=6$}
\end{subfigure}
\\
\caption{The $\tilde{\gamma}$-dependence of the real and imaginary parts of the scaled energy eigenvalues $\tilde{E}_N$ for $1\leq N\leq6$.
The blue solid curves indicate the real eigenvalues, while the red chain curves indicate the pure imaginary eigenvalues.}
\label{fig7}
\end{figure}

Figure~\ref{fig7} shows the scaled eigenvalues $\tilde{E}_N=-2\cos k$.
They behave similarly to the solution of $k$.
We can again see in Fig.~\ref{fig7} that the only two eigenvalues in the middle become imaginary eigenvalues, while the other eigenvalues remain real.
The imaginary eigenvalues in the $\PT$-broken region are given by
\begin{align}\label{eq620-1}
\tilde{E}_N^{\pm}=-2\cos\qty(\frac{\pi}{2}\pm\ii\kappa)=\pm2\ii\sinh\kappa,
\end{align}
where we take the double signs in the same order.
Since we estimate $\kappa\simeq\ln\tilde{\gamma}$ for large $\tilde{\gamma}$, the complex eigenvalues are estimated at $\pm\ii\tilde{\gamma}$ for large $\tilde{\gamma}$.
This linear behavior is confirmed, for example, in Fig.~\ref{fig7}(e) and~(f).
For large $N$, we will have a denser eigenvalue distribution, forming the cosine band $\tilde{E}_N=-2\cos k$ for $\tilde{\gamma}=0$;
still only the two in the band center will become imaginary as we increase $\tilde{\gamma}$.

There is also a difference between the cases of odd $N$ and even $N$.
For odd $N$, two of the $(N+1)$ pieces of real eigenvalues in the $\PT$-unbroken region $0\leq\tilde{\gamma}<1$ coalesce at $\tilde{E}_N=0$ at the second-order exceptional point $\tilde{\gamma}=\tilde{\gamma}_\textrm{ex}=1$ and become pure imaginary eigenvalues in the $\PT$-broken region $\tilde{\gamma}>1$.
For even $N$, the collision occurs at $\tilde{\gamma}=\tilde{\gamma}_\textrm{ex}=\sqrt{(N+2)/N}$ together with the additional eigenvalue $\tilde{E}_N=0$, which constitutes the third-order exceptional point.

\subsection{Eigenfunctions}
\label{subsec4.3}

\begin{figure}
\begin{subfigure}{0.45\textwidth}
\includegraphics[width=\textwidth]{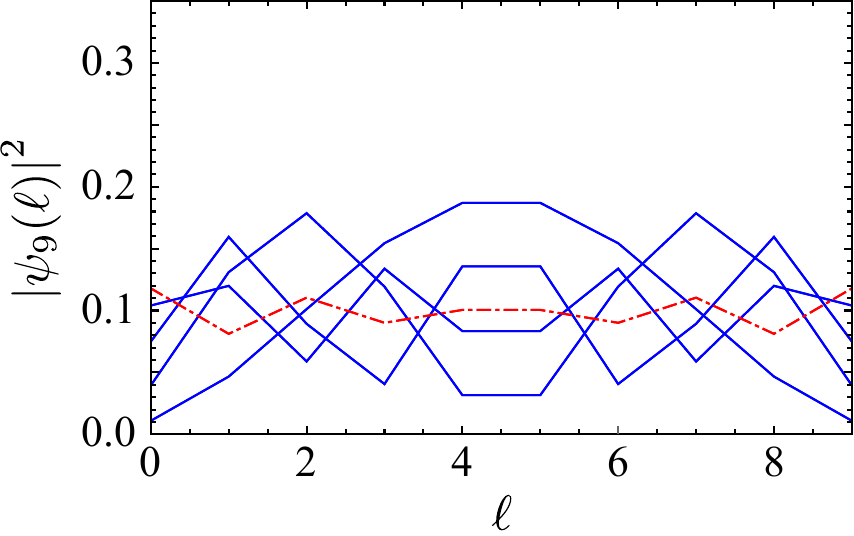}
\caption{$\tilde{\gamma}=0.8$}
\end{subfigure}
\hfill
\begin{subfigure}{0.45\textwidth}
\includegraphics[width=\textwidth]{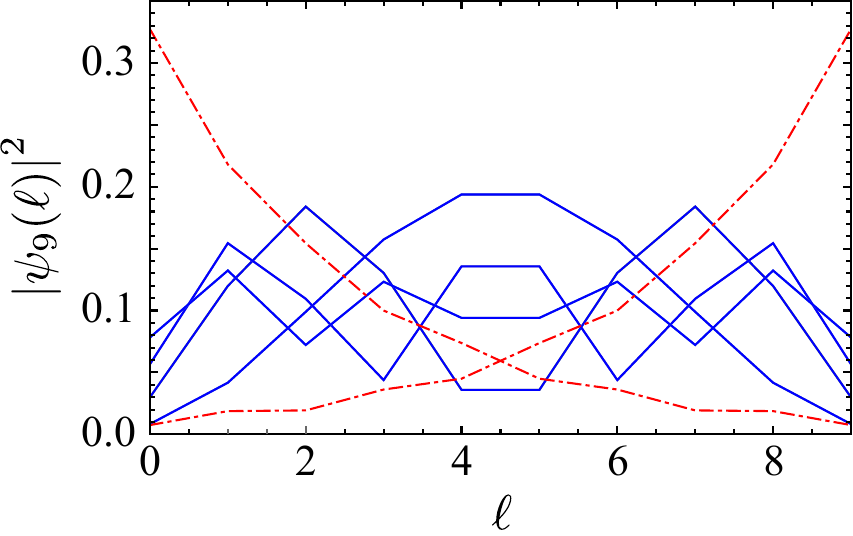}
\caption{$\tilde{\gamma}=1.2$}
\end{subfigure}
\\[\baselineskip]
\begin{subfigure}{0.45\textwidth}
\includegraphics[width=\textwidth]{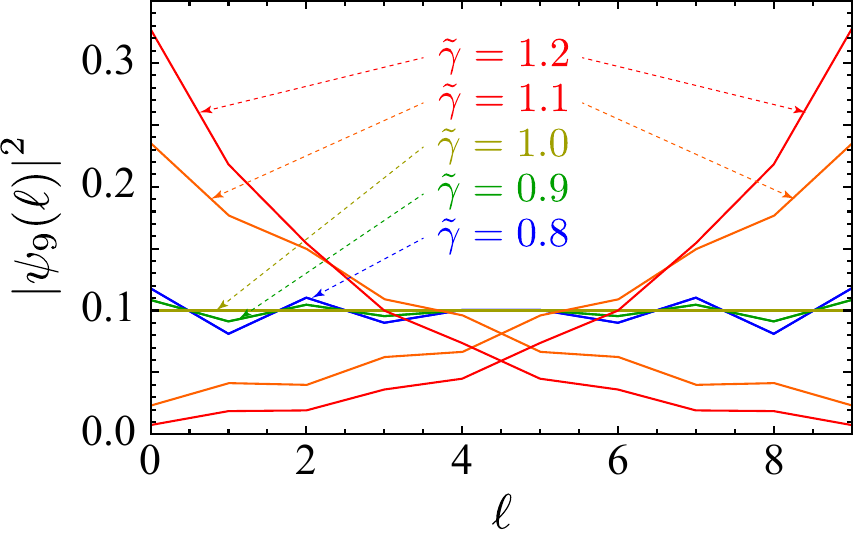}
\caption{$0.8\leq\tilde{\gamma}\leq1.2$}
\end{subfigure}
\hfill
\begin{subfigure}{0.45\textwidth}
\includegraphics[width=\textwidth]{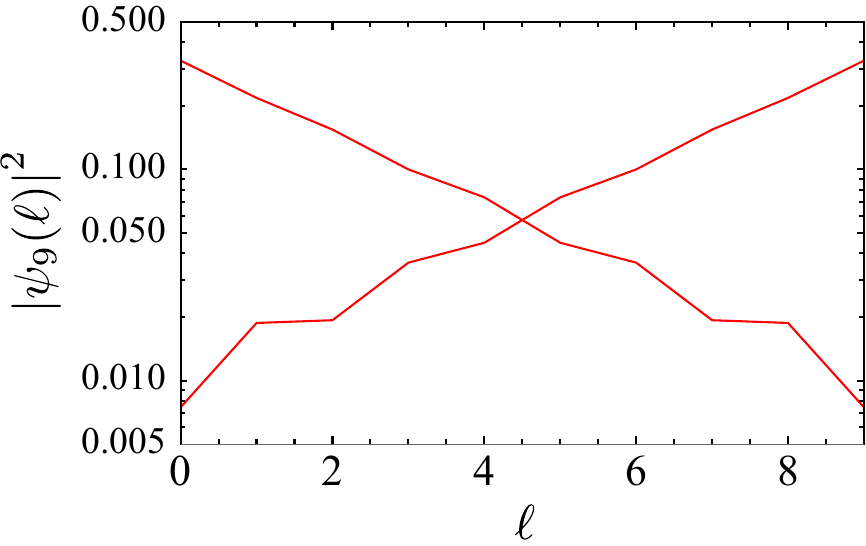}
\caption{$\tilde{\gamma}=1.2$}
\end{subfigure}
\\
\caption{All eigenfunctions in the case of $N=9$ for (a) $\tilde{\gamma}=0.8$ and for (b) $\tilde{\gamma}=1.2$.
The red chain curves indicate the two particular eigenfunctions that coalesce at the exceptional point $\tilde{\gamma}=1$.
Note that in panel (a), and generally for $\tilde{\gamma}<1.0$, the two overlap with each other.
Panel (c) shows the variation of the two particular eigenfunctions for $\tilde{\gamma}=0.8,0.9,1.0,1.1,1.2$.
Panel (d) is a semi-logarithmic plot of the two for $\tilde{\gamma}=1.2$.}
\label{fig8}
\end{figure}

Once we fix the values of $k$ for the eigenvalues $E_N=-2\sqrt{n}\cos k$, we also fix the eigenfunctions~\eqref{eq470} by fixing 
the coefficients $A$ and $B$ from Eq.~\eqref{eq570} in the forms
\begin{align}\label{eq619}
\begin{pmatrix}A \\ B\end{pmatrix} \propto
\begin{pmatrix}
\ee^{\ii k} -\ii\tilde{\gamma} \\
-\qty(\ee^{-\ii k}-\ii\tilde{\gamma})
\end{pmatrix}
\propto
\begin{pmatrix}
\ee^{-\ii Nk}\qty(\ee^{-\ii k}+\ii\tilde{\gamma}) \\
-\ee^{\ii Nk}\qty(\ee^{\ii k}+\ii\tilde{\gamma}) 
\end{pmatrix}
\end{align}
except for normalization.
The first expression for the coefficients $A$ and $B$ in Eq.~\eqref{eq619} yields the eigenfunction explicitly in the form
\begin{align}
\psi_N(\ell)&\propto\qty(\ee^{\ii k} -\ii\tilde{\gamma})\ee^{\ii k \ell}
-\qty(\ee^{-\ii k}-\ii\tilde{\gamma})\ee^{-\ii k \ell} 
    \label{eq620}
=2\ii\sin k(\ell+1)+2\tilde{\gamma}\sin k\ell
\end{align}
for $\ell=0,1,2,\ldots,N$.
See Fig.~\ref{fig8}(a) and~(b) for examples in the case of $N=9$ with $\tilde{\gamma}=0.8$ in the $\PT$-unbroken region and $\tilde{\gamma}=1.2$ in the $\PT$-broken region, respectively.
We find the normalization constant for real $k$ as follows.
For real $k$, we have
\begin{align}
\abs{\psi_N(\ell)}^2
&\propto
\abs{2\ii\sin k(\ell+1)+2\tilde{\gamma}\sin k\ell}^2
=4\sin^2 k(\ell+1)+4\tilde{\gamma}^2\sin^2 k\ell.
\end{align}
Using the formulae
\begin{align}
&\sum_{\ell=0}^N 4\sin^2 k\ell=2N+1-\frac{\sin k(2N+1)}{\sin k},\\
&\sum_{\ell=0}^N 4\sin^2 k(\ell+1)=2N+3-\frac{\sin k(2N+3)}{\sin k}, 
\end{align}
we arrive at
\begin{align}\label{eq800}
\mathcal{N}^2&\coloneqq\sum_{\ell=0}^N\abs{\psi_N(\ell)}^2
=2N(1+\tilde{\gamma}^2)+4,
\end{align}
where we used Eq.~\eqref{eq580} twice.

The eigenfunctions at the exceptional point are found as follows.
For odd $N$, the eigenfunction at the second-order exceptional point $\tilde{\gamma}=1$ has the wave number $k=\pi/2$, and hence Eq.~\eqref{eq620} with the normalization~\eqref{eq800} yields
\begin{align}
\psi_N^\textrm{EP}(\ell)
&=\frac{\ii}{\sqrt{N+1}}  \sin\qty[\frac{\pi}{2}(\ell+1)]+ \frac{1}{\sqrt{N+1}}\sin\qty(\frac{\pi}{2}\ell) \nonumber \\
&=
\begin{cases}
\frac{\ii}{\sqrt{N+1}} (-1)^{\ell/2} &\quad\mbox{for even $\ell$}, \\
\frac{1}{\sqrt{N+1}} (-1)^{(\ell-1)/2}&\quad\mbox{for odd $\ell$} ,
\end{cases}
\end{align}
and hence $\abs{\psi_N^\textrm{EP}(\ell)}^2\equiv 1/(N+1)$, which is indeed demonstrated for $N=9$ in Fig.~\ref{fig8}(c).
For even $N$, the eigenfunction at the third-order exceptional point $\tilde{\gamma}=\sqrt{(N+2)/N}$ has the wave number $k=\pi/2$, and hence
\begin{align}
\psi_N^\textrm{EP}(\ell)&=\frac{\ii}{\sqrt{N+2}}  \sin\qty[\frac{\pi}{2}(\ell+1)]+ \frac{1}{\sqrt{N}}\sin\qty(\frac{\pi}{2}\ell) \nonumber \\
&=
\begin{cases}
\frac{\ii}{\sqrt{N+2}} (-1)^{\ell/2} &\quad\mbox{for even $\ell$}, \\
\frac{1}{\sqrt{N}} (-1)^{(\ell-1)/2}&\quad\mbox{for odd $\ell$} ,
\end{cases}
\end{align}
and its square modulus oscillates slightly:
\begin{align}
\abs{\psi_N^\textrm{EP}(\ell)}^2=\begin{cases}
1/(N+2)&\quad\mbox{for even $\ell$,}\\
1/N&\quad\mbox{for odd $\ell$} .
\end{cases}
\end{align}
Note that since the eigenfunction~\eqref{eq620} is uniquely determined when we fix $k$ and $\tilde{\gamma}$,  all eigenvalues for $k=\pi/2$ at the exceptional point, $\tilde{\gamma}=1$ for odd $N$ and $\tilde{\gamma}=\sqrt{(N+2)/N}$ for even $N$, certainly coalesce with an identical eigenvector, not degenerate with mutually orthogonal eigenvectors.

When $k$ is complex in the $\PT$-broken region, we put $k=\pi/2\pm\ii\kappa$ into
the eigenvector~\eqref{eq620}, obtaining
\begin{align}\label{eq850}
\psi_N^\pm(\ell)&\propto\ii^{\ell+1}\qty(\ee^{\mp\kappa}-\tilde{\gamma})\ee^{\mp\kappa\ell}
-(-\ii)^{\ell+1}\qty(\ee^{\pm\kappa}+\tilde{\gamma})\ee^{\pm\kappa\ell},
\end{align}
for which we would avoid writing down the complicated normalization.
The double signs are taken in the same order here and Eq.~\eqref{eq620-1}.
In fact, the expression~\eqref{eq850} is convenient in estimating the eigenfunction for the lower sign.
Since we estimate at $\ee^{\kappa}\simeq\tilde{\gamma}$ for large $\tilde{\gamma}$, the first term on the right-hand side of Eq.~\eqref{eq850} is suppressed, and we find 
\begin{align}\label{eq73}
\abs{\psi_N^{-}(\ell)}\propto \ee^{-\kappa\ell}
\quad\mbox{for $\tilde{E}_N^{-}=-2\ii\sinh\kappa$},
\end{align}
which grows towards the central site, decaying towards the peripheral sites.
For the upper sign, on the other hand, the second expression of the coefficients $A$ and $B$ in Eq.~\eqref{eq619} is more convenient:
\begin{align}
\psi_N^\pm(\ell)&\propto(-\ii)^{-(N+1-\ell)}\qty(\ee^{\pm\kappa}-\tilde{\gamma})\ee^{\pm\kappa(N-\ell)}
-\ii^{N+1-\ell}\qty(\ee^{\mp\kappa}+\tilde{\gamma})\ee^{\mp\kappa(N-\ell)}.
\end{align}
The first term on the right-hand side is suppressed for large $\tilde{\gamma}$ because of the estimate $\ee^{\kappa}\simeq\tilde{\gamma}$, and hence we find  
\begin{align}\label{eq75}
\abs{\psi_N^{+}(\ell)}\propto \ee^{-\kappa(N-\ell)}
\quad\mbox{for $\tilde{E}_N^{+}=+2\ii\sinh\kappa$},
\end{align}
which grows towards the peripheral sites, decaying towards the central site.

Figure~\ref{fig8}(c) shows how the eigenvectors for the specific two eigenvalues change 
as we increase $\tilde{\gamma}$ from the $\PT$-unbroken region to the $\PT$-broken region.
We saw in Eq.~\eqref{eq520-0} that the probability distributions of the eigenvectors in the $\PT$-broken region 
are
spatially imbalanced, and now we see that this is a general feature.
The exponential behavior in Eqs.~\eqref{eq73} and~\eqref{eq75} is numerically confirmed in Fig.~\ref{fig8}(d) for $N=9$ with $\tilde{\gamma}=1.2$.

\subsection{Implications for the dynamics on the Bethe lattice}
At this point, let us remember that the limited number of states that span the effective Hamiltonian~\eqref{eq460-1} are related to all the basis states on the original Bethe lattice as in Eq.~\eqref{eq330}.
This reduction in the number of states was possible because only the eigenstates in which all sites in each generation have identical amplitudes can penetrate to the central site and contribute to transport from the peripheral sites to the central site.
Going back from the reduced set of states to the original basis states, we find that wave amplitudes of all sites for an extended eigenstate are given by
\begin{align}
\label{eq330rev}
&\braket{[0]|\psi}=\braket{(0)|\psi},
\qquad
\braket{[\mu]|\psi}=\frac{1}{\sqrt{\ntot_1}}\braket{(1)|\psi}\quad\mbox{for all $1\leq \mu\le n_1$},
\nonumber\\
&\braket{[\mu,\nu]|\psi}=\frac{1}{\sqrt{\ntot_2}}\braket{(2)|\psi}\quad\mbox{for all $1\leq \mu\le n_1$ and $1\le\nu\le n_2$},\cdots
\nonumber\\
&\braket{[\mu_1, \mu_2, \mu_3, \cdots,\mu_{N-1},\mu_N]|\psi}=\frac{1}{\sqrt{\ntot_N}}\braket{(N)|\psi}
\nonumber\\
&\qquad\mbox{for all $1 \le \mu_1 \le n_1$, $1 \le \mu_2 \le n_2$, $\cdots$, $1\le\mu_N\le n_N$.}
\end{align}
We can thus infer the wave amplitudes of all sites of an extended state on the Bethe lattice from Eq.~\eqref{eq620}, which is exemplified in Fig.~\ref{fig8}.

We then mention implications regarding the dynamics of the initial-value problem.
The initial condition should be a superposition of many eigenstates, and hence the dynamics in the initial-value problem should be expanded in terms of the eigenstates that we find here.
For the initial state on the peripheral sites to reach the central site, we need to excite a substantial part of the peripheral sites, in order to have a sufficient overlap with the state $\ket{(N)}/\sqrt{\ntot_N}$.
Excitation of a single peripheral site would have an overlap of only the amplitude $1/\sqrt{n_N^\textrm{tot}}$.

In the $\PT$-symmetric region up to the exceptional point, 
the zero mode should carry the greatest current across all channels.
As another point, the phase velocity is greatest at the zero mode, which has almost linear dispersion for large $N$.
Hence, in the initial-value problem, the current on the zero mode should run fastest towards the central site.

In the $\PT$-broken region, on the other hand, the state with the positive imaginary eigenvalue in Eq.~\eqref{eq75} will excel in time.
Therefore, the current should be dominated by this single channel after a certain point in time.

\section{Currents carried by extended eigenstates}
\label{sec5}

In the present section, we finally define and compute the current carried by the extended eigenstates given in Sec.~\ref{sec4}.
We here analyze the expectation values of the current operator with respect to the extended eigenstates.
Another way to analyze quantum transport in the present model is to attach leads to the $\PT$-symmetric linear chain in Eq.~\eqref{eq420} and compute the conductance using the Landauer formula; see \ref{appB}.

We discover numerically for general $N$ that the current expectation value is maximum at the exceptional point for the eigenstates that coalesce at zero energy.
In other words, the current generally decreases as we increase the amplitude of the source potential at the peripheral sites beyond the exceptional point.

\subsection{Case of $N=1$}
\label{subsec5.1}
Let us first consider the current operator for the solutions found in Subsec.~\ref{subsec4.1} for $N=1$ as a tutorial example.
To make the current flowing from a peripheral site $\ket{[\mu]}$ on the first generation to the origin site $\ket{[0]}$ positive,
we define
\begin{align}
J([0];[\mu])\coloneqq\ii\qty(\dyad{[0]}{[\mu]}-\dyad{[\mu]}{[0]}).
\end{align}
Since the amplitudes on all the peripheral sites must be equal to each other for the extended eigenstates, as we found in Secs.~\ref{sec3} and~\ref{sec4}, the expectation values of the current operator for all $\mu=1,2,3,\ldots,n_1$ with respect to any extended eigenstate should be equal to each other.
We can thereby use the state $\ket{(1)}$ in Eq.~\eqref{eq330} instead of each $\ket{[\mu]}$, and define the scaled current operator from the first generation to the zeroth generation:
\begin{align}
    \label{eq870}
\tilde{J}_1(0)&\coloneqq\ii\qty(\dyad{(0)}{(1)}-\dyad{(1)}{(0)})
=\frac{1}{\sqrt{n_1}}\sum_{\mu=1}^{n_1}J([0];[\mu]),
\end{align}
where the subscript $1$ of the current operator indicates the case $N=1$ while the argument $(0)$ is added to make the notation consistent with the general case described in Subsec.~\ref{subsec5.2}.
In the basis of the effective Hamiltonian~\eqref{eq500}, the current operator~\eqref{eq870} is given in the form of a Hermitian matrix 
\begin{align}\label{eq890}
\tilde{J}_1(0)=\begin{pmatrix}
0 & \ii\\
-\ii & 0 
\end{pmatrix}.
\end{align}

The expectation values of this scaled current operator with respect to the eigenvectors in Eq.~\eqref{eq520-2} in the $\PT$-unbroken region $\tilde{\gamma}<1$ are equal to each other:
\begin{align}\label{eq900}
\ev{\tilde{J}_1(0)}{\psi_1^\pm}=\tilde{\gamma}.
\end{align}
Note that in calculating the expectation value, we used $\bra{\psi_1^\pm}\coloneqq\ket{\psi_1^\pm}^\dag$ instead of the left eigenvectors $\bra{\phi_1^\pm}$ in Eq.~\eqref{eq520-1}.
Since the current operators~\eqref{eq870} and \eqref{eq890} are Hermitian operators, the expectation value~\eqref{eq900} must be real. 
See \ref{appA} for this point.

We understand Eq.~\eqref{eq900} in the following manner.
In the $\PT$-symmetric effective Hamiltonian~\eqref{eq500},
the wave-function amplitude is injected to the first generation at the rate $\tilde{\gamma}$ and removed from the zeroth generation at the same rate.
Equation~\eqref{eq900} shows that the amplitude is indeed transported from the first generation to the zeroth generation at the rate $\tilde{\gamma}$.

\begin{figure}
\begin{subfigure}{0.45\textwidth}
\includegraphics[width=\textwidth]{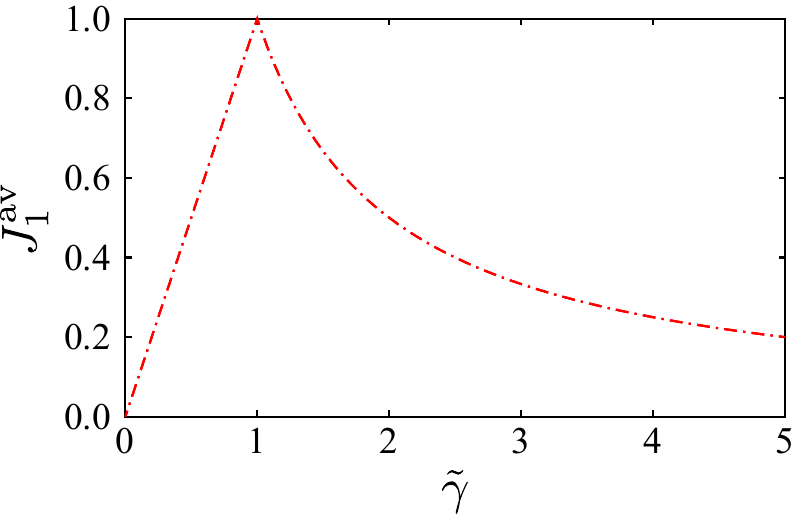}
\caption{$N=1$}
\end{subfigure}
\hfill
\begin{subfigure}{0.45\textwidth}
\includegraphics[width=\textwidth]{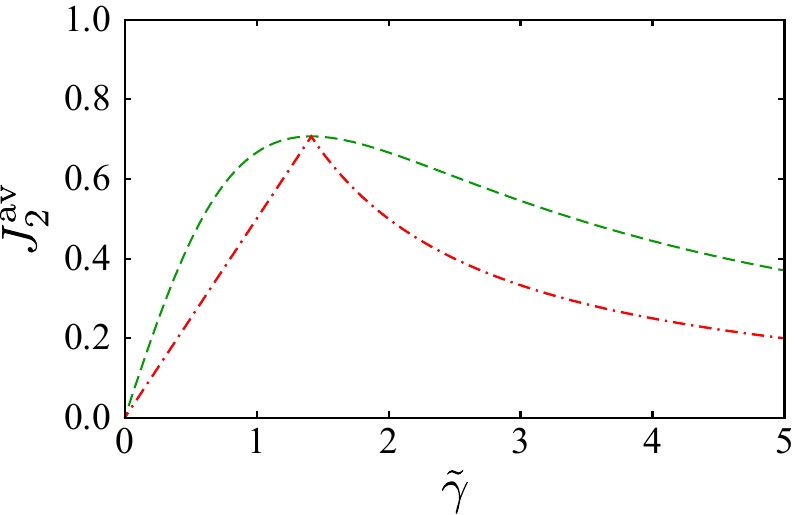}
\caption{$N=2$}
\end{subfigure}
\\[\baselineskip]
\begin{subfigure}{0.45\textwidth}
\includegraphics[width=\textwidth]{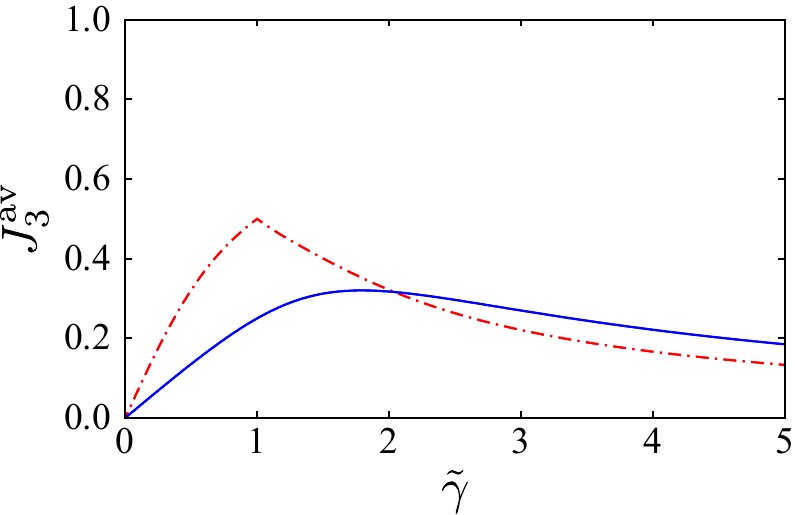}
\caption{$N=3$}
\end{subfigure}
\hfill
\begin{subfigure}{0.45\textwidth}
\includegraphics[width=\textwidth]{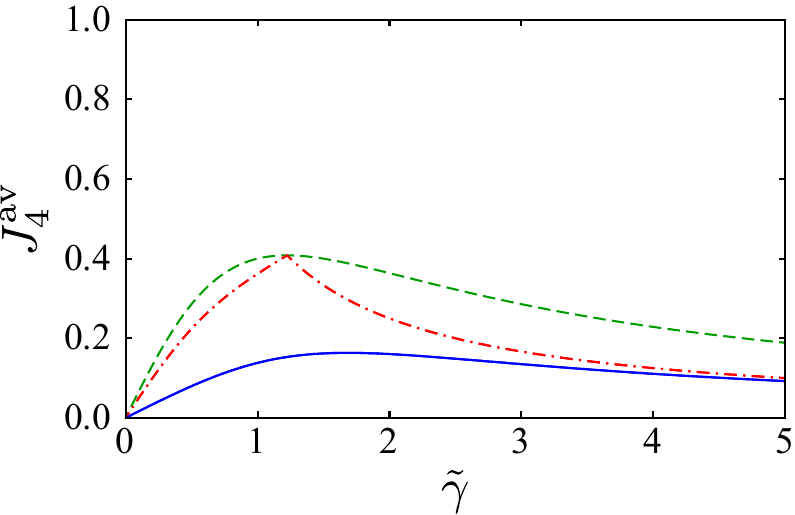}
\caption{$N=4$}
\end{subfigure}
\\[\baselineskip]
\begin{subfigure}{0.45\textwidth}
\includegraphics[width=\textwidth]{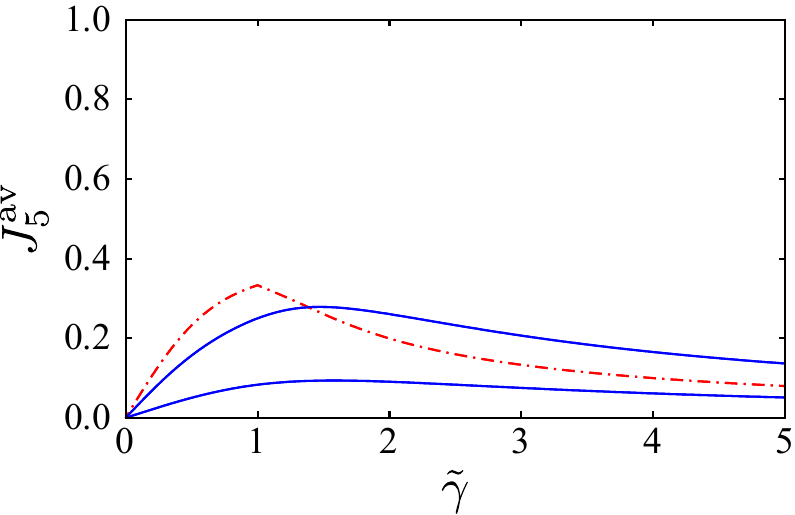}
\caption{$N=5$}
\end{subfigure}
\hfill
\begin{subfigure}{0.45\textwidth}
\includegraphics[width=\textwidth]{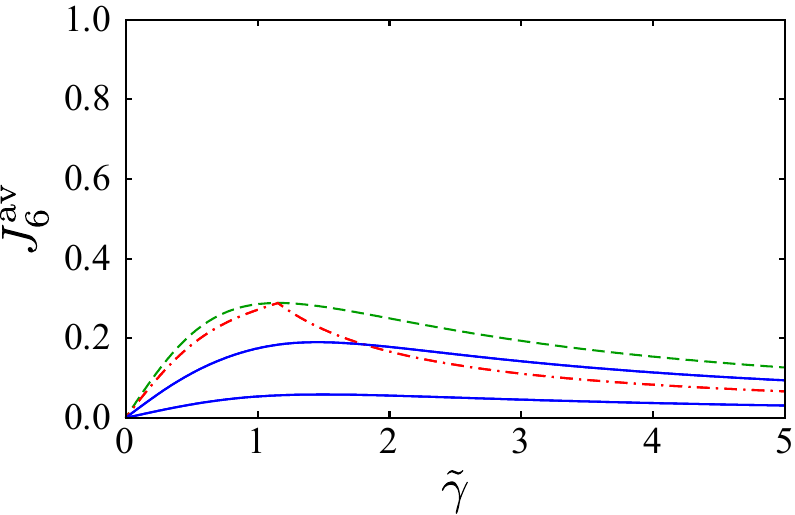}
\caption{$N=6$}
\end{subfigure}
\\
\caption{The $\tilde{\gamma}$-dependence of the average current expectation values with respect to all the eigenstates for $1\leq N\leq6$.
In each panel, the red chain curve indicates the one with respect to the eigenstates that coalesce at the exceptional point and acquire complex eigenvalues beyond it, 
while the green broken curve indicates the one with respect to the additional eigenstate with $E=0$ for even $N$.
The blue solid curves indicate the ones with respect to the other eigenstates with nonzero real eigenvalues.
}
\label{fig9}
\end{figure}

We now move over to the $\PT$-broken region $\tilde{\gamma}>1$.
The expectation values of the current operator~\eqref{eq890} with respect to the eigenvectors in Eq.~\eqref{eq520-0} are again equal to each other:
\begin{align}\label{eq920}
\ev{\tilde{J}_1(0)}{\psi_1^\pm}=\frac{1}{\tilde{\gamma}}.
\end{align}
This result is different from Eq.~\eqref{eq900} in the $\PT$-unbroken region because one element in each eigenvector is pure imaginary in the $\PT$-broken region.
We plot Eqs.~\eqref{eq900} and~\eqref{eq920} in Fig.~\ref{fig9}(a).
The $\tilde{\gamma}$-dependence of the current has a cusp at the exceptional point between the $\PT$-unbroken and $\PT$-broken regions.

Equation~\eqref{eq920} means that the current counterintuitively decreases as we inject the amplitude at a rate beyond the exceptional point, $\tilde{\gamma}>1$, and vanishes in the limit of infinite strengths of the sources and the drain.
We see in Fig.~\ref{fig9} and in Subsec.~\ref{subsec5.2} that this is a general feature of the $\PT$-broken region for general $N$.
We also see the counterintuitive decrease of current \textit{e.g.}\ in a molecular junction.
Reference~\cite{Toroker09} computed the current using the Landauer formula and found that the current decreases when the connection to leads is strong; see its Fig.~7.

We can understand it schematically in the present simple case of $N=1$.
For $\tilde{\gamma}=0$, the Hamiltonian $\tilde{H}^\textrm{eff}_1$ in Eq.~\eqref{eq500} reduces to $-\sigma_x$ and
the current expectation value~\eqref{eq900} vanishes there. As we turn on $\tilde{\gamma}$, the current emerges as a result of the first-order perturbation of $\tilde{\gamma}$, which is indeed exact in the particular case of $N=1$, as in Eq.~\eqref{eq900}. 

In the limit of $\tilde{\gamma}\to\infty$, on the other hand, the Hamiltonian $\tilde{H}^\textrm{eff}_1$ reduces to a diagonal one with the eigenvectors $(1,0)^T$ and $(0,1)^T$.
The current expectation value vanishes there too.
As we turn on the off-diagonal elements, the current emerges this time as a result of the first-order perturbation of $1/\tilde{\gamma}$, which is again exact in the particular case of $N=1$, as in Eq.~\eqref{eq920}. 

Some would call the vanishing current in the limit of $\tilde{\gamma}\to\infty$ a result of the quantum Zeno effect; see \textit{e.g.}\ Ref.~\cite{Syassen08}. 
Although the ``quantum Zeno effect"~\cite{Misra77, Nakazato96-1, Nakazato96-2} originally refers to the prohibition of decay due to frequently repeated projection measurements, the present suppression of the current is due to excelling diagonal elements that eliminate off-diagonal coherence, and hence the suppression of the current in the limit of large $\tilde{\gamma}$ may be called the quantum Zeno effect in a broad sense.

\subsection{Case of general $N$}
\label{subsec5.2}

For general $N$, we follow the argument for $N=1$ in Subsec.~\ref{subsec5.1}.
The current operator on a link connecting $\ket{[\mu,\nu]}$ and $\ket{[\mu,\nu,\kappa]}$, for example, is defined by
\begin{align}\label{eq740}
J([\mu,\nu];[\mu,\nu,\kappa])=\ii\qty(
\dyad{[\mu,\nu]}{[\mu,\nu,\kappa]}-\dyad{[\mu,\nu,\kappa]}{[\mu,\nu]}).
\end{align}
The convention is to make it positive when the current runs toward the origin site.

Since all amplitudes on the sites of the outer generation (the third generation in the example above) should be equal to each other in the extended eigenstates, summing up with respect to the sites does not matter, and we thereby define the scaled current operator
\begin{align}
\tilde{J}([\mu,\nu])\coloneqq\frac{1}{\sqrt{n_3}}\sum_{\kappa=1}^{n_3}J([\mu,\nu];[\mu,\nu,\kappa]).
\end{align}
Again, since the current expectation value should be the same for all $[\mu,\nu]$, we average $\tilde{J}([\mu,\nu])$ over all $\mu$ and $\nu$, and rewrite it in terms of the states in Eq.~\eqref{eq330}:
\begin{align}
\tilde{J}_N(2)&\coloneqq\ii\qty(\dyad{(2)}{(3)}-\dyad{(3)}{(2)})
=\frac{1}{n_1n_2}\sum_{\mu=1}^{n_1}\sum_{\nu=1}^{n_2}\tilde{J}([\mu,\nu]).
\end{align}
For general $\ell=0,1,2,\ldots,N-1$, we define the scaled current from the $(\ell+1)$th generation to the $\ell$th generation in the form
\begin{align}\label{eq970}
\tilde{J}_N(\ell)&\coloneqq\ii\qty(\dyad{(\ell)}{(\ell+1)}-\dyad{(\ell+1)}{(\ell)}).
\end{align}

Let us now focus on the simplest case specified by Eq.~\eqref{eq430}; all $n_\ell$ are set to $n$ and both $\gamma_0$ and $\gamma_N$ are set to $\gamma$.
The expectation value of the scaled current operator~\eqref{eq970} with respect to the eigenfunction~\eqref{eq620} for real $k$ with the normalization~\eqref{eq800} is explicitly given by
\begin{align}\label{eq760}
\ev{\tilde{J}_N(\ell)}{\psi_N}&=-2\Im \qty[\psi_N(\ell+1) \psi_N(\ell)^\ast]
 =\frac{8\tilde{\gamma}}{\mathcal{N}^2}\qty[\sin^2k(\ell+1) -\sin k\ell \sin k(\ell+2)]
\nonumber\\
&=\frac{4\tilde{\gamma}\sin^2k}{N(1+\tilde{\gamma}^2)+2},
\end{align}
which is independent of $\ell$.
At the exceptional point, we find the following values. 
For odd $N$, the exceptional point is given by $\tilde{\gamma}=1$, at which two eigenstates coalesce to $k=\pi/2$.
Hence, Eq.~\eqref{eq760} is reduced to
\begin{align}
\ev{\tilde{J}_N(\ell)}{\psi_N^\textrm{EP}}=\frac{2}{N+1} \quad\mbox{for odd $N$}.
\end{align}
For even $N$, on the other hand, the exceptional point is given by $\tilde{\gamma}=\sqrt{(N+2)/N}$, at which three eigenstates coalesce to $k=\pi/2$.
Therefore, Eq.~\eqref{eq760} is reduced to
\begin{align}
\ev{\tilde{J}_N(\ell)}{\psi_N^\textrm{EP}}=\frac{2}{\sqrt{N(N+2)}} \quad\mbox{for even $N$}.
\end{align}
We find in Fig.~\ref{fig9} that these values give the maximum current for a fixed value of $N$ forming a cusp at the exceptional point.

\begin{figure}
\begin{subfigure}{0.45\textwidth}
\includegraphics[width=\textwidth]{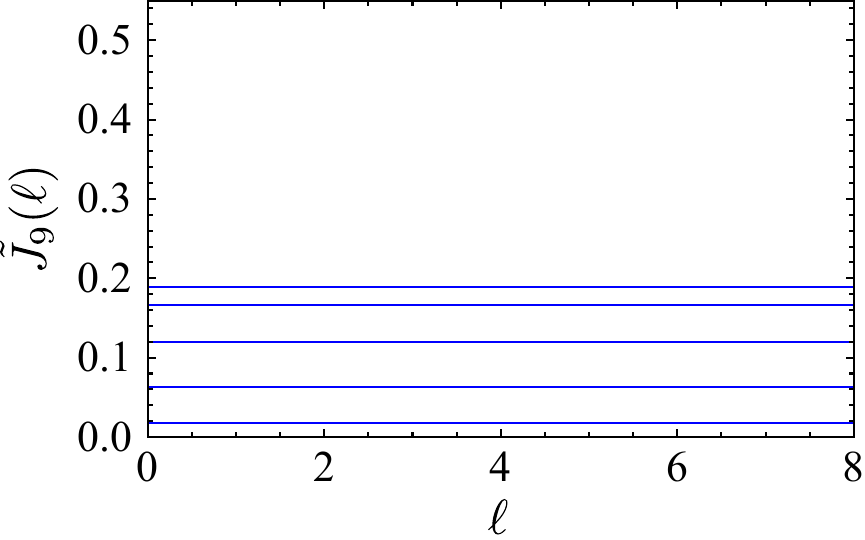}
\caption{$\tilde{\gamma}=0.8$}
\end{subfigure}
\hfill
\begin{subfigure}{0.45\textwidth}
\includegraphics[width=\textwidth]{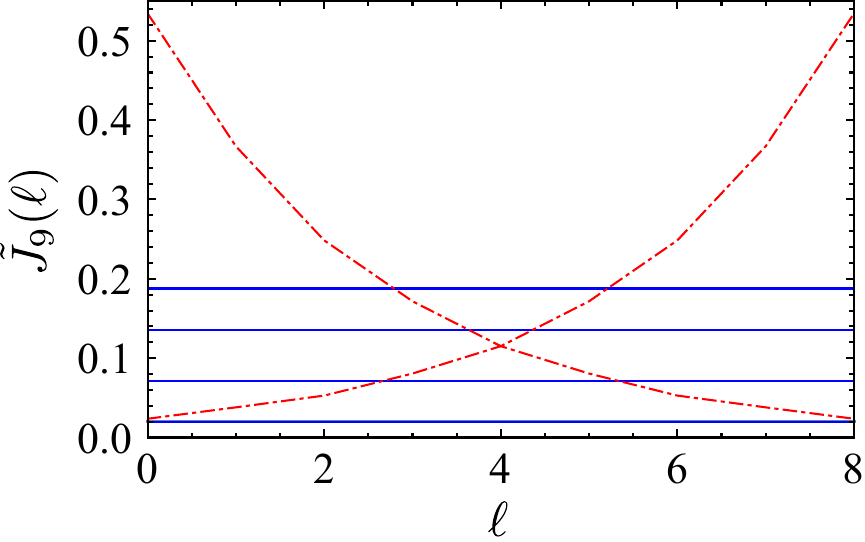}
\caption{$\tilde{\gamma}=1.2$}
\end{subfigure}
\\[\baselineskip]
\begin{subfigure}{0.45\textwidth}
\includegraphics[width=\textwidth]{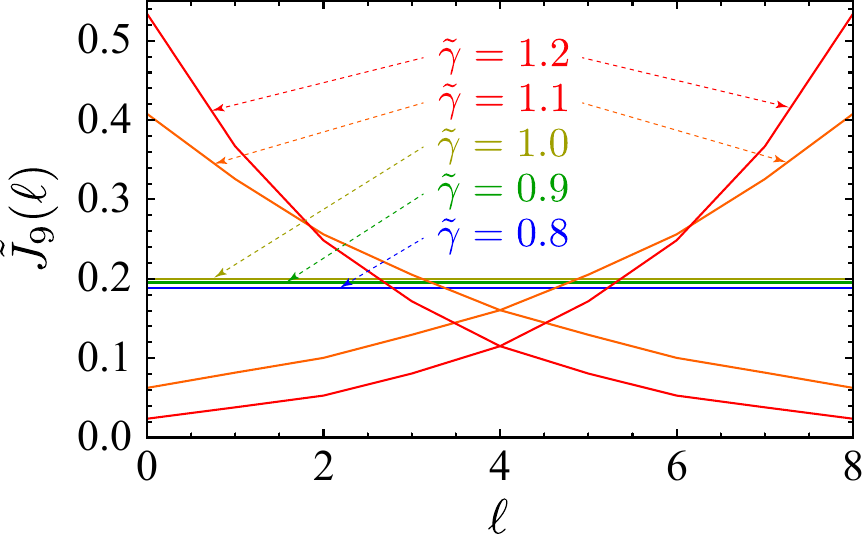}
\caption{$0.8\leq\tilde{\gamma}\leq1.2$}
\end{subfigure}
\hfill
\begin{subfigure}{0.45\textwidth}
\includegraphics[width=\textwidth]{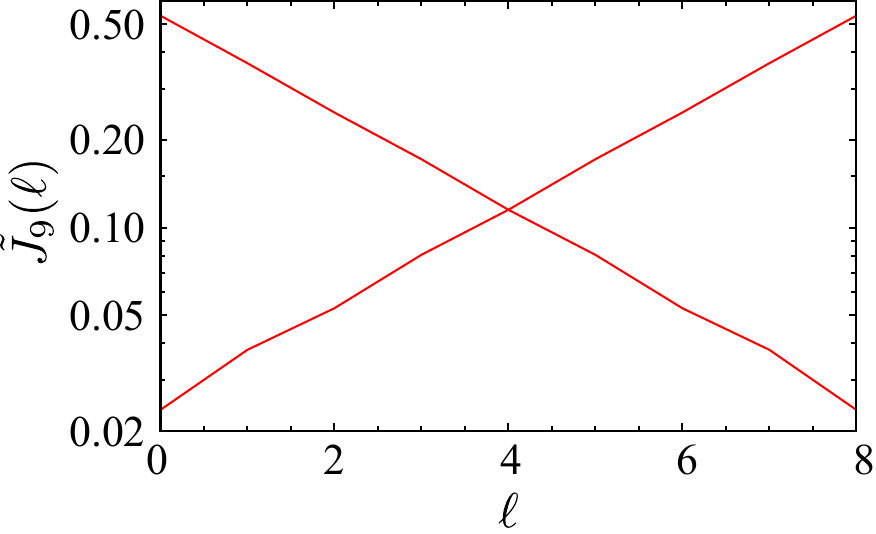}
\caption{$\tilde{\gamma}=1.2$}
\end{subfigure}
\\
\caption{Current expectation values with respect to all eigenfunctions in the case of $N=9$ for (a) $\tilde{\gamma}=0.8$ and for (b) $\tilde{\gamma}=1.2$.
The red chain curves indicate the one with respect to the two particular eigenfunctions that coalesce at the exceptional point $\tilde{\gamma}=1$.
Note that in the panel (a), and generally for $\tilde{\gamma}<1.0$, the two overlap with each other.
The panel (c) shows the variation of the current expectation values with respect to the two particular eigenfunctions for $\tilde{\gamma}=0.8,0.9,1.0,1.1,1.2$.
The panel (d) is a semilogarithmic plot of the two for $\tilde{\gamma}=1.2$.}
\label{fig10}
\end{figure}

In the $\PT$-broken region, the current expectation value is not conserved spatially.
Figure~\ref{fig10} shows the current expectation values for $N=9$.
It demonstrates that the current expectation values for the two eigenstates that coalesce at the exceptional point depend on $\ell$ in the $\PT$-broken region, and behave similarly to the eigenfunctions shown in Fig.~\ref{fig8}.

For this reason, we use the average current
\begin{align}\label{eq1010}
\tilde{J}_N^\textrm{av}\coloneqq\frac{1}{N}\sum_{\ell=0}^{N-1}\tilde{J}_N(\ell)
\end{align}
to plot in Fig.~\ref{fig9}, which shows the current expectation values with respect to all the eigenstates.
Except for the ones for the eigenstates with the complex eigenvalues, it is indeed equal to $\tilde{J}_N(\ell)$ for any $\ell$.
We again note that the current is suppressed in the limit of large $\tilde{\gamma}$, and hence there is always a current maximum in the middle. 
The maximum current is achieved at the exceptional point by all the zero-energy eigenstates, namely the zero modes, which coalesce there.
We also note that for even $N$, the additional zero mode gives the current maximal for a fixed value of $\tilde{\gamma}$.

\section{Case of random tight-binding model}
\label{sec6}

In this section, we analyze a random-hopping tight-binding model, rather than the uniform-hopping $\PT$-symmetric one in ~\eqref{eq420}.
We can randomly distribute the hopping amplitudes in the following two ways.
First, we can make the hopping on the original Bethe lattice random.
Second, we can remove the assumption~\eqref{eq430} and randomly distribute the link numbers. 
In either case, a slight generalization of the recursive method discussed in Ref.~\cite{Mahan01} allows us to reduce the model to an effective tight-binding chain of the form~\eqref{eq420} with random hopping on the superdiagonal and subdiagonal elements.

A random-hopping tight-binding model without the present complex potentials is known to exhibit a unique feature.
In one-dimensional random systems, almost all eigenstates are localized due to the Anderson localization.
The zero mode of the random-hopping tight-binding model is an exception due to chiral symmetry.
It is widely accepted that the inverse localization length of the model diverges as $\kappa(E)\sim\abs{\ln E^2}^{-1}$~\cite{Dyson53, Theodorou76, Eggarter78, Ziman82, Brouwer02}, and hence only the zero mode is extended.

For the random-hopping model with the source and drain, the problem is reduced 
to the Hamiltonian~\eqref{eq460-1} with the random hopping term, for which we again assume $\gamma \coloneqq \gamma_0=\gamma_N>0$ as in Eq.~\eqref{eq430b} throughout the present section.
We find numerically the following deviation from the results of the uniform case $n \coloneqq n_1=n_2=\cdots=n_N$ presented in the previous section~\ref{sec4}.
In the uniform case, Figs.~\ref{fig5} and~\ref{fig7} demonstrate that all eigenvalues are real before two real eigenvalues coalesce at $E=0$ at the exceptional point $\tilde{\gamma}_\textrm{ex}$ and split into two pure-imaginary eigenvalues.
In the random case, the reality of the eigenvalues is broken for any non-zero value of $\tilde{\gamma}$, although the reflection symmetry of the complex spectrum with respect to the imaginary axis is preserved, and accordingly, the exceptional-point coalescence of two eigenvalues occurs on the imaginary axis but not generally at $E=0$; 
see Fig.~\ref{fig11} below, for example.

In the present section, we introduce randomness to the hopping of the tight-binding model in terms of a box distribution of the form
 $   n_\ell=n\qty(1+\Delta)$,
where we choose $\Delta$ independently for each $\ell$ and randomly from the range $[-\delta,\delta]$ with $0<\delta<1$.
We then scale the Hamiltonian and the current operator with the factor $1/\sqrt{n}$, and hence the hopping elements randomly deviate from the negative unity given in Eq.~\eqref{eq420}.
We numerically diagonalize each random sample of the $(N+1)\times(N+1)$ Hamiltonian matrix to find $N+1$ pieces of eigenstates, and compute the current expectation values with respect to these eigenstates.

\subsection{Case of odd $N$}
\label{subsec6.1}

\begin{figure}
\begin{subfigure}{0.45\textwidth}
\includegraphics[width=\textwidth]{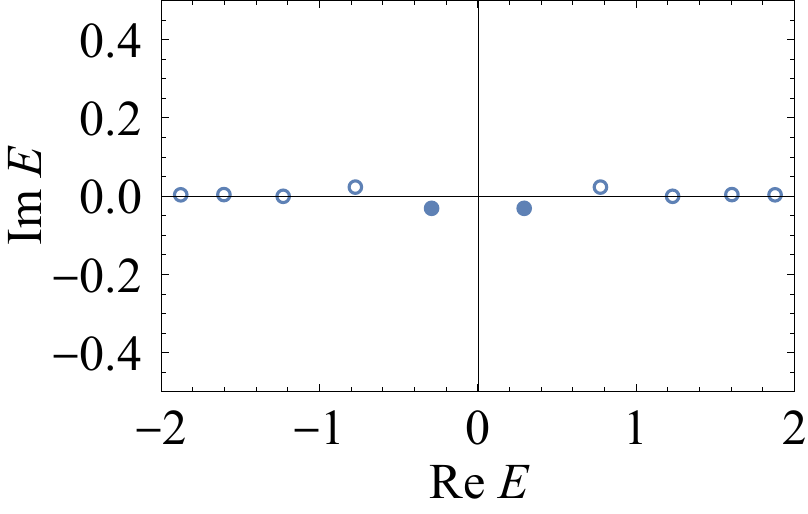}
\caption{$\tilde{\gamma}=0.5$}
\end{subfigure}
\hfill
\begin{subfigure}{0.45\textwidth}
\includegraphics[width=\textwidth]{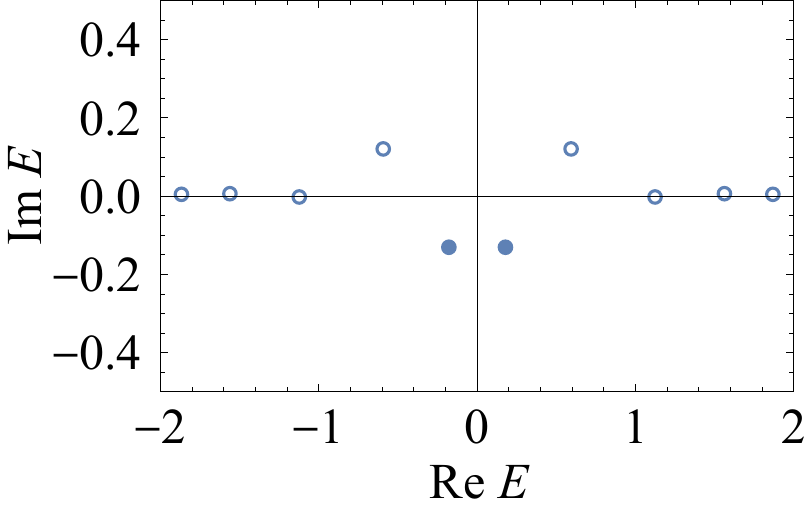}
\caption{$\tilde{\gamma}=1.0$}
\end{subfigure}
\\[\baselineskip]
\begin{subfigure}{0.45\textwidth}
\includegraphics[width=\textwidth]{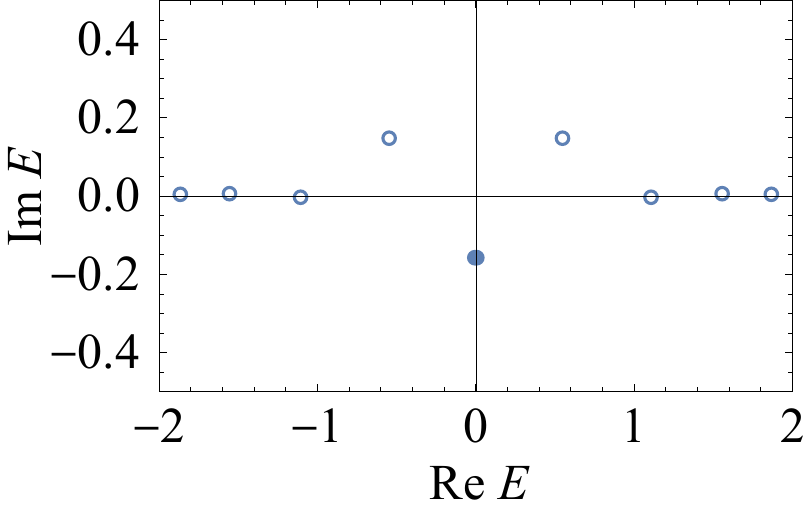}
\caption{$\tilde{\gamma}=1.077$}
\end{subfigure}
\hfill
\begin{subfigure}{0.45\textwidth}
\includegraphics[width=\textwidth]{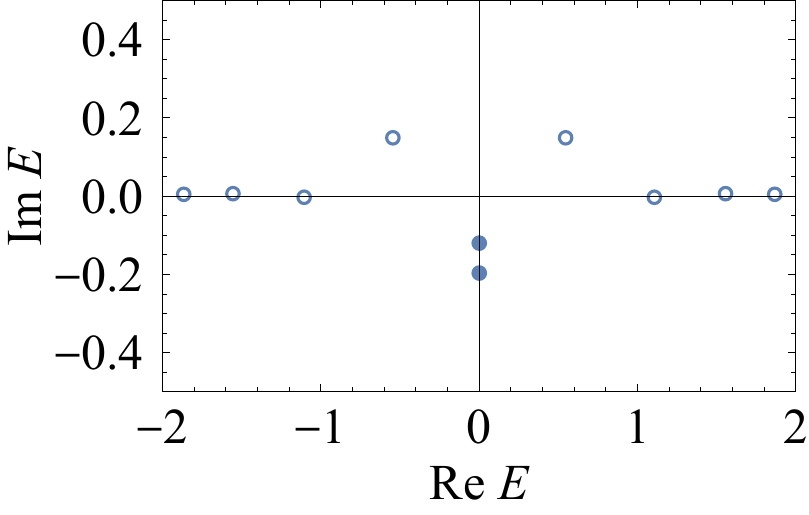}
\caption{$\tilde{\gamma}=1.08$}
\end{subfigure}
\\[\baselineskip]
\begin{subfigure}{0.45\textwidth}
\includegraphics[width=\textwidth]{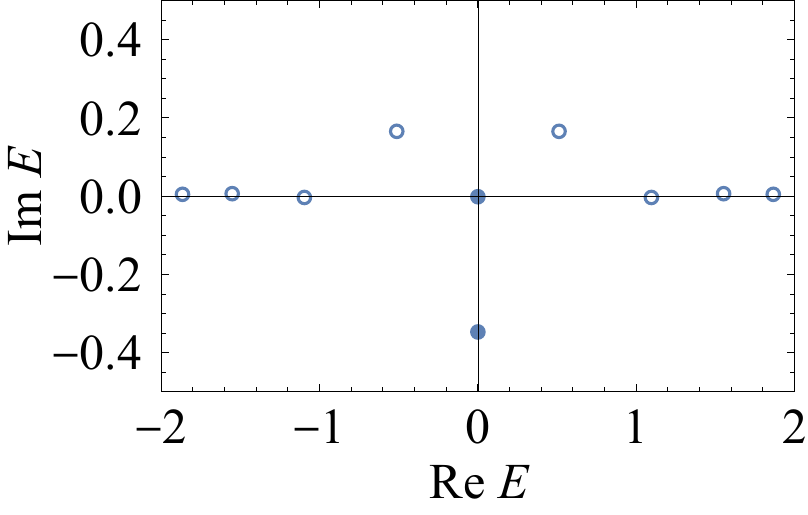}
\caption{$\tilde{\gamma}=1.13$}
\end{subfigure}
\hfill
\begin{subfigure}{0.45\textwidth}
\includegraphics[width=\textwidth]{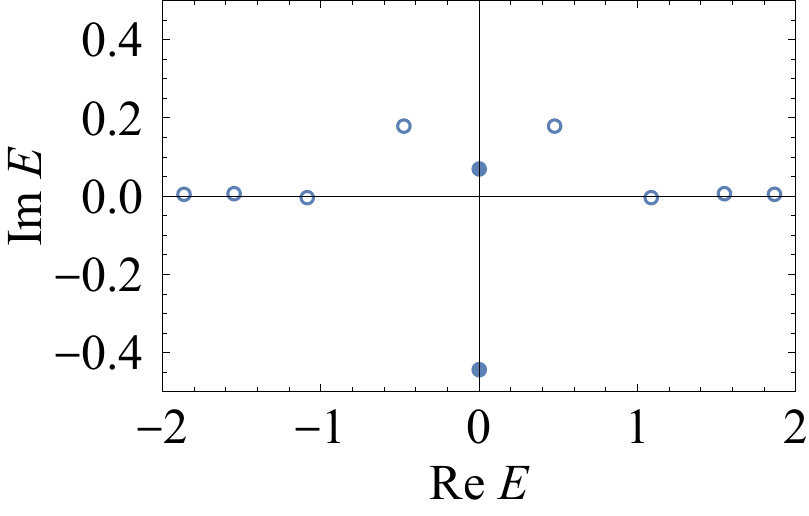}
\caption{$\tilde{\gamma}=1.18$}
\end{subfigure}
\\
\caption{Evolution of the scaled eigenvalues of the random-hopping Hamiltonian with source and drain for a specific random sample chosen with the randomness parameter $\delta=0.1$, as we change the parameter $\tilde{\gamma}$.
We set $N=9$, and hence there are ten eigenvalues indicated by the circles, two of which are especially marked by the solid circles, in order to emphasize the exceptional-point coalescence on the imaginary axis.}
\label{fig11}
\end{figure}

Figure~\ref{fig11} shows an example of the evolution of the eigenvalues in the complex energy plane for a specific random sample with $N=9$.
The two eigenvalues indicated by the closed circles coalesce at a point on the imaginary axis in the panel (c), or more accurately at the exceptional point $\tilde{\gamma}_\textrm{ex}\simeq 1.077077819323$, and one of them that splits from the coalescence point climbs the imaginary axis, passing the origin $E=0$ on its way in the panel (e), or more accurately at $\tilde{\gamma}_0\simeq 1.1308194251143$.
The coalescence at the exceptional point can occur on either of the positive or negative side of the imaginary axis, depending on the random sample, but one eigenvalue always passes the origin at a point $\tilde{\gamma}=\tilde{\gamma}_0$ either from above or below, respectively.
This means that we now generally have the inequality $\tilde{\gamma}_\textrm{ex}\leq \tilde{\gamma}_0$.
The exceptional point is a remnant of the $\PT$-symmetry breaking transition in this random system, but perturbed from the occurrence of the zero mode.

\begin{figure}
\centering
\includegraphics[width=0.6\textwidth]{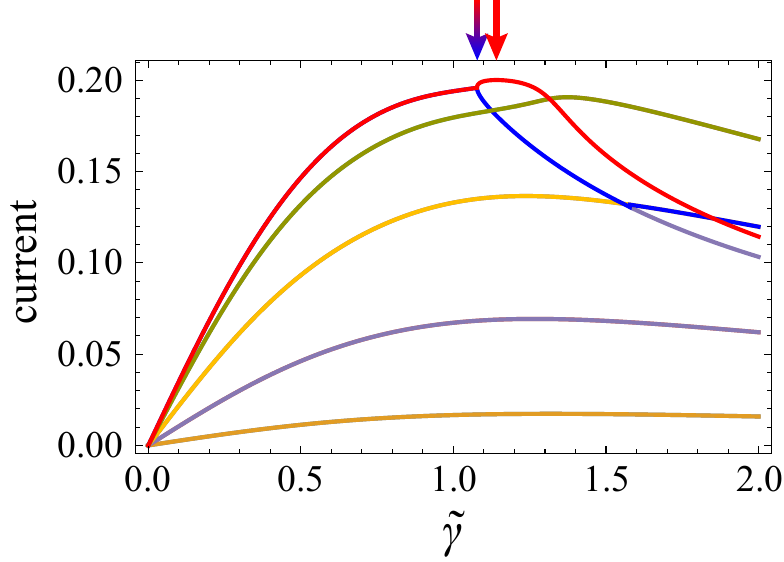}
\caption{The $\tilde{\gamma}$-dependence of the current averaged over the system for the same specific random sample as in Fig.~\ref{fig11}; we set $N=9$ and $\delta=0.1$. The left arrow indicates the exceptional point $\tilde{\gamma}_\textrm{ex}\simeq1.077$, which corresponds to Fig.~\ref{fig10}(c).
The right arrow indicates the maximum of the current at $\tilde{\gamma}_\textrm{max}\simeq1.140727$.}
\label{fig12}
\end{figure}

The relevant question to ask at this point is as to which of the remnant of the $\PT$-symmetry breaking transition and the occurrence of the zero mode is more significant to the current maximum.
The answer is the latter.
Figure~\ref{fig12} shows the $\tilde{\gamma}$-dependence of the current (averaged over the system) for each eigenvalue of the same random sample as in Fig.~\ref{fig11}.
In this particular sample, the current expectation value reaches its maximum at $\tilde{\gamma}=\tilde{\gamma}_\textrm{max}\simeq1.140727$, which is close to the case of Fig.~\ref{fig11}(e), $\tilde{\gamma}_0\simeq 1.130819$.

\begin{figure}
\centering
\includegraphics[width=0.6\textwidth]{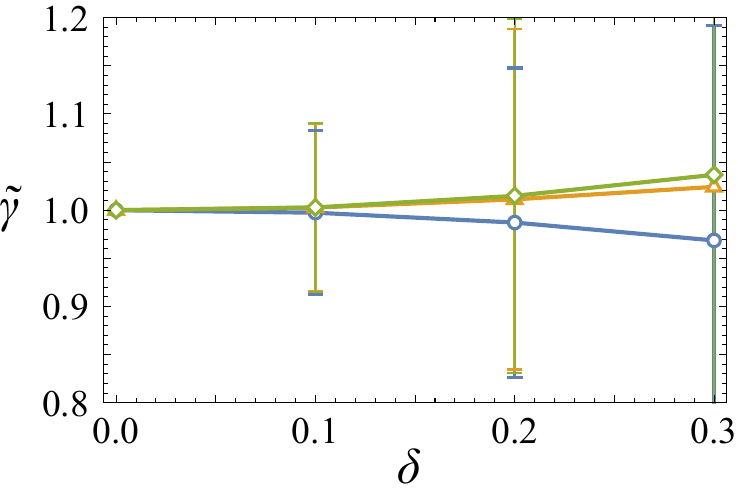}
\caption{The $\delta$-dependence of the values of $\tilde{\gamma}$ at the exceptional point $\tilde{\gamma}_\textrm{ex}$ (the lowest, blue circles), at the point where one of the eigenvalues becomes zero, $\tilde{\gamma}_0$,  (the second lowest, orange triangles), and at the point where the current expectation becomes maximum, $\tilde{\gamma}_\textrm{max}$ (highest, green squares). We set $N=9$ and used $10^6$ samples. The vertical line attached to each point is the standard deviation of the random distribution, \textit{not} an error bar.}
\label{fig13}
\end{figure}

Figure~\ref{fig13} shows the randomness dependence of the sample average of the exceptional point $\tilde{\gamma}_\textrm{ex}$, the point of zero eigenvalue, $\tilde{\gamma}_0$, and the point of the maximum current expectation value, $\tilde{\gamma}_\textrm{max}$.
For odd $N$, the average value of $\tilde{\gamma}$ at the exceptional point $\tilde{\gamma}_\textrm{ex}$ decreases from the clean 
case as we increase the randomness parameter $\delta$, whereas that at the point of the maximum current increases.
The maximum of the current expectation value is achieved by the eigenstate whose eigenvalue passes the origin $E=0$ after going through the exceptional point.
Indeed, the point of the zero eigenvalue is quite close to the point of the maximum current expectation value.
We thereby conclude that the current maximum is achieved by the approximate zero-eigenvalue eigenstate.

\subsection{Case of even $N$}
\label{subsec6.2}

\begin{figure}
\begin{subfigure}{0.45\textwidth}
\includegraphics[width=\textwidth]{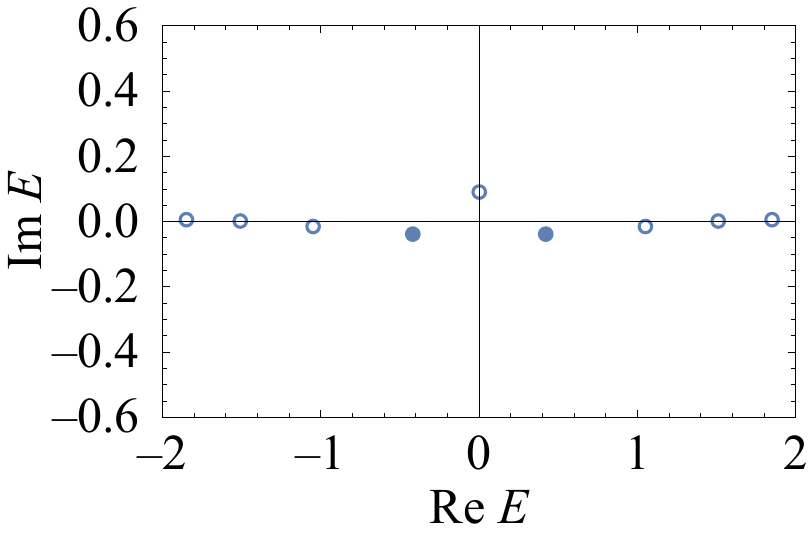}
\caption{$\tilde{\gamma}=0.75$}
\end{subfigure}
\hfill
\begin{subfigure}{0.45\textwidth}
\includegraphics[width=\textwidth]{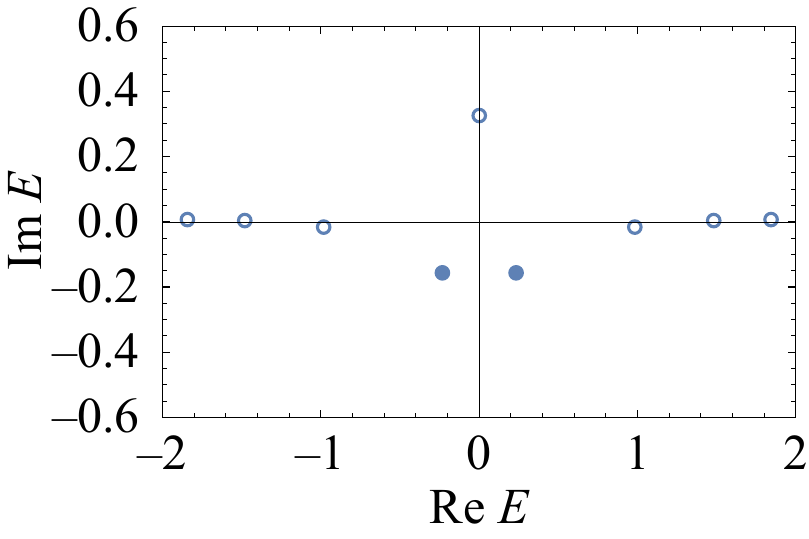}
\caption{$\tilde{\gamma}=1.0$}
\end{subfigure}
\\[\baselineskip]
\begin{subfigure}{0.45\textwidth}
\includegraphics[width=\textwidth]{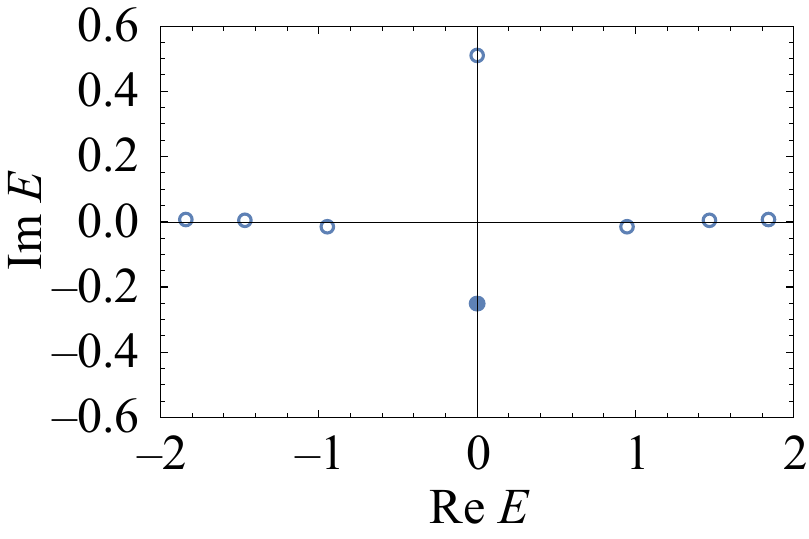}
\caption{$\tilde{\gamma}=1.1367$}
\end{subfigure}
\hfill
\begin{subfigure}{0.45\textwidth}
\includegraphics[width=\textwidth]{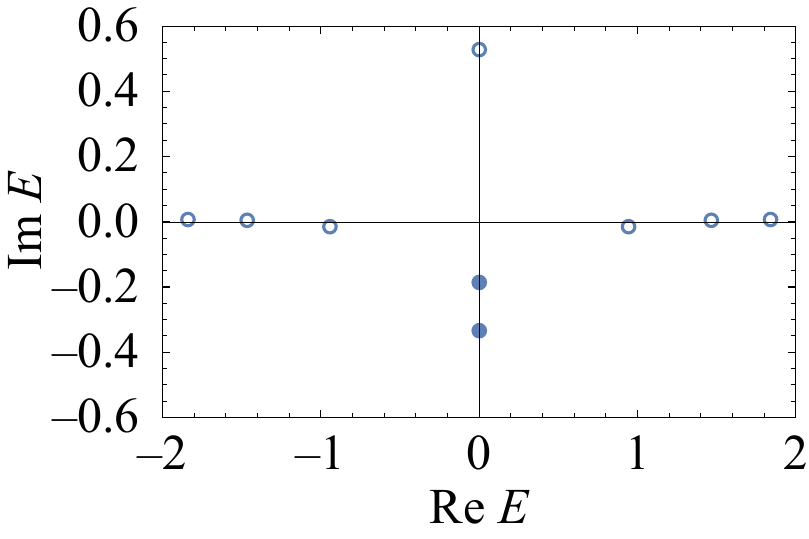}
\caption{$\tilde{\gamma}=1.15$}
\end{subfigure}
\\[\baselineskip]
\begin{subfigure}{0.45\textwidth}
\includegraphics[width=\textwidth]{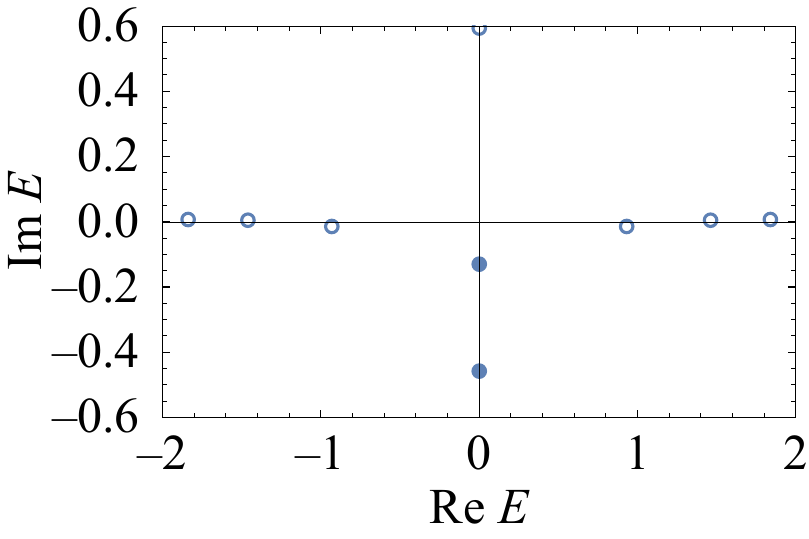}
\caption{$\tilde{\gamma}=1.2$}
\end{subfigure}
\hfill
\begin{subfigure}{0.45\textwidth}
\includegraphics[width=\textwidth]{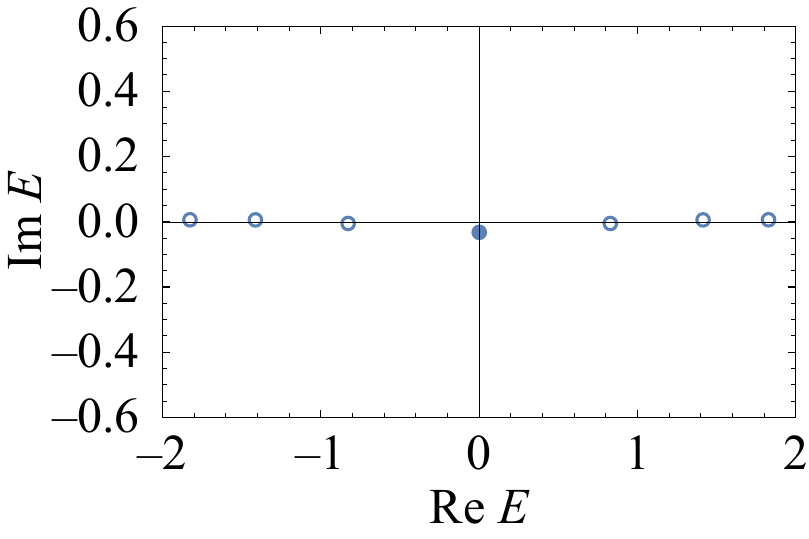}
\caption{$\tilde{\gamma}=2.0$}
\end{subfigure}
\\
\caption{Evolution of the scaled eigenvalues of the random-hopping Hamiltonian with source and drain for a specific random sample chosen with $\delta=0.1$, as we change the parameter $\tilde{\gamma}$.
We set $N=8$, and hence there are nine eigenvalues indicated by the circles, two of which are especially marked by the solid circles, in order to emphasize the exceptional-point coalescence on the imaginary axis. In panel (f), two eigenvalues went out of the plot region.}
\label{fig14}
\end{figure}

Figure~\ref{fig14} shows an example of the evolution of the eigenvalues in the complex energy plane for a specific random sample with $N=8$.
The difference from the case of odd $N$ is the fact that for a nonzero value of $\tilde{\gamma}$, there is always one purely imaginary eigenvalue, but a zero eigenvalue never exists.
The eigenvalue, which is $E=0$ for $\tilde{\gamma}=0$, departs the origin for finite values of $\tilde{\gamma}$ and never comes back. On the other hand, the two eigenvalues indicated by the closed circles coalesce at a point on the imaginary axis in the panel (c), or more accurately at the exceptional point $\tilde{\gamma}_\textrm{ex}\simeq 1.136734964524$, and one of them that splits from the coalescence point climbs the imaginary axis, but only approaches the origin asymptotically, never passing there.
Note that the coalescence at the exceptional point can occur on either of the positive or negative side of the imaginary axis, depending on the random sample.

The following point, however, does not change from the case of odd $N$.
The exceptional point, which is again the remnant of the $\PT$-symmetry breaking transition, is perturbed from the zero mode and occurs before the current maximum as we increase $\tilde{\gamma}$.
In short, the random case reveals that the zero mode is more relevant than the exceptional point due to the $\PT$-symmetry.

\begin{figure}
\centering
\includegraphics[width=0.6\textwidth]{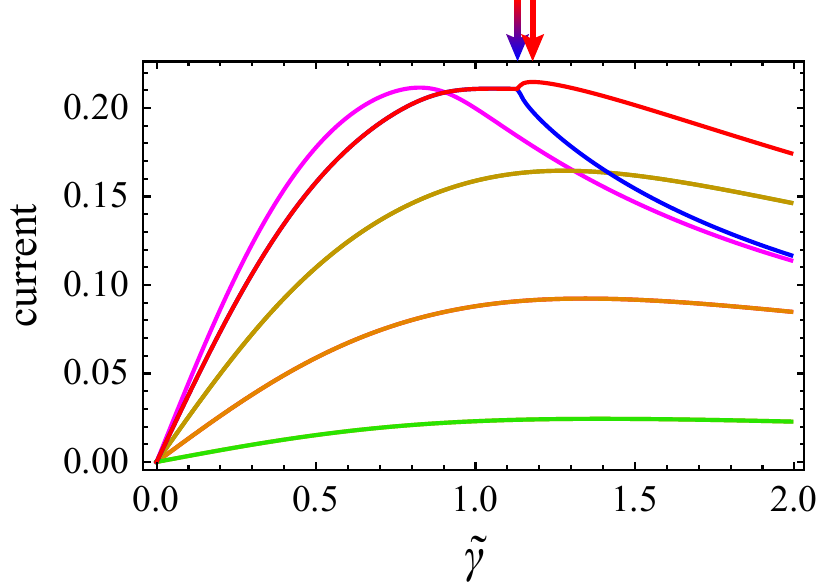}
\caption{The $\tilde{\gamma}$-dependence of the current averaged over the system for the same specific random sample as in Fig.~\ref{fig13}; we set $N=8$ and $\delta=0.1$.
The left arrow indicates the exceptional point $\tilde{\gamma}_\textrm{ex}\simeq1.1367$, which corresponds to Fig.~\ref{fig14}(c).
The right arrow indicates the maximum of the current at $\tilde{\gamma}_\textrm{max}\simeq1.179$.}
\label{fig15}
\end{figure}

Figure~\ref{fig15} shows the $\tilde{\gamma}$-dependence of the current (averaged over the system) for each
eigenstate of the same random sample as in Fig.~\ref{fig14}.
In this particular sample, the current expectation value reaches its maximum at $\tilde{\gamma}=\tilde{\gamma}_\textrm{max}\simeq1.179220$ by the eigenvalue approaching the origin.

\begin{figure}
\centering
\includegraphics[width=0.6\textwidth]{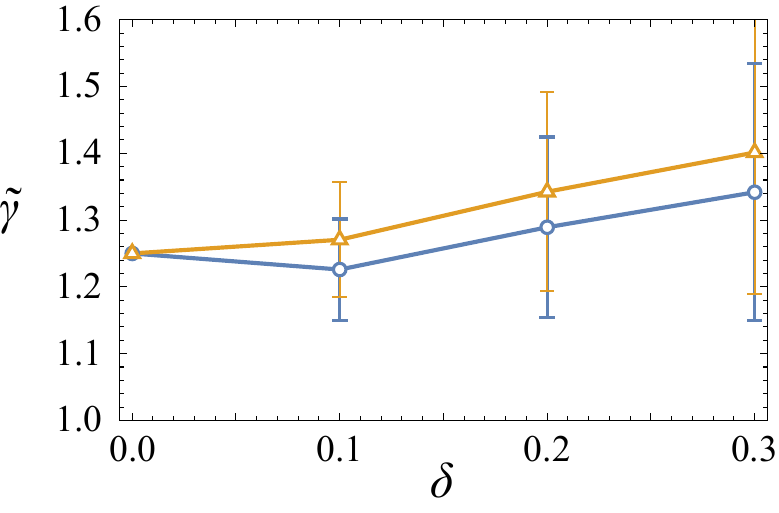}
\caption{The $\delta$-dependence of the values of $\tilde{\gamma}$ at the exceptional point (the lowest, blue circles) and at the point where the current expectation becomes maximum (highest, green squares).  We set $N=8$ and used $10^6$ samples. 
The vertical line attached to each point is the standard deviation of the random distribution, \textit{not} an error bar.}
\label{fig16}
\end{figure}

Figure~\ref{fig16} shows the randomness dependence of the sample average of the exceptional point $\tilde{\gamma}_\textrm{ex}$ and the point of the maximum current expectation value, $\tilde{\gamma}_\textrm{max}$.
For even $N$, the former first decreases from the clean case, but then increases as we increase the randomness parameter $\delta$. 
The latter, on the other hand, increases from the beginning.

The maximum of the current expectation value is achieved by the eigenstate whose eigenvalue approaches the origin $E=0$.
Although there is no zero eigenvalue for even $N$, we still conclude that the current maximum is achieved by the approximate zero-eigenvalue eigenstate.

\begin{figure}
\centering
\includegraphics[width=0.6\textwidth]{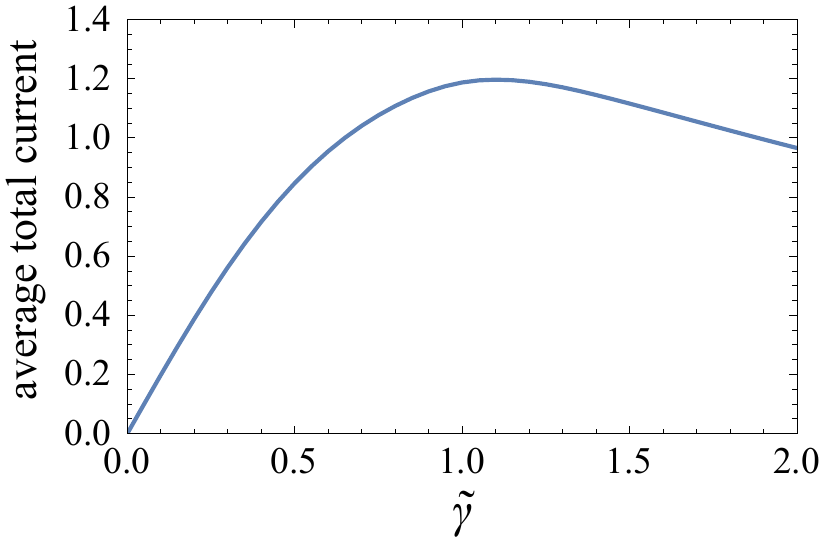}
\caption{The $\tilde{\gamma}$-dependence of the sample average of the total current summed over all eigenstates for $N=9$ and $\delta=0.1$. We used $10^4$ samples. Statistical errors of data points are of order $10^{-3}$ to $10^{-4}$, and hence are not plotted here.}
\label{fig17}
\end{figure}

Finally, we show in Fig.~\ref{fig17} a sample average of the current summed over all eigenmodes.
We numerically observed that the behavior of the plot in Fig.~\ref{fig17} is not essentially different from that for other values of $N$ and for binary probability distributions; hence, we do not show them here.
We see in Fig.~\ref{fig17} that the cusp of the maximum current is rounded due to the randomness, and now looks similar to Fig.~7 of Ref.~\cite{Toroker09}.

\section{Summary}
\label{sec7}

In the present paper, we have considered a quantum transport problem on a truncated Bethe lattice with sources on the 
peripheral sites and the drain on the central site.
We solved the problem as an eigenvalue problem and sought the condition for the maximum current.
We revealed that the maximum current is carried by the exactly zero mode in the pure case and a nearly zero mode even in the random case.

We solved the eigenvalue problem mostly analytically, thanks to the model's simplicity.
We first proved that most eigenstates are localized on the side of the periphery, not penetrating into the central site.
Correspondingly, the imaginary parts of the eigenvalues are positive, which indicates that the probability input from the source accumulates outside.
The number of extended eigenstates that do penetrate to the drain on the central site is only $N+1$.
We handle them by mapping the entire system to a one-dimensional segment of $N+1$ sites within the restricted Hilbert space of the extended eigenstates.

The limitation of the conducting channels arises from destructive interference among the conductive pathways from peripheral sites to the central site.
This suggests that the limitation can be universal across fractal systems, with the number of surface sites much greater than that of central sites.

When the number of links in each generation of the Bethe lattice is uniform, the maximum current is always achieved by the zero modes, which also lie at the exceptional point.
When the number of links is not uniform, the zero mode no longer produces the current at exactly the maximum, but still approximately.
In the case of odd $N$, the maximum current is achieved by the eigenstate whose eigenvalue passes the origin $E=0$, closely when the eigenvalue becomes zero. 
In the case of even $N$, the maximum current is achieved by the eigenstate whose eigenvalue converges to the origin.
This implies that the occurrence of the zero mode is more relevant to the current maximum than the exceptional point.

One possibly universal feature of the present model is the presence of a maximum in the current profile as we increase the strengths of the sources and the drain.
In other words, as we try to increase the current further and further, we are betrayed.
The current vanishes in the limit of infinite strengths of the sources and the drain.
As we mentioned at the end of Sec.~\ref{sec6}, the same feature was indeed observed in electronic conduction in a molecular junction~\cite{Toroker09}.
We thereby conjecture that this is a feature common to various cases of electronic transport, which may be even observable experimentally.

Another possibly universal feature is the fact that the maximum current is carried by the zero modes.
As we mentioned in the Introduction, the zero mode occupies a special position in various physics due to symmetry; see \textit{e.g.} Refs.~\cite{Gade93, Sugiyama93, Avishai93, Motoi95, Frohlich00, Tanaka24}.
Given its universality, we conjecture that the zero mode carries the maximum current in other systems as well.
Possible generalizations include the one in which the number of children is different even within each generation.
Fractal lattices are also interesting candidates.
Reference~\cite{Manna23} considers a Sierpinski carpet with asymmetric hopping, but a transport problem with complex potentials may be worth pursuing.

{\it Note added}.---While preparing the present work, we notice other related works for non-Hermitian models on a similar lattice~\cite{Sun24, Hamanaka24, ShimomuraPrivate24}. 
In these works, non-Hermiticity comes from the off-diagonal asymmetric hopping of the type of the Hatano-Nelson model~\cite{HN96, HN97} rather than the diagonal complex potentials in the present work.

\appendix

\section{Open and closed non-Hermitian systems and use of right and left eigenvectors}
\label{appA}

The purpose of the present Appendix is to justify the definition of the current expectation value of the form in Eqs.~\eqref{eq900} and~\eqref{eq760}. 
In non-Hermitian quantum mechanics, there are two options for defining the expectation value of a physical quantity: (i) using a right eigenvector and its Hermitian conjugate as in the standard quantum mechanics for an observable of a Hermitian operator; (ii) using both right and left eigenvectors for an observable, which is possibly non-Hermitian.

The choice is deeply related to the distinction of open and closed non-Hermitian systems~\cite{Brody16}.
We here review this point from our own perspectives.
For other references, see Refs.~\cite{Plenio98, Daley14} for the option (i) and see Ref.~\cite{Brody02} for the option (ii), for example.
A related issue is also discussed in Ref.~\cite{Meden23}.

\subsection{Open quantum systems and use of right eigenvectors}

\begin{figure}
    \centering
    \includegraphics[width=0.5\linewidth]{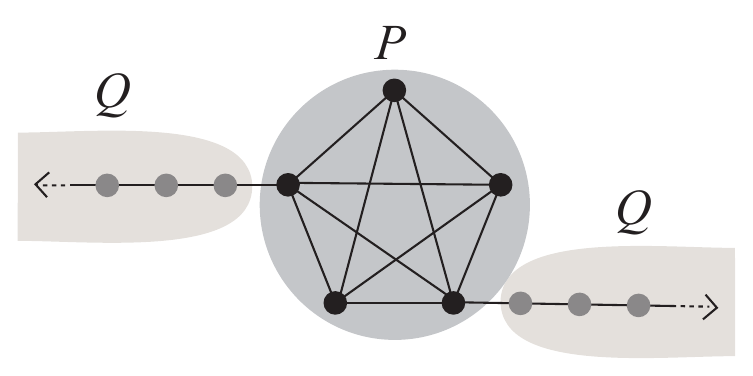}
    \caption{A schematic view of an open quantum system of a one-body problem.
    The operator $P$ projects on the Hilbert space of only the central system, while the operator $Q$ projects on the other parts of the system.}
    \label{figA1}
\end{figure}

We start with a brief derivation of an open non-Hermitian system.
Consider a one-body problem in an infinite system, schematically shown in Fig.~\ref{figA1};
the real space therefore spans the Hilbert space.
We divide the system into a central system of interest with a finite number of degrees of freedom with a discrete spectrum and an environmental system with an infinite number of degrees of freedom with at least one continuum spectrum.
In the following, we argue that the effective Hamiltonian of the central system, which is an open quantum system, is non-Hermitian.

Let us assume that the Hamiltonian $H$ that governs the dynamics of the entire system is Hermitian in the whole Hilbert space.
Therefore, the solution of the eigenequation
\begin{align}\label{eqA10}
H\ket{\psi}=E\ket{\psi}
\end{align}
is an eigenstate $\ket{\psi_n}$ with a real eigenvalue $E_n$.
The eigenstates satisfy the orthogonality $\braket{\psi_m|\psi_n}=\delta_{mn}$.

Let $P$ denote the projection operator onto the Hilbert space of the central system.
Hence, the projection operator $Q=I-P$ with the identity operator $I$ projects states onto the Hilbert space of the environmental system.
The two projection operators satisfy $P^2=P$, $Q^2=Q$, and $PQ=QP=0$.

We now find an effective Hamiltonian for the projected state $P\ket{\psi}$.
Applying the projection operators $P$ and $Q$ to Eq.~\eqref{eqA10}, we respectively have 
\begin{align}\label{eqA20}
PHP\qty(P\ket{\psi})+PHQ\qty(Q\ket{\psi})&=E\qty(P\ket{\psi}),
\\\label{eqA30}
QHP\qty(P\ket{\psi})+QHQ\qty(Q\ket{\psi})&=E\qty(Q\ket{\psi}).
\end{align}
The second equation~\eqref{eqA30} produces
\begin{align}
Q\ket{\psi}=\frac{1}{E-QHQ}QHP\qty(P\ket{\psi}).
\end{align}
We put this into the first equation~\eqref{eqA20} and find
\begin{align}\label{eqA50}
\qty(PHP+PHQ\frac{1}{E-QHQ}QHP)\qty(P\ket{\psi})=E\qty(P\ket{\psi}),
\end{align}
from which we deduce that the effective Hamiltonian for $P\ket{\psi}$ is~\cite{Hatano14,Hatano21}
\begin{align}\label{eqA60}
H_\textrm{eff}(E)=PHP+PHQ\frac{1}{E-QHQ}QHP.
\end{align}
We can prove an equality~\cite{Hatano14}
\begin{align}
P\frac{1}{E-H}P=P\frac{1}{E-H_\textrm{eff}(E)}P,
\end{align}
which implies that the point spectra of the total Hamiltonian $H$ are carried over, unchanged, to the point spectra of the effective Hamiltonian.
Note, however, that the eigenequation $H_\textrm{eff}(E)\qty(P\ket{\psi})=E\qty(P\ket{\psi})$ is nonlinear in $E$ because the effective Hamiltonian is energy-dependent, which is related to non-Markovianity of the dynamics of open quantum systems~\cite{Hatano21}.

Since we assume that $QHQ$ has an infinite number of degrees of freedom and has a continuum spectrum, its Green's function $(E-QHQ)^{-1}$ has two branches of the retarded and advanced Green's functions in the form $(E-QHQ\pm\ii\eta)^{-1}$ with an infinitesimal constant $\eta > 0$.
This imaginary part $\pm\ii\eta$ makes the effective Hamiltonian~\eqref{eqA60} non-Hermitian~\cite{Hatano21}.
Because of the operators $PHQ$ and $QHP$ on both sides of the Green's function, non-Hermiticity arises at the interface between the central system and the environment, typically in the form of imaginary potentials. 
Thus, this projection procedure can produce non-Hermiticity of the Hamiltonians~\eqref{eq50} and~\eqref{eq420} of the present model.

Equation~\eqref{eqA50} shows that $P\ket{\psi}$ is an eigenstate of the effective non-Hermitian Hamiltonian~\eqref{eqA60}.
The set of thus obtained eigenstates $\{P\ket{\psi_n}\}$ is not orthogonal to each other.
In fact, obviously $\qty(P\ket{\psi_m})^\dag \qty(P\ket{\psi_n})=\mel{\psi_m}{P}{\psi_n}\neq\delta_{mn}$, in general.
This demonstrates that the absence of the orthogonality of the eigenstates~\eqref{eq520-0} is a legitimate feature of open quantum systems.
Reference~\cite{Wiersig18} also points out the importance of the usage of the non-orthogonal right eigenvectors.

Nonetheless, we can use $\qty(P\ket{\psi_m})^\dag$ to define the expectation value of a physical quantity $A$ defined as a Hermitian operator inside the central system, for which $A^\dag=A$ and $PAP=A$ should hold.
Therefore, we have
\begin{align}\label{eqA80-1}
\qty(P\ket{\psi_m})^\dag A\qty(P\ket{\psi_n})=\ev{PAP}{\psi_m}=\ev{A}{\psi_m},
\end{align}
which must be real.

This demonstrates that in computing the expectation value in open quantum systems, we should use right eigenvectors of the non-Hermitian Hamiltonian and its Hermitian conjugate for an observable of a Hermitian operator. 
This is why we used the right eigenvectors in computing the current expectation value~\eqref{eq900}, instead of the left eigenvectors given in Eq.~\eqref{eq520-1}.
Let us stress again that the expectation value~\eqref{eq900} is guaranteed to be real because the current operator is Hermitian.

\subsection{Closed non-Hermitian systems and use of left eigenvectors}

The computation of the expectation value using both left and right eigenvectors, as in
\begin{align}\label{eqA80}
\mel{\phi_n}{A}{\psi_n},
\end{align}
is legitimate when we assume the total Hamiltonian $H$ to be non-Hermitian.
This means that the entire system is closed, despite the Hamiltonian's non-Hermiticity.
In fact, the original motivation of the authors of Ref.~\cite{Bender98} was to define a novel formulation of quantum mechanics for closed systems~\cite{Bender99}.
They found that Hermiticity is too mathematical a condition to guarantee the reality of energy eigenvalues.
Indeed, $\PT$ symmetry makes energy eigenvalues real in the $\PT$-unbroken phase.
The universe might be a closed non-Hermitian system in a $\PT$-unbroken phase.
In this mindset, it is legitimate to compute the expectation value using a biorthogonal set of left and right eigenvectors.

As a tutorial example, let us compute for $N=1$ the expectation value of the current operator~\eqref{eq890} using the right eigenvectors in Eqs.~\eqref{eq520-2}--\eqref{eq520-0} and the left eigenvectors in Eqs.~\eqref{eq520-1}--\eqref{eq520-3}.
A straightforward algebra shows that in both $\PT$-unbroken and $\PT$-broken regions, we have
\begin{align}
\mel{\phi_1^\pm}{\tilde{J}_1}{\psi_1^\pm}=0.
\end{align}
This is a plausible result in the formulation of closed non-Hermitian systems.
There should be no net current in a closed system.

\section{Scattering through a non-Hermitian dot}
    \label{appB}

We here consider transport properties of a non-Hermitian dot [{\it i.e.} Eq.~(\ref{eq500})]
\begin{align}
    H = \begin{pmatrix}
        - \ii \gamma & -1 \\
        -1 & \ii \gamma
    \end{pmatrix},
\end{align}
where $\gamma \in \mathbb{R}$ describes the degree of non-Hermiticity and hence balanced gain and loss.
While we calculated the expectation value of the current operator for each eigenstate in Subsec.~\ref{subsec5.1}, we here employ the scattering approach in a similar manner to the Landauer formula~\cite{Landauer57, Datta95}.

Let us attach two ideal leads to each of the two sites of the above non-Hermitian dot.
The single-particle eigenequation reads
\begin{align}
    - \psi_{n-1} + V_n \psi_n + - \psi_{n+1} = E \psi_{n}
\end{align}
with the single-particle eigenenergy $E$ and the corresponding single-particle wave function $\psi = \left( \cdots \psi_n \cdots \right)$.
Here, we assume that the balanced gain and loss are put on the sites $n=1$ and $n=2$, respectively, and hence the onsite potentials are given as
\begin{align}
    V_n = \begin{cases}
        - \ii \gamma & \mbox{for $n=1$}; \\
        \ii \gamma & \mbox{for $n=2$}; \\
        0 & \mbox{otherwise} .
    \end{cases}
\end{align}

We define the transfer matrix $M_n$ from the left (smaller $n$) to the right (larger $n$) by
\begin{align}
    \begin{pmatrix}
        \psi_{n+1} \\ \psi_n
    \end{pmatrix} = M_n \begin{pmatrix}
        \psi_{n} \\ \psi_{n-1}
    \end{pmatrix},
\end{align}
which reads
\begin{align}
    M_n = \begin{pmatrix}
        V_n - E & -1 \\
        1 & 0
    \end{pmatrix}.
\end{align}
The transmission and reflection amplitudes are calculated from the product of the transfer matrices.
In fact, the transmission probability from the left lead ($n\leq 0$) to the right lead ($n \geq 3$) is given as~\cite{Datta95}
\begin{align}
    T = \frac{1}{\left| \mathcal{M}_{22} \right|^2},
\end{align}
where $\mathcal{M}$ is the transfer matrix of the entire system, 
\begin{align}
    \mathcal{M} \coloneqq Q^{-1} \cdots M_3 M_2 M_1 M_0 M_{-1} M_{-2} \cdots Q
\end{align}
with 
\begin{align}
    Q \coloneqq \begin{pmatrix}
        1 & 1 \\
        e^{-\ii k} & e^{\ii k}
    \end{pmatrix}.
\end{align}
Here, since we assume that the leads are ideal, the wave number $k$ is determined by 
\begin{align}
    E = - 2 \cos k.
\end{align}

Intuitively, the transmission probability $T$ does not depend on the length of the leads, although the transmission amplitude does.
In fact, we have
\begin{align}
    T = \left| 1 - \frac{e^{2\ii k}}{e^{2\ii k} - 1} \gamma^2 \right|^{-2}
    = \left[ \left( 1- \frac{\gamma^2}{2} \right)^2 + \left( \frac{\gamma^2}{2\tan k} \right)^2 \right]^{-1}.
\end{align}
For clarity, let us focus on the band center $E=0$, \textit{i.e.}, $k = \pm \pi/2$.
Then, the transmission probability is simplified to
\begin{align}
    T \left( E=0 \right) = \frac{1}{\left( 1 - \gamma^2/2 \right)^2}.
\end{align}
Thus, the transmission probability $T$ increases for $\left| \gamma \right| < \sqrt{2}$ and decreases for $\left| \gamma \right| > \sqrt{2}$.
Notably, $T$ diverges at $\left| \gamma \right| = \sqrt{2}$, which seems to be unique to $E=0$ (\textit{i.e.} $T$ does not diverge for $E \neq 0$).
This divergence for $E=0$ may originate from a certain symmetry (\textit{e.g.} chiral symmetry).

The above behavior is qualitatively consistent with the results in Subsec.~\ref{subsec5.1} that focus on the expectation value of the current operator for each eigenstate.
However, the two approaches differ quantitatively.
Specifically, the transmission probability $T$ does not exhibit any singularities around the exceptional point $\left| \gamma \right| = 1$.
Such a singular point shifts to $\left| \gamma \right| = \sqrt{2}$ for $T \left( E = 0 \right)$.
This discrepancy seems to be due to the presence of the attached leads.
In fact, the attached leads of a tight-binding model may generate additional energy-dependent complex potentials at sites $n=1$ and $n=2$.
Furthermore, it is worth noting that $T$ is usually not related to the current itself but to the conductance, whereas we explicitly calculate the current in Subsec.~\ref{subsec5.1}.

\section*{Acknowledgments}
N.H.\ was supported by JSPS KAKENHI Grand Numbers JP19H00658, JP21H01005, JP22H01140, JP23K22411, and JP24K00545.
H.K.\ was supported by JSPS KAKENHI Grant Numbers JP23K25783, JP23K25790, 
and Grant-in-Aid for Transformative Research Areas A “Extreme Universe” No. JP21H05191.
K.K.\ was supported by  JSPS KAKENHI Grant Numbers JP24H00945, JP26H02015, JP26K06970, and JP26K17046.

\section*{References}
\bibliographystyle{unsrt}
\bibliography{hatano}

@article{Bueno20,
  title = {Null-eigenvalue localization of quantum walks on complex networks},
  author = {Bueno, R. and Hatano, N.},
  journal = {Phys. Rev. Research},
  volume = {2},
  issue = {3},
  pages = {033185},
  numpages = {8},
  year = {2020},
  publisher = {American Physical Society},
  doi = {10.1103/PhysRevResearch.2.033185},
  url = {https://link.aps.org/doi/10.1103/PhysRevResearch.2.033185}
}

@article{Tanaka24,
    author = {Tanaka, Y. and Tamura, S. and Cayao, J.},
    title = {{Theory of Majorana Zero Modes in Unconventional Superconductors}},
    journal = {Prog. Theo. Exp. Phys.},
    volume = {2024},
    pages = {08C105},
    year = {2024},
    issn = {2050-3911},
    doi = {10.1093/ptep/ptae065},
    url = {https://doi.org/10.1093/ptep/ptae065},
    eprint = {https://academic.oup.com/ptep/article-pdf/2024/8/08C105/58919082/ptae065.pdf},
}

@article{Avishai93,
  title = {Localization problem of a two-dimensional lattice in a random magnetic field},
  author = {Avishai, Y. and Hatsugai, Y. and Kohmoto, M.},
  journal = {Phys. Rev. B},
  volume = {47},
  issue = {15},
  pages = {9561},
  numpages = {0},
  year = {1993},
  publisher = {American Physical Society},
  doi = {10.1103/PhysRevB.47.9561},
  url = {https://link.aps.org/doi/10.1103/PhysRevB.47.9561}
}

@article{Sugiyama93,
  title = {{Localization in a random magnetic field in 2D}},
  author = {Sugiyama, T. and Nagaosa, N.},
  journal = {Phys. Rev. Lett.},
  volume = {70},
  issue = {13},
  pages = {1980},
  numpages = {0},
  year = {1993},
  publisher = {American Physical Society},
  doi = {10.1103/PhysRevLett.70.1980},
  url = {https://link.aps.org/doi/10.1103/PhysRevLett.70.1980}
}

@article{Gade93,
title = {Anderson localization for sublattice models},
journal = {Nucl. Phys. B},
volume = {398},
pages = {499},
year = {1993},
issn = {0550-3213},
doi = {https://doi.org/10.1016/0550-3213(93)90601-K},
url = {https://www.sciencedirect.com/science/article/pii/055032139390601K},
author = {Renate Gade}
}

@article{Motoi95,
  title = {{Quantum mechanics of the dynamical zero mode in (1+1)-dimensional QCD on the light cone}},
  author = {Tachibana, M.},
  journal = {Phys. Rev. D},
  volume = {52},
  issue = {10},
  pages = {6008},
  numpages = {0},
  year = {1995},
  publisher = {American Physical Society},
  doi = {10.1103/PhysRevD.52.6008},
  url = {https://link.aps.org/doi/10.1103/PhysRevD.52.6008}
}

@article{Frohlich00,
title = {Asymptotic form of zero energy wave functions in supersymmetric matrix models},
journal = {Nucl. Phys. B},
volume = {567},
pages = {231},
year = {2000},
issn = {0550-3213},
doi = {https://doi.org/10.1016/S0550-3213(99)00649-5},
url = {https://www.sciencedirect.com/science/article/pii/S0550321399006495},
author = {J. Fr\"{o}hlich and G. M. Graf and D. Hasler and J. Hoppe and S.-T. Yau}
}

@article{Mulken06,
  title = {Efficiency of quantum and classical transport on graphs},
  author = {M\"ulken, O. and Blumen, A.},
  journal = {Phys. Rev. E},
  volume = {73},
  issue = {6},
  pages = {066117},
  numpages = {5},
  year = {2006},
  publisher = {American Physical Society},
  doi = {10.1103/PhysRevE.73.066117},
  url = {https://link.aps.org/doi/10.1103/PhysRevE.73.066117}
}

@article{MaquinBatalha22,
	author = {Maquin{\'e} Batalha, G. and Volta, A. and Strunz, W. T. and Galiceanu, M.},
	date = {2022/04/27},
	doi = {10.1038/s41598-022-10537-w},
	id = {Maquin{\'e}Batalha2022},
	isbn = {2045-2322},
	journal = {Sci. Rep.},
	pages = {6896},
	title = {Quantum transport on honeycomb networks},
	url = {https://doi.org/10.1038/s41598-022-10537-w},
	volume = {12},
	year = {2022}}

@article{Rebentrost09,
doi = {10.1088/1367-2630/11/3/033003},
url = {https://doi.org/10.1088/1367-2630/11/3/033003},
year = {2009},
publisher = {},
volume = {11},
pages = {033003},
author = {Rebentrost, P. and Mohseni, M. and Kassal, I. and Lloyd, S. and Aspuru-Guzik, A.},
title = {Environment-assisted quantum transport},
journal = {New J. Phys.}
}

@article{Silva24,
    author = {Silva, A. A. and Bazeia, D. and Andrade, F. M.},
    title = {Quantum transport in randomized quantum graphs},
    journal = {APL Quantum},
    volume = {1},
    pages = {046126},
    year = {2024},
    month = {12},
    issn = {2835-0103},
    doi = {10.1063/5.0239742},
    url = {https://doi.org/10.1063/5.0239742},
    eprint = {https://pubs.aip.org/aip/apq/article-pdf/doi/10.1063/5.0239742/20312391/046126_1_5.0239742.pdf},
}

@article{Jackson12,
  title = {{Quantum walks on trees with disorder: Decay, diffusion, and localization}},
  author = {Jackson, S. R. and Khoo, T. J. and Strauch, F. W.},
  journal = {Phys. Rev. A},
  volume = {86},
  issue = {2},
  pages = {022335},
  numpages = {11},
  year = {2012},
  publisher = {American Physical Society},
  doi = {10.1103/PhysRevA.86.022335},
  url = {https://link.aps.org/doi/10.1103/PhysRevA.86.022335}
}

@article{Collini10,
	author = {Collini, E. and Wong, C. Y. and Wilk, K. E. and Curmi, P. M. G. and Brumer, P. and Scholes, G. D.},
	date = {2010/02/01},
	doi = {10.1038/nature08811},
	id = {Collini2010},
	isbn = {1476-4687},
	journal = {Nature},
	pages = {644},
	title = {Coherently wired light-harvesting in photosynthetic marine algae at ambient temperature},
	url = {https://doi.org/10.1038/nature08811},
	volume = {463},
	year = {2010}}

@book{Mohseni14,
	address = {Cambridge},
	db = {Cambridge Core},
	doi = {DOI: 10.1017/CBO9780511863189},
	dp = {Cambridge University Press},
	editor = {Mohseni, M. and Omar, Y. and Engel, G. S. and Plenio, M. B.},
	isbn = {9781107010802},
	publisher = {Cambridge University Press},
	title = {{Quantum Effects in Biology}},
	url = {https://www.cambridge.org/core/product/EE21E9E35A7F0CDC8455118B23762533},
	year = {2014}}

@article{Panitchayangkoon10,
	author = {G. Panitchayangkoon and D. Hayes and K. A. Fransted and J. R. Caram and E. Harel and J. Wen and R. E. Blankenship and G. S. Engel},
	doi = {10.1073/pnas.1005484107},
	eprint = {https://www.pnas.org/doi/pdf/10.1073/pnas.1005484107},
	journal = {Proc. Natl. Acad. Sci. U. S. A.},
	pages = {12766},
	title = {Long-lived quantum coherence in photosynthetic complexes at physiological temperature},
	url = {https://www.pnas.org/doi/abs/10.1073/pnas.1005484107},
	volume = {107},
	year = {2010}}

@article{Engel07,
	author = {Engel, G. S. and Calhoun, T. R. and Read, E. L. and Ahn, T.-K. and Man{\v c}al, T. and Cheng, Y.-C. and Blankenship, R. E. and Fleming, G. R.},
	date = {2007/04/01},
	doi = {10.1038/nature05678},
	id = {Engel2007},
	isbn = {1476-4687},
	journal = {Nature},
	pages = {782},
	title = {Evidence for wavelike energy transfer through quantum coherence in photosynthetic systems},
	url = {https://doi.org/10.1038/nature05678},
	volume = {446},
	year = {2007}}

@article{Bender05,
	author = {C. M. Bender},
	doi = {10.1080/00107500072632},
	eprint = {https://doi.org/10.1080/00107500072632},
	journal = {Contemp. Phys.},
	pages = {277},
	publisher = {Taylor & Francis},
	title = {{Introduction to $\mathcal{PT}$-symmetric quantum theory}},
	url = {https://doi.org/10.1080/00107500072632},
	volume = {46},
	year = {2005}}

@article{Bender98,
	author = {Bender, C. M. and Boettcher, S.},
	doi = {10.1103/PhysRevLett.80.5243},
	issue = {24},
	journal = {Phys. Rev. Lett.},
	numpages = {0},
	pages = {5243},
	publisher = {American Physical Society},
	title = {{Real Spectra in Non-Hermitian Hamiltonians Having $\mathcal{PT}$ Symmetry}},
	url = {https://link.aps.org/doi/10.1103/PhysRevLett.80.5243},
	volume = {80},
	year = {1998}}

@article{Bender99,
	author = {C. M. Bender and S. Boettcher and P. N. Meisinger},
	journal = {J. Math. Phys.},
	pages = {2201},
	title = {{$\mathcal{PT}$-symmetric quantum mechanics}},
	volume = {40},
	year = 1999}

@book{BenderBook,
	author = {Bender, C. M. and Dorey, P. E. and Dunning, C. and Fring, A. and Hook, D. W. and Jones, H. F. and Kuzhel, S. and L\'{e}vai, G. and Tateo, R.},
	doi = {10.1142/q0178},
	eprint = {https://www.worldscientific.com/doi/pdf/10.1142/q0178},
	publisher = {World Scientific},
	title = {PT Symmetry},
	url = {https://www.worldscientific.com/doi/abs/10.1142/q0178},
	year = {2019}}

@article{bergman2008band,
	author = {D. L. Bergman and C. Wu and L. Balents},
	journal = {Phys. Rev. B},
	pages = {125104},
	publisher = {APS},
	title = {Band touching from real-space topology in frustrated hopping models},
	volume = {78},
	year = {2008}}

@article{Brody02,
	author = {Bender, C. M. and Brody, D. C. and Jones, H. F.},
	doi = {10.1103/PhysRevLett.89.270401},
	issue = {27},
	journal = {Phys. Rev. Lett.},
	numpages = {4},
	pages = {270401},
	publisher = {American Physical Society},
	title = {{Complex Extension of Quantum Mechanics}},
	url = {https://link.aps.org/doi/10.1103/PhysRevLett.89.270401},
	volume = {89},
	year = {2002}}

@article{Brody14,
	author = {D. C. Brody},
	doi = {10.1088/1751-8113/47/3/035305},
	journal = {J. Phys. A},
	pages = {035305},
	publisher = {IOP Publishing},
	title = {Biorthogonal quantum mechanics},
	url = {https://dx.doi.org/10.1088/1751-8113/47/3/035305},
	volume = {47},
	year = {2013}}

@article{Brody16,
	author = {Brody, D. C.},
	doi = {10.1088/1751-8113/49/10/10LT03},
	journal = {J. Phys. A},
	pages = {10LT03},
	title = {{Consistency of PT-symmetric quantum mechanics}},
	volume = {49},
	year = {2016}}

@article{Brouwer02,
	author = {P. W. Brouwer and E. Racine and A. Furusaki and Y. Hatsugai and Y. Morita and C. Mudry},
	journal = {Phys. Rev. B},
	pages = {014204},
	title = {{Zero modes in the random hopping model}},
	volume = {66},
	year = {2002}}

@article{Daley14,
	author = {A. J. Daley},
	doi = {10.1080/00018732.2014.933502},
	eprint = {https://doi.org/10.1080/00018732.2014.933502},
	journal = {Adv. Phys.},
	pages = {77},
	publisher = {Taylor \& Francis},
	title = {Quantum trajectories and open many-body quantum systems},
	url = {https://doi.org/10.1080/00018732.2014.933502},
	volume = {63},
	year = {2014}}

@article{Dyson53,
	author = {F. J. Dyson},
	journal = {Phys. Rev.},
	pages = {1331},
	title = {{The Dynamics of a Disordered Linear Chain}},
	volume = {92},
	year = {1953}}

@article{Eggarter78,
	author = {T. P. Eggarter and R. Riedinger},
	journal = {Phys. Rev. B},
	pages = {569},
	title = {{Singular behavior of tight-binding chains with off-diagonal disorder}},
	volume = {18},
	year = {1978}}

@article{Hamanaka24,
  title = {{Multifractal statistics of non-Hermitian skin effect on the Cayley tree}},
  author = {Hamanaka, S. and Iliasov, A. A. and Neupert, T. and Bzdu\ifmmode \check{s}\else \v{s}\fi{}ek, T. and Yoshida, T.},
  journal = {Phys. Rev. B},
  volume = {111},
  issue = {7},
  pages = {075162},
  numpages = {19},
  year = {2025},
  publisher = {American Physical Society},
  doi = {10.1103/PhysRevB.111.075162},
  url = {https://link.aps.org/doi/10.1103/PhysRevB.111.075162}
}

@article{Hatano08,
	author = {Hatano, N. and Sasada, K. and Nakamura, H. and Petrosky, T.},
	doi = {10.1143/PTP.119.187},
	issn = {0033-068X},
	journal = {Prog. Theo. Phys.},
	month = {02},
	pages = {187},
	title = {{{{Some Properties of the Resonant State in Quantum Mechanics and Its Computation}}}},
	url = {https://doi.org/10.1143/PTP.119.187},
	volume = {119},
	year = {2008}}

@article{Hatano14,
	author = {Hatano, N. and Ordonez, G.},
	doi = {10.1063/1.4904200},
	journal = {J. Math. Phys.},
	pages = {122106},
	title = {{Time-reversal symmetric resolution of unity without background integrals in open quantum systems}},
	url = {https://doi.org/10.1063/1.4904200},
	volume = {55},
	year = {2014}}

@article{Hatano21,
	author = {N. Hatano},
	doi = {10.1088/1742-6596/2038/1/012013},
	journal = {J. Phys.: Conf. Ser.},
	pages = {012013},
	publisher = {{IOP} Publishing},
	title = {{What is the resonant state in open quantum systems?}},
	url = {https://doi.org/10.1088/1742-6596/2038/1/012013},
	volume = {2038},
	year = 2021}

@article{HN96,
	author = {Hatano, N. and Nelson, D. R.},
	doi = {10.1103/PhysRevLett.77.570},
	issue = {3},
	journal = {Phys. Rev. Lett.},
	numpages = {0},
	pages = {570},
	publisher = {American Physical Society},
	title = {{Localization Transitions in Non-Hermitian Quantum Mechanics}},
	url = {https://link.aps.org/doi/10.1103/PhysRevLett.77.570},
	volume = {77},
	year = {1996}}

@article{HN97,
	author = {Hatano, N. and Nelson, D. R.},
	doi = {10.1103/PhysRevB.56.8651},
	issue = {14},
	journal = {Phys. Rev. B},
	numpages = {0},
	pages = {8651},
	publisher = {American Physical Society},
	title = {{Vortex pinning and non-Hermitian quantum mechanics}},
	url = {https://link.aps.org/doi/10.1103/PhysRevB.56.8651},
	volume = {56},
	year = {1997}}

@article{Kawabata19,
	author = {Kawabata, K. and Shiozaki, K. and Ueda, M. and Sato, M.},
	doi = {10.1103/PhysRevX.9.041015},
	issue = {4},
	journal = {Phys. Rev. X},
	numpages = {52},
	pages = {041015},
	publisher = {American Physical Society},
	title = {{Symmetry and Topology in Non-Hermitian Physics}},
	url = {https://link.aps.org/doi/10.1103/PhysRevX.9.041015},
	volume = {9},
	year = {2019}}

@article{Kunst18,
	author = {Kunst, F. K. and Edvardsson, E. and Budich, J. C. and Bergholtz, E. J.},
	doi = {10.1103/PhysRevLett.121.026808},
	issue = {2},
	journal = {Phys. Rev. Lett.},
	numpages = {6},
	pages = {026808},
	publisher = {American Physical Society},
	title = {{Biorthogonal Bulk-Boundary Correspondence in Non-Hermitian Systems}},
	url = {https://link.aps.org/doi/10.1103/PhysRevLett.121.026808},
	volume = {121},
	year = {2018}}

@article{Landauer57,
	author = {Landauer, R.},
	doi = {10.1147/rd.13.0223},
	journal = {IBM J. Res. Dev.},
	pages = {223},
	title = {{{Spatial Variation of Currents and Fields Due to Localized Scatterers in Metallic Conduction}}},
	volume = {1},
	year = {1957}}

@article{Lee16,
	author = {Lee, T. E.},
	doi = {10.1103/PhysRevLett.116.133903},
	issue = {13},
	journal = {Phys. Rev. Lett.},
	numpages = {5},
	pages = {133903},
	publisher = {American Physical Society},
	title = {{Anomalous Edge State in a Non-Hermitian Lattice}},
	url = {https://link.aps.org/doi/10.1103/PhysRevLett.116.133903},
	volume = {116},
	year = {2016}}

@article{loghin2006bounds,
	author = {Loghin, D. and van Gijzen, M. and Jonkers, E.},
	journal = {J. Comput. Appl. Math.},
	pages = {304},
	publisher = {Elsevier},
	title = {{Bounds on the eigenvalue range and on the field of values of non-Hermitian and indefinite finite element matrices}},
	volume = {189},
	year = {2006}}

@article{Mahan01,
	author = {Mahan, G. D.},
	doi = {10.1103/PhysRevB.63.155110},
	issue = {15},
	journal = {Phys. Rev. B},
	numpages = {5},
	pages = {155110},
	publisher = {American Physical Society},
	title = {{Energy bands of the Bethe lattice}},
	url = {https://link.aps.org/doi/10.1103/PhysRevB.63.155110},
	volume = {63},
	year = {2001}}

@article{Manna23,
	author = {Manna, S. and Roy, B.},
	date = {2023/01/16},
	doi = {10.1038/s42005-023-01130-2},
	id = {Manna2023},
	isbn = {2399-3650},
	journal = {Commun. Phys.},
	pages = {10},
	title = {{Inner skin effects on non-Hermitian topological fractals}},
	url = {https://doi.org/10.1038/s42005-023-01130-2},
	volume = {6},
	year = {2023}}

@article{Meden23,
	author = {V. Meden and L. Grunwald and D. M. Kennes},
	doi = {10.1088/1361-6633/ad05f3},
	journal = {Rep. Prog. Phys.},
	pages = {124501},
	publisher = {IOP Publishing},
	title = {{$\mathcal{PT}$-symmetric, non-Hermitian quantum many-body physics---a methodological perspective}},
	url = {https://dx.doi.org/10.1088/1361-6633/ad05f3},
	volume = {86},
	year = {2023}}

@article{Misra77,
	author = {Misra, B. and Sudarshan, E. C. G.},
	doi = {10.1063/1.523304},
	eprint = {https://pubs.aip.org/aip/jmp/article-pdf/18/4/756/19182345/756¥_1¥_online.pdf},
	issn = {0022-2488},
	journal = {J. Math. Phys.},
	pages = {756},
	title = {{The Zeno's paradox in quantum theory}},
	url = {https://doi.org/10.1063/1.523304},
	volume = {18},
	year = {1977}}

@misc{Nakabayashi25,
      title={Fractional decay in the spontaneous emission of a two-level system}, 
      author={H. Nakabayashi and H. Kinkawa and T. Taira and N. Hatano},
      year={2025},
      eprint={2512.13817},
      archivePrefix={arXiv},
      primaryClass={quant-ph},
      url={https://arxiv.org/abs/2512.13817}, 
}

@article{Nakazato96-1,
	author = {H. Nakazato and M. Namiki and S. Pascazio},
	doi = {10.1142/S0217979296000118},
	eprint = {https://doi.org/10.1142/S0217979296000118},
	journal = {Int. J. Mod. Phys. B},
	pages = {247},
	title = {Temporal behavior of quantum mechanical systems},
	url = {https://doi.org/10.1142/S0217979296000118},
	volume = {10},
	year = {1996}}

@article{Nakazato96-2,
	author = {H. Nakazato and M. Namiki and S. Pascazio and H. Rauch},
	doi = {https://doi.org/10.1016/0375-9601(96)00350-7},
	issn = {0375-9601},
	journal = {Phys. Lett. A},
	pages = {203},
	title = {{Understanding the quantum Zeno effect}},
	url = {https://www.sciencedirect.com/science/article/pii/0375960196003507},
	volume = {217},
	year = {1996}}

@article{Nishino24,
	author = {A. Nishino and N. Hatano},
	doi = {10.1088/1751-8121/ad4d31},
	journal = {J. Phys. A},
	pages = {245302},
	publisher = {IOP Publishing},
	title = {Exact time-evolving scattering states in open quantum-dot systems with an interaction: discovery of time-evolving resonant states},
	url = {https://dx.doi.org/10.1088/1751-8121/ad4d31},
	volume = {57},
	year = {2024}}

@article{NoneqReview24,
	author = {{\v S}pi{\v c}ka, V. and Keefe, P. D. and Nieuwenhuizen, T. M.},
	date = {2023/12/01},
	doi = {10.1140/epjs/s11734-023-01072-4},
	id = {{\v S}pi{\v c}ka2023},
	isbn = {1951-6401},
	journal = {Eur. Phys. J. Spec. Top.},
	pages = {3185},
	title = {Non-equilibrium quantum physics, many body systems, and foundations of quantum physics},
	url = {https://doi.org/10.1140/epjs/s11734-023-01072-4},
	volume = {232},
	year = {2023}}

@article{Novo16,
	author = {Novo, L. and Mohseni, M. and Omar, Y.},
	da = {2016/01/04},
	doi = {10.1038/srep18142},
	id = {Novo2016},
	isbn = {2045-2322},
	journal = {Sci. Rep.},
	pages = {18142},
	title = {{Disorder-assisted quantum transport in suboptimal decoherence regimes}},
	ty = {JOUR},
	url = {https://doi.org/10.1038/srep18142},
	volume = {6},
	year = {2016}}

@article{Plenio98,
	author = {Plenio, M. B. and Knight, P. L.},
	doi = {10.1103/RevModPhys.70.101},
	issue = {1},
	journal = {Rev. Mod. Phys.},
	numpages = {0},
	pages = {101},
	publisher = {American Physical Society},
	title = {The quantum-jump approach to dissipative dynamics in quantum optics},
	url = {https://link.aps.org/doi/10.1103/RevModPhys.70.101},
	volume = {70},
	year = {1998}}

@MastersThesis{ShimomuraPrivate24,
    author = "K. Shimomura",
    title = {},
    type = "Master's thesis",
    school = "Kyoto University",
    year = 2024,
}

@article{Sun24,
  title = {{Non-Hermitian quantum fractals}},
  author = {Sun, J. and Li, C.-A. and Guo, Q. and Zhang, W. and Feng, S. and Zhang, X. and Guo, H. and Trauzettel, B.},
  journal = {Phys. Rev. B},
  volume = {110},
  issue = {20},
  pages = {L201103},
  numpages = {6},
  year = {2024},
  publisher = {American Physical Society},
  doi = {10.1103/PhysRevB.110.L201103},
  url = {https://link.aps.org/doi/10.1103/PhysRevB.110.L201103}
}

@article{sutherland1986localization,
	author = {B. Sutherland},
	journal = {Phys. Rev. B},
	pages = {5208},
	publisher = {APS},
	title = {Localization of electronic wave functions due to local topology},
	volume = {34},
	year = {1986}}

@article{Syassen08,
	author = {N. Syassen and D. M. Bauer and M. Lettner and T. Volz and D. Dietze and J. J. Garc{\'{i}}a-Ripoll and J. I. Cirac and G. Rempe and S. D\"{u}rr},
	doi = {10.1126/science.1155309},
	eprint = {https://www.science.org/doi/pdf/10.1126/science.1155309},
	journal = {Science},
	pages = {1329},
	title = {Strong Dissipation Inhibits Losses and Induces Correlations in Cold Molecular Gases},
	url = {https://www.science.org/doi/abs/10.1126/science.1155309},
	volume = {320},
	year = {2008}}

@unpublished{Taira24,
	archiveprefix = {arXiv},
	author = {T. Taira and N. Hatano and A. Nishino},
	eprint = {2406.17436},
	note = {arXiv:2406.17436},
	title = {{Markovianity and non-Markovianity of Particle Bath with Dirac Dispersion Relation}},
	year = {2024}}

@book{tasaki2020physics,
	address = {Cham},
	author = {Tasaki, H.},
	publisher = {Springer},
	title = {Physics and Mathematics of Quantum Many-Body Systems},
	year = {2020}}

@article{Theodorou76,
	author = {G. Theodorou and M. H. Cohen},
	journal = {Phys. Rev. B},
	pages = {4597},
	title = {{Extended states in a one-dimensional system with off-diagonal disorder}},
	volume = {13},
	year = {1976}}

@article{Toroker09,
	author = {M. C. Toroker and U. Peskin},
	doi = {10.1088/0953-4075/42/4/044013},
	journal = {J. Phys. B},
	pages = {044013},
	title = {On the relation between steady-state currents and resonance states in molecular junctions},
	url = {https://dx.doi.org/10.1088/0953-4075/42/4/044013},
	volume = {42},
	year = {2009}}

@article{Wiersig18,
	author = {Wiersig, J.},
	doi = {10.1103/PhysRevA.98.052105},
	issue = {5},
	journal = {Phys. Rev. A},
	numpages = {7},
	pages = {052105},
	publisher = {American Physical Society},
	title = {Role of nonorthogonality of energy eigenstates in quantum systems with localized losses},
	url = {https://link.aps.org/doi/10.1103/PhysRevA.98.052105},
	volume = {98},
	year = {2018}}

@article{YaoWang18,
	author = {Yao, S. and Wang, Z.},
	doi = {10.1103/PhysRevLett.121.086803},
	issue = {8},
	journal = {Phys. Rev. Lett.},
	numpages = {8},
	pages = {086803},
	publisher = {American Physical Society},
	title = {{Edge States and Topological Invariants of Non-Hermitian Systems}},
	url = {https://link.aps.org/doi/10.1103/PhysRevLett.121.086803},
	volume = {121},
	year = {2018}}

@article{Ziman82,
	author = {T. A. L. Ziman},
	journal = {Phys. Rev. Lett.},
	pages = {337},
	title = {{Localization and Spectral Singularities in Random Chains}},
	volume = {49},
	year = {1982}}

@book{Datta95,
	author = {Datta, S.},
	collection = {Cambridge Studies in Semiconductor Physics and Microelectronic Engineering},
	doi = {10.1017/CBO9780511805776},
	place = {Cambridge},
	publisher = {Cambridge University Press},
	series = {Cambridge Studies in Semiconductor Physics and Microelectronic Engineering},
	title = {{{Electronic Transport in Mesoscopic Systems}}},
	year = {1995}}

\end{document}